\documentclass[acmsmall]{acmart}

\usepackage{tabularx}
\usepackage{hyperref}
\usepackage{array}
\usepackage{makecell}
\usepackage{tablefootnote}

\AtBeginDocument{%
  \providecommand\BibTeX{{%
    \normalfont B\kern-0.5em{\scshape i\kern-0.25em b}\kern-0.8em\TeX}}}

\setcopyright{acmcopyright}
\copyrightyear{2021}
\acmYear{2022}
\acmDOI{[DOI]}

\acmConference[CSCW '22]{CSCW '22: }{November 12--16, 2022}{Taipei, Taiwan}
\acmBooktitle{Kaviani and Salehi '21: Proceedings of the ACM on Human Computer Interaction,
  November 12—16, 2022, Taipei, Taiwan}
\acmPrice{15.00}
\acmISBN{978-1-4503-XXXX-X/21/XX}

\acmSubmissionID{XXX-XXX-XXX}

\begin{document}


\title[Instagram Infographics in U.S. Ethnic Movements]{Bridging Action Frames: Instagram Infographics in U.S. Ethnic Movements}

%
\author{Darya Kaviani}
\email{daryakaviani@berkeley.edu}
\affiliation{%
  \institution{UC Berkeley EECS \& Ethnic Studies}
  \city{Berkeley}
  \state{CA}
  \country{United States}
  \postcode{94720}
}
\author{Niloufar Salehi}
\email{nsalehi@berkeley.edu}
\affiliation{%
  \institution{UC Berkeley School of Information}
  \city{Berkeley}
  \state{CA}
  \country{United States}
  \postcode{94720}
}

%
\renewcommand{\shortauthors}{Darya Kaviani, Niloufar Salehi}

%
\begin{abstract}
Instagram infographics are a digital activism tool that have redefined action frames for technology-facilitated social movements. From the 1960s through the 1980s, United States ethnic movements practiced collective action: ideologically unified, resource-intensive traditional activism. Today, technologically enabled movements have been categorized as practicing connective action: individualized, low-resource online activism. Yet, we argue that Instagram infographics are both connective \textit{and} collective. This paper juxtaposes the insights of past and present U.S. ethnic movement activists and analyzes Black Lives Matter Instagram data over the course of 7 years (2014-2020). We find that Instagram infographic activism bridges connective and collective action in three ways: (1) \textit{Scope for Education}: Visually enticing and digestible infographics reduce the friction of information dissemination, facilitating collective movement education while preserving customizability. (2) \textit{Reconciliation for Credibility}: Activists use connective features to combat infographic misinformation and resolve internal differences, creating a trusted collective movement front. (3) \textit{High-Resource Efforts for Transformative Change}: Instagram infographic activism has been paired with boots on the ground and action-oriented content, curating a connective-to-collective pipeline that expends movement resources. Our work unveils the vitality of evaluating digital activism action frames at the movement integration level, exemplifies the powerful coexistence of connective and collective action, and offers meaningful design implications for activists seeking to leverage this novel tool.
\end{abstract}

%
%


%

\maketitle

\begin{figure}[h]
  \caption{Black Lives Matter Instagram Infographics (blklivesmatter) \protect\footnotemark}
  \centering
  \includegraphics[width=3.4cm]{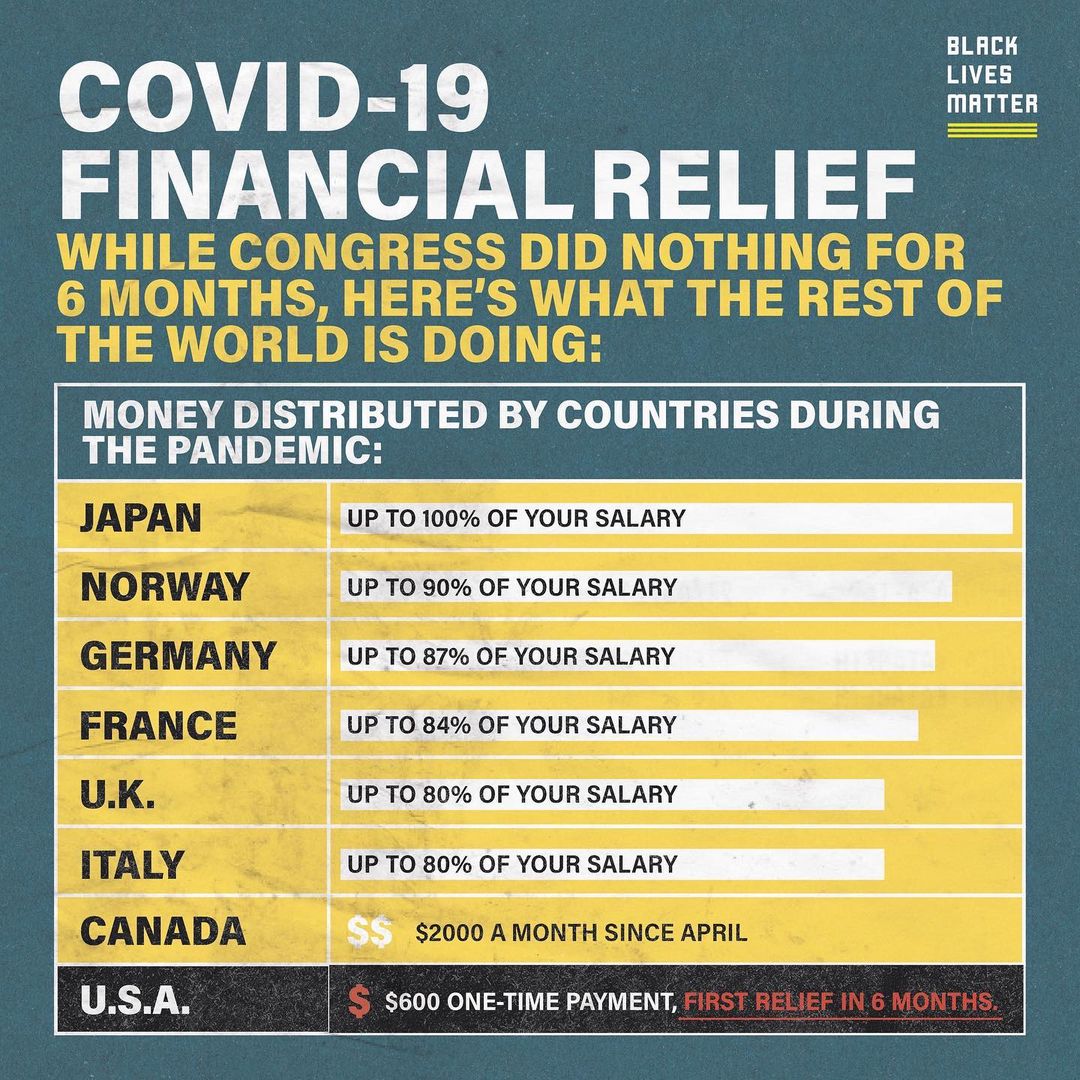}
  \includegraphics[width=3.4cm]{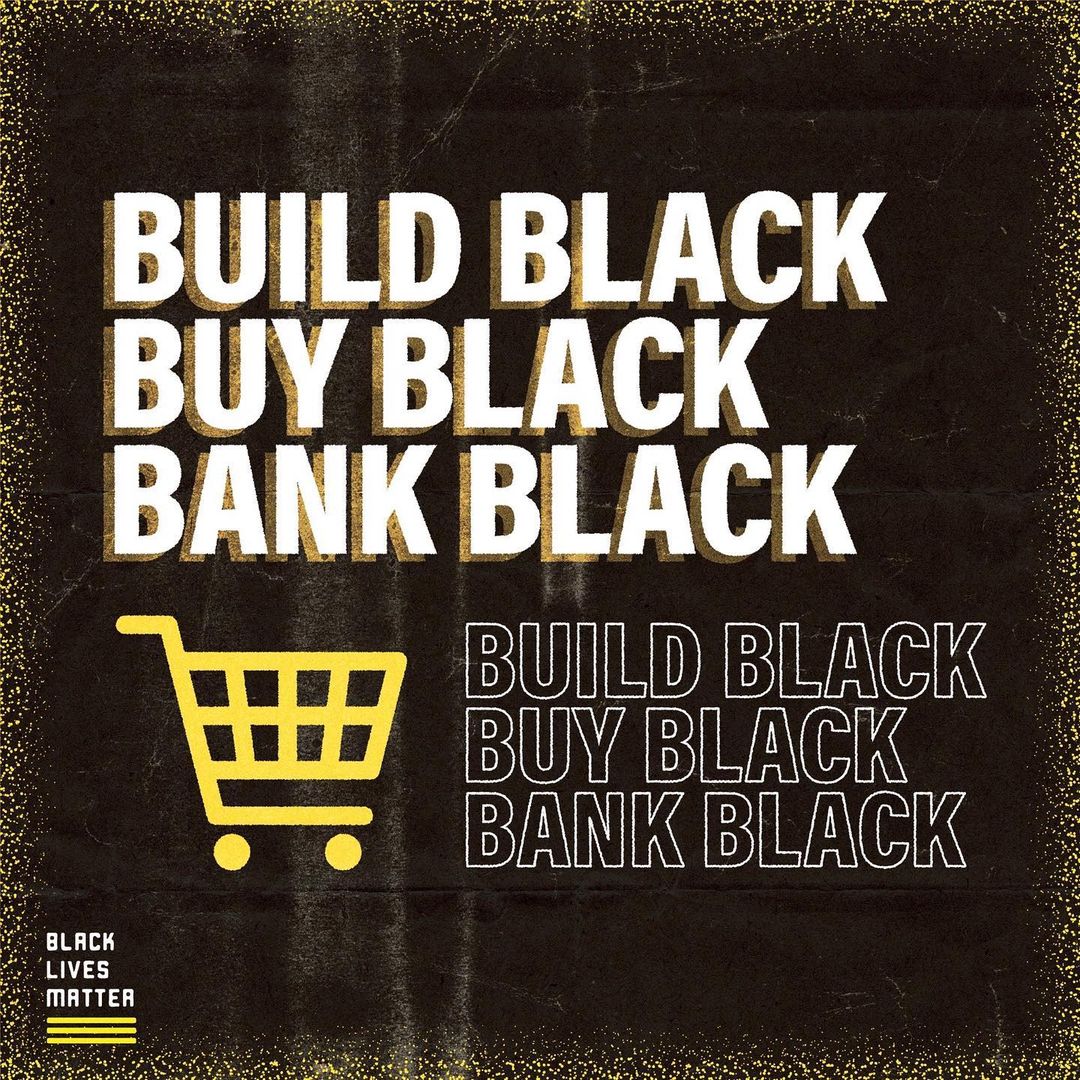}
  \includegraphics[width=3.4cm]{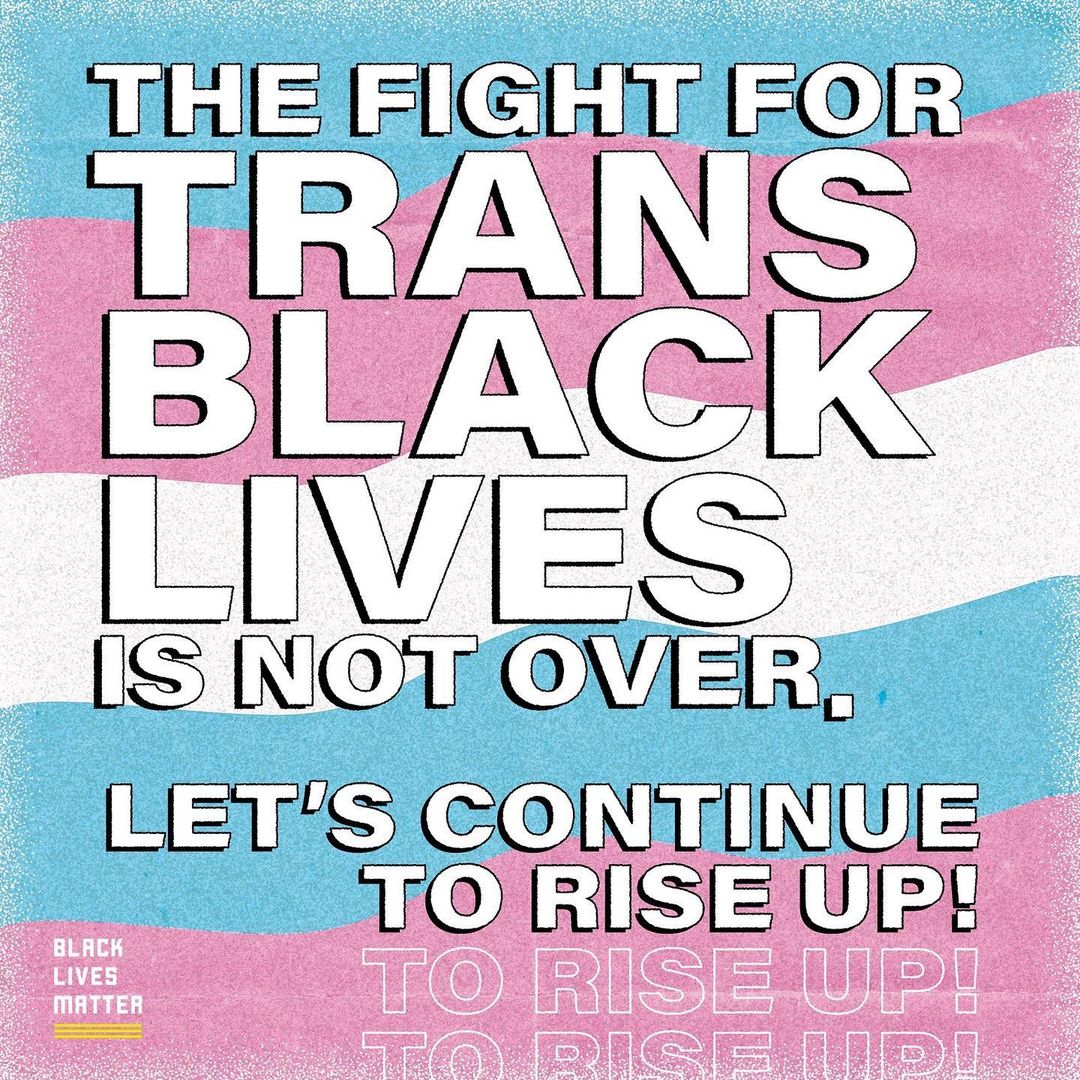}
  \includegraphics[width=3.4cm]{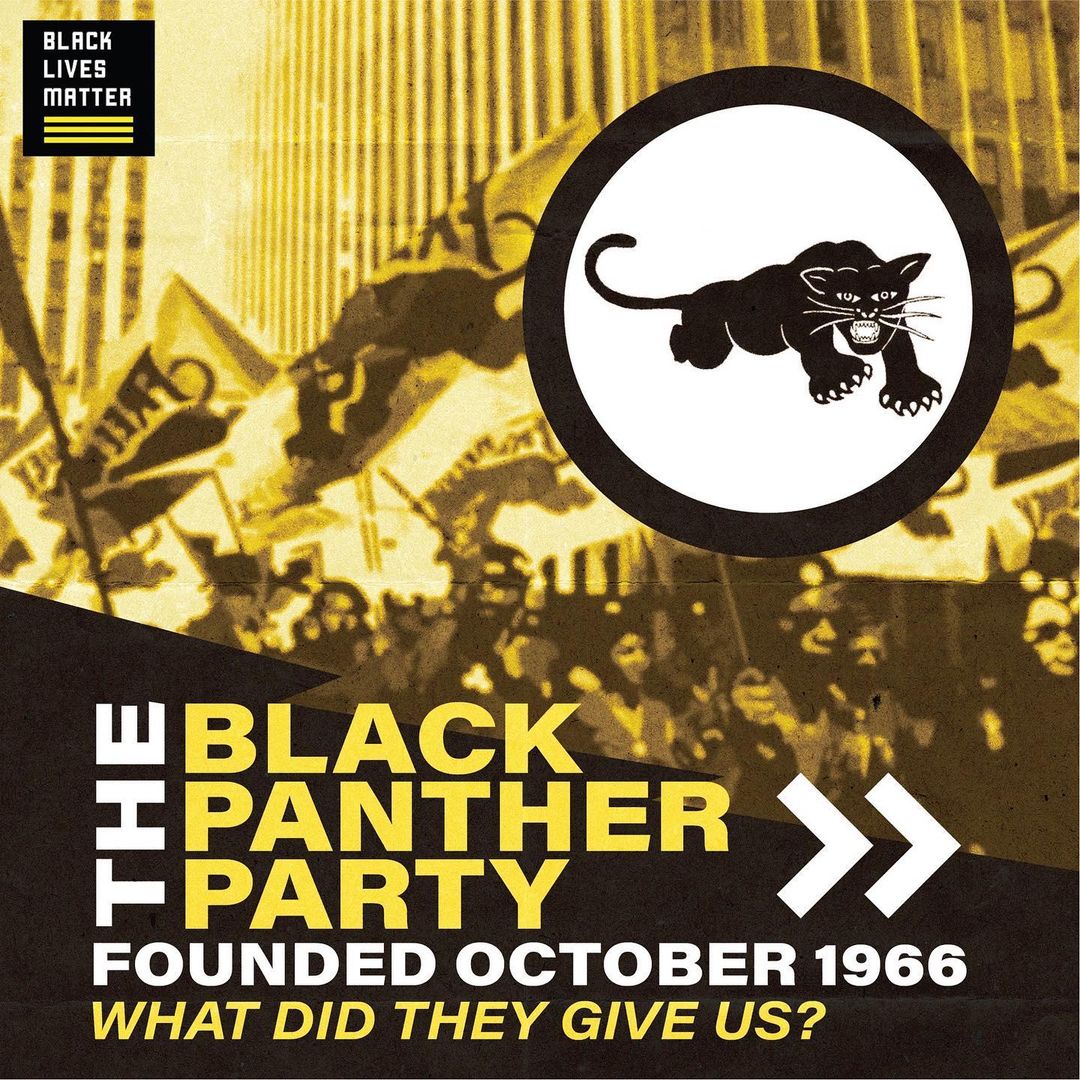}
  \includegraphics[width=3.4cm]{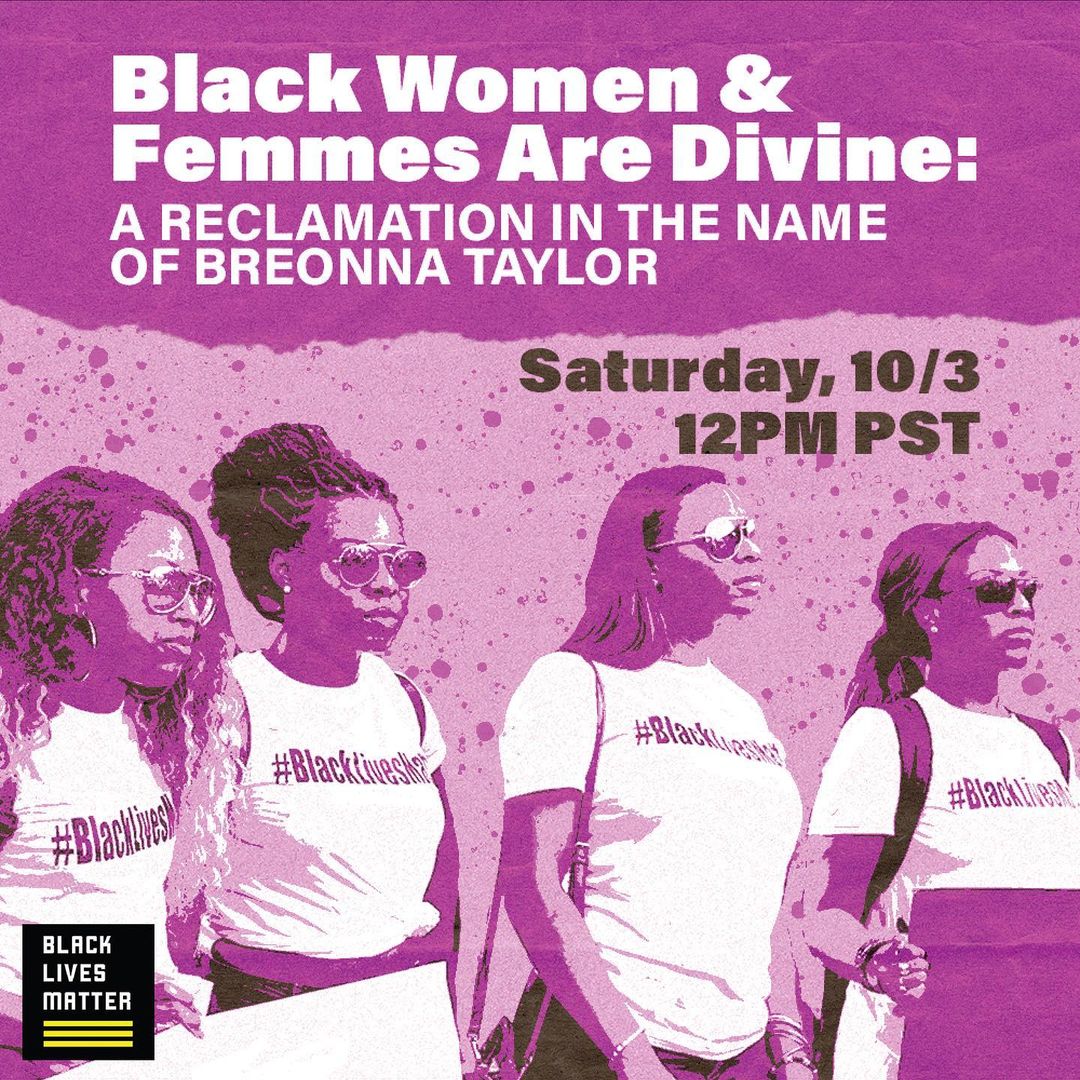}
  \includegraphics[width=3.4cm]{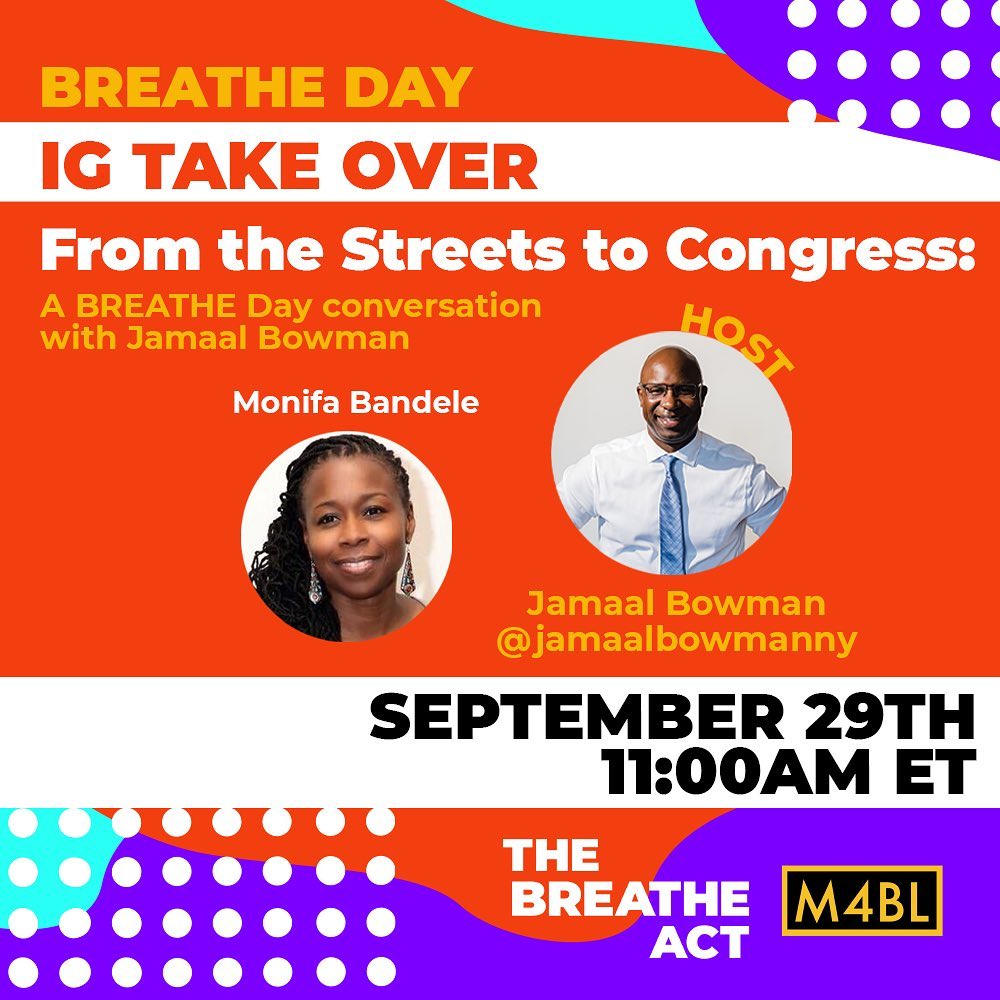}
    \includegraphics[width=3.4cm]{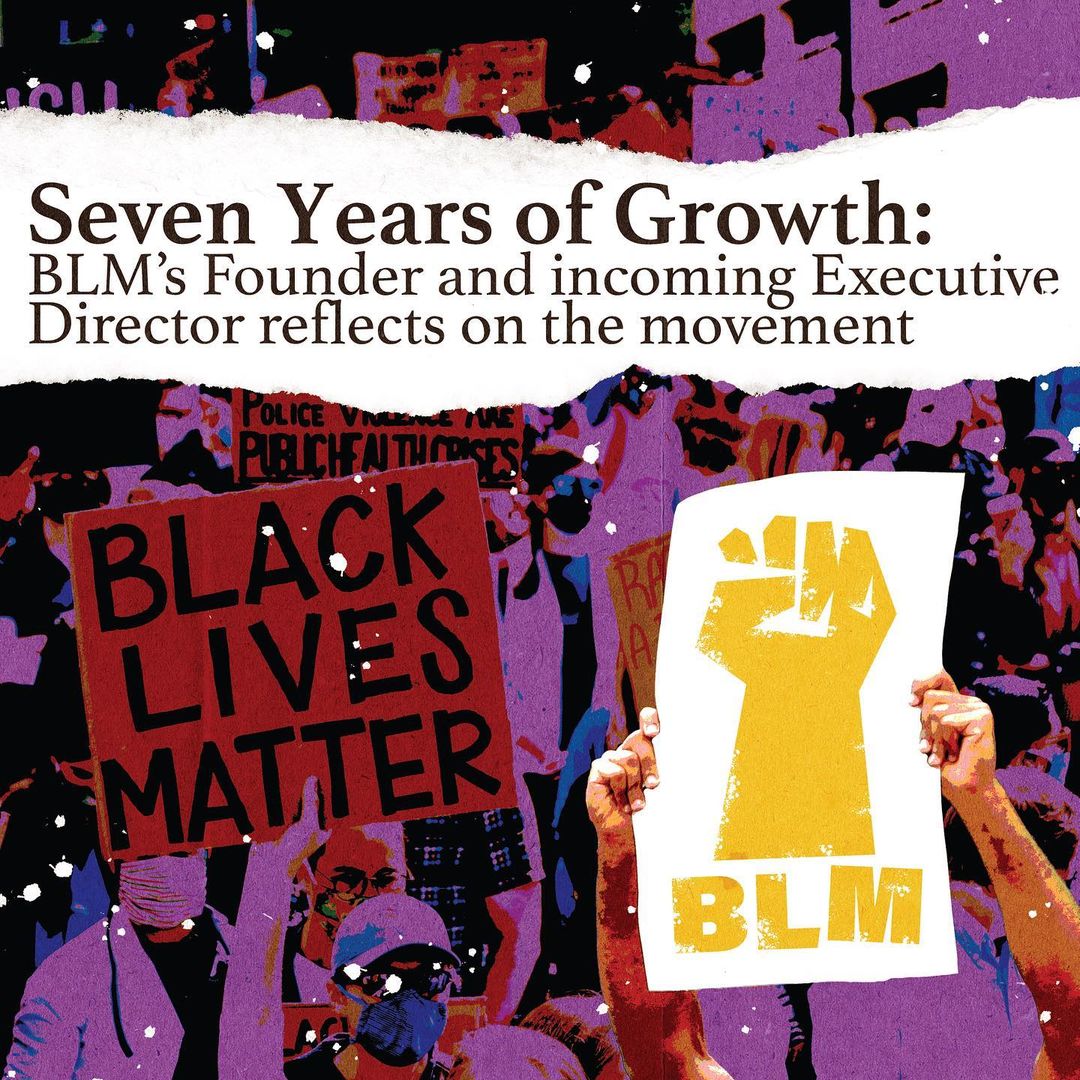}
  \includegraphics[width=3.4cm]{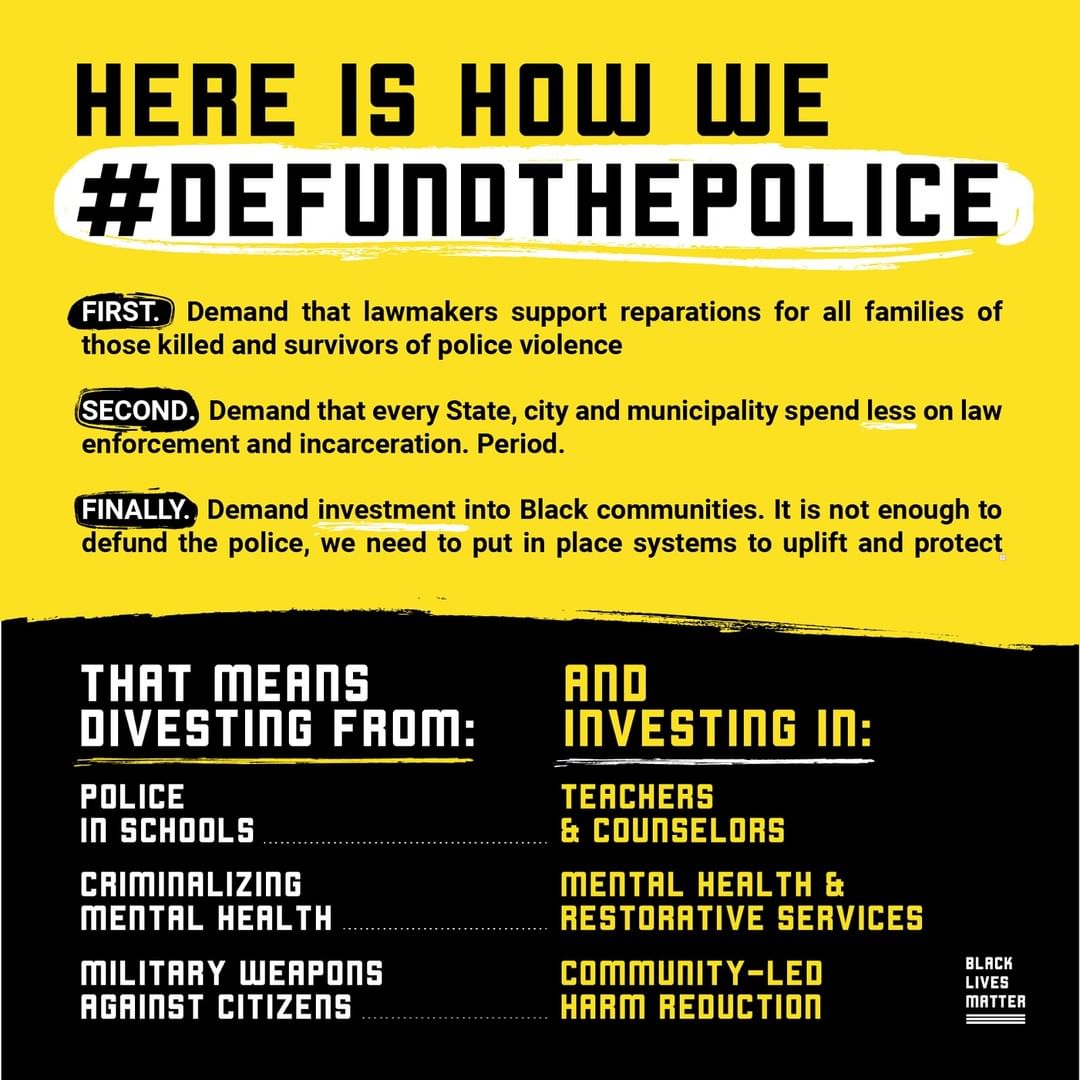}
    \includegraphics[width=3.4cm]{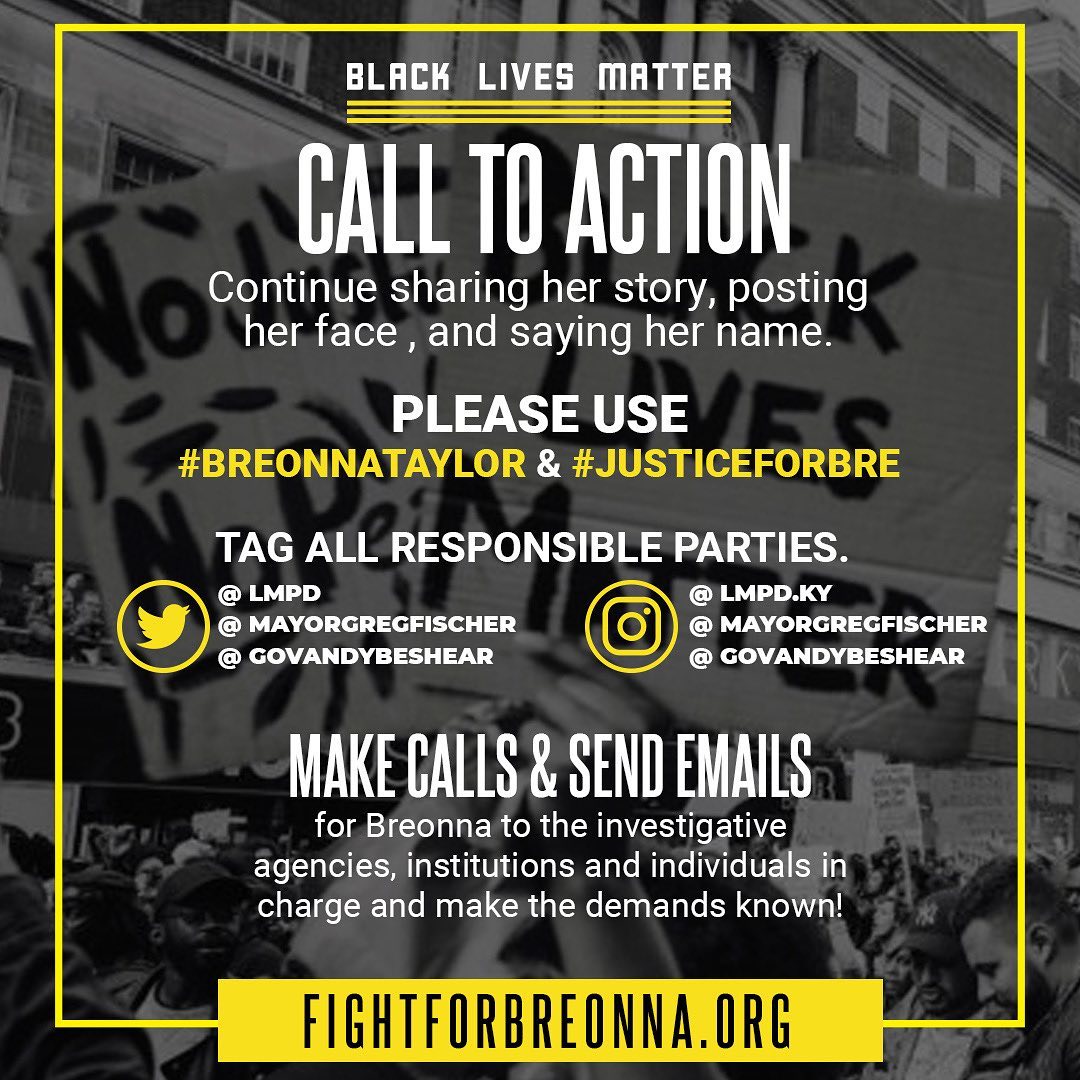}
  \includegraphics[width=3.4cm]{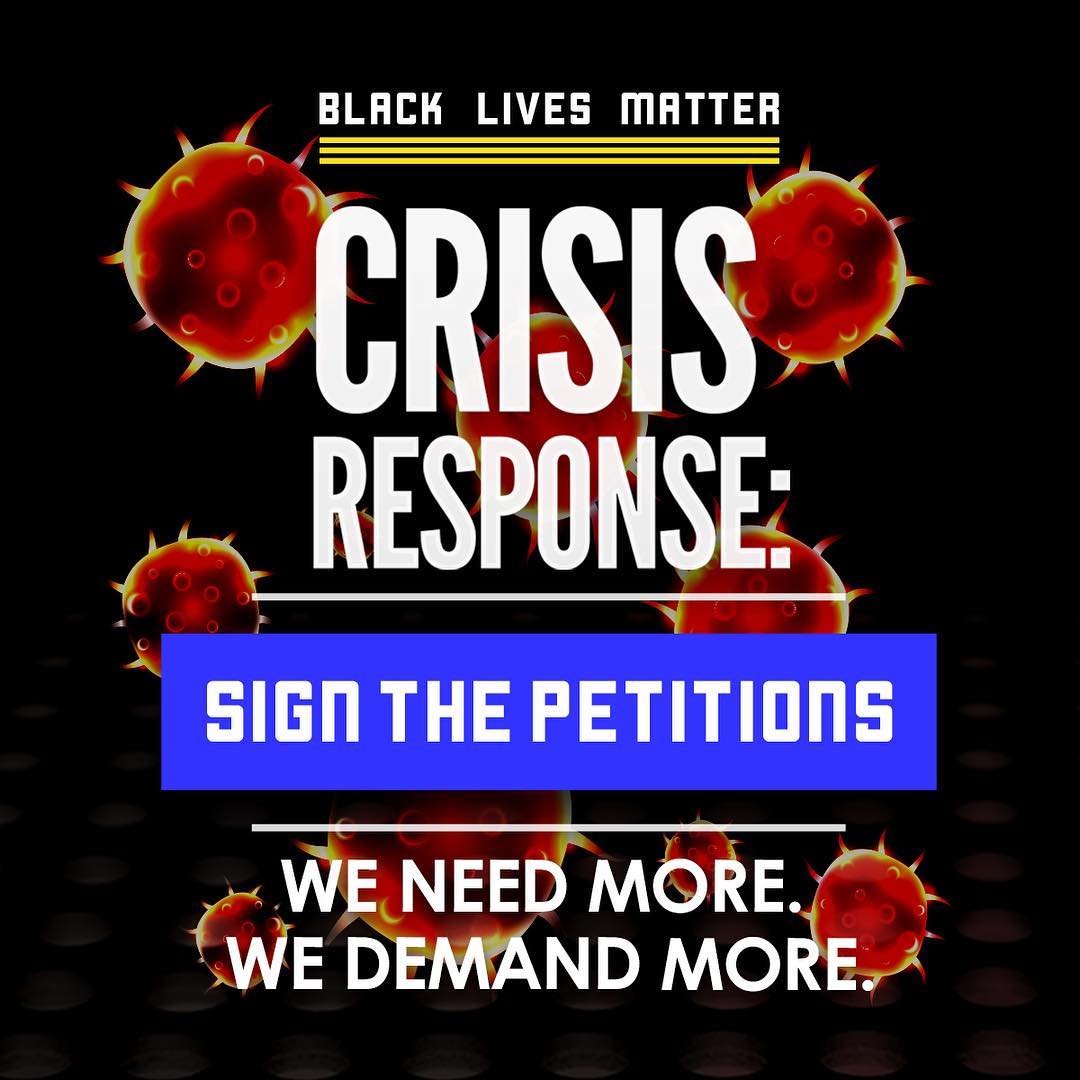}
  \includegraphics[width=3.4cm]{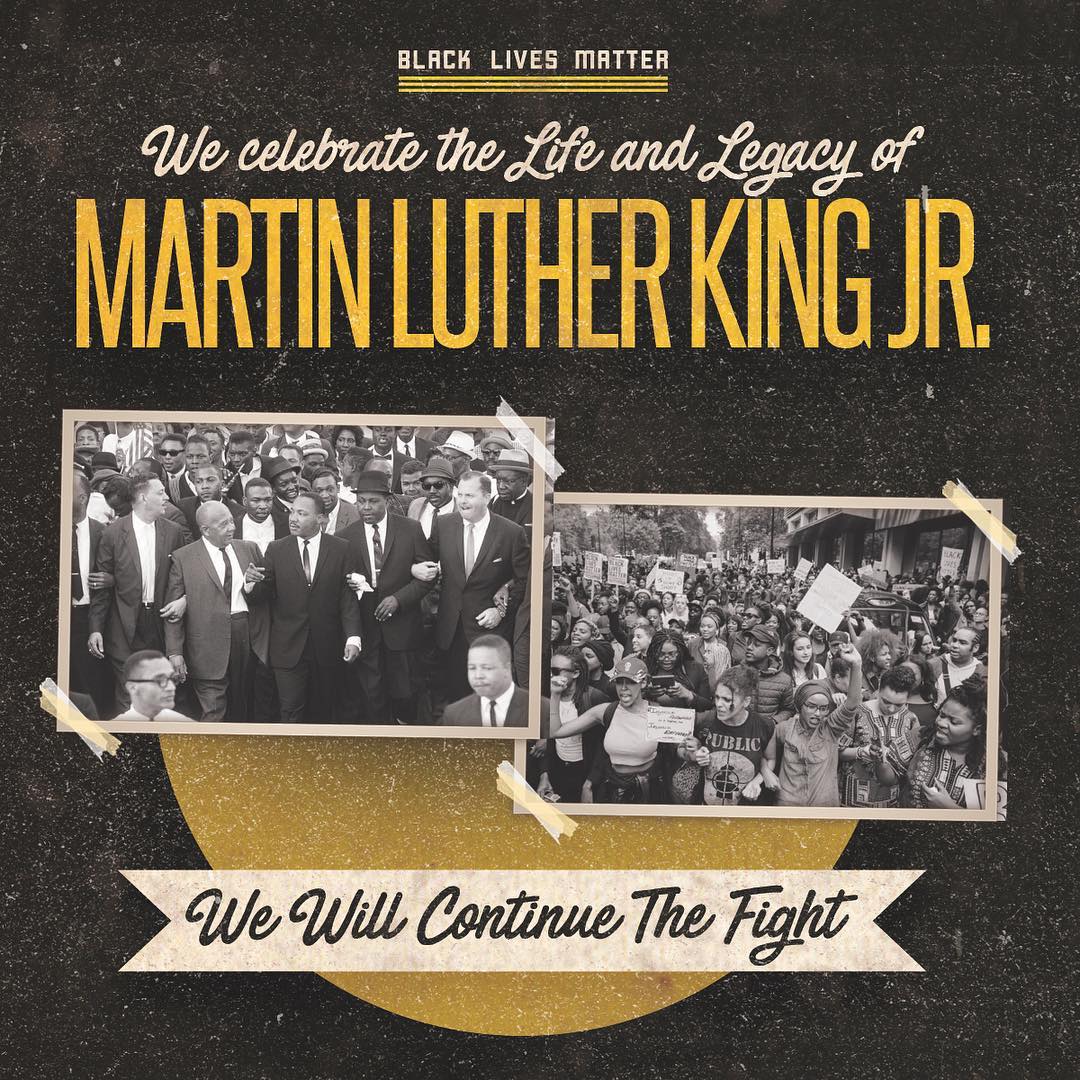}
  \includegraphics[width=3.4cm]{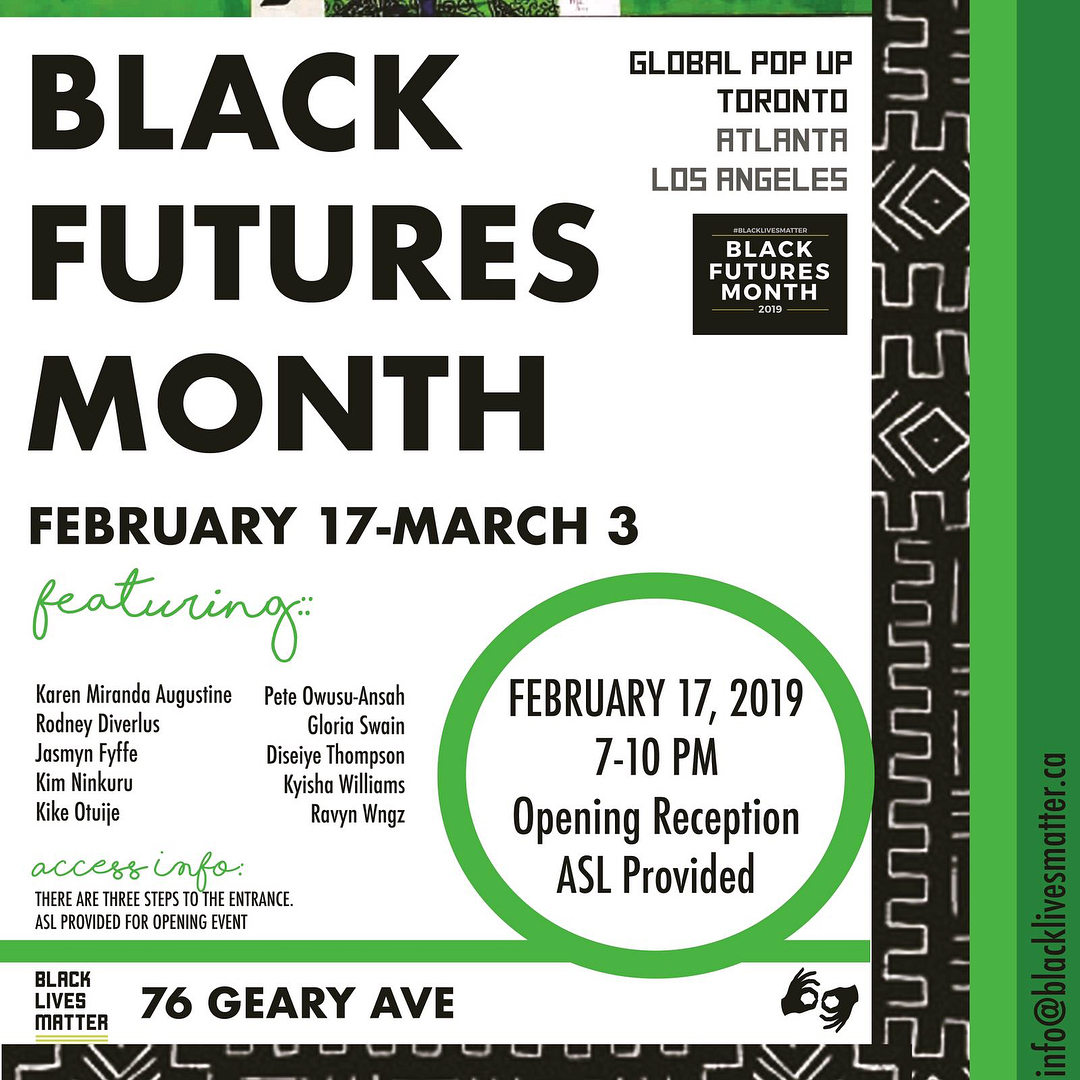}
  \includegraphics[width=3.4cm]{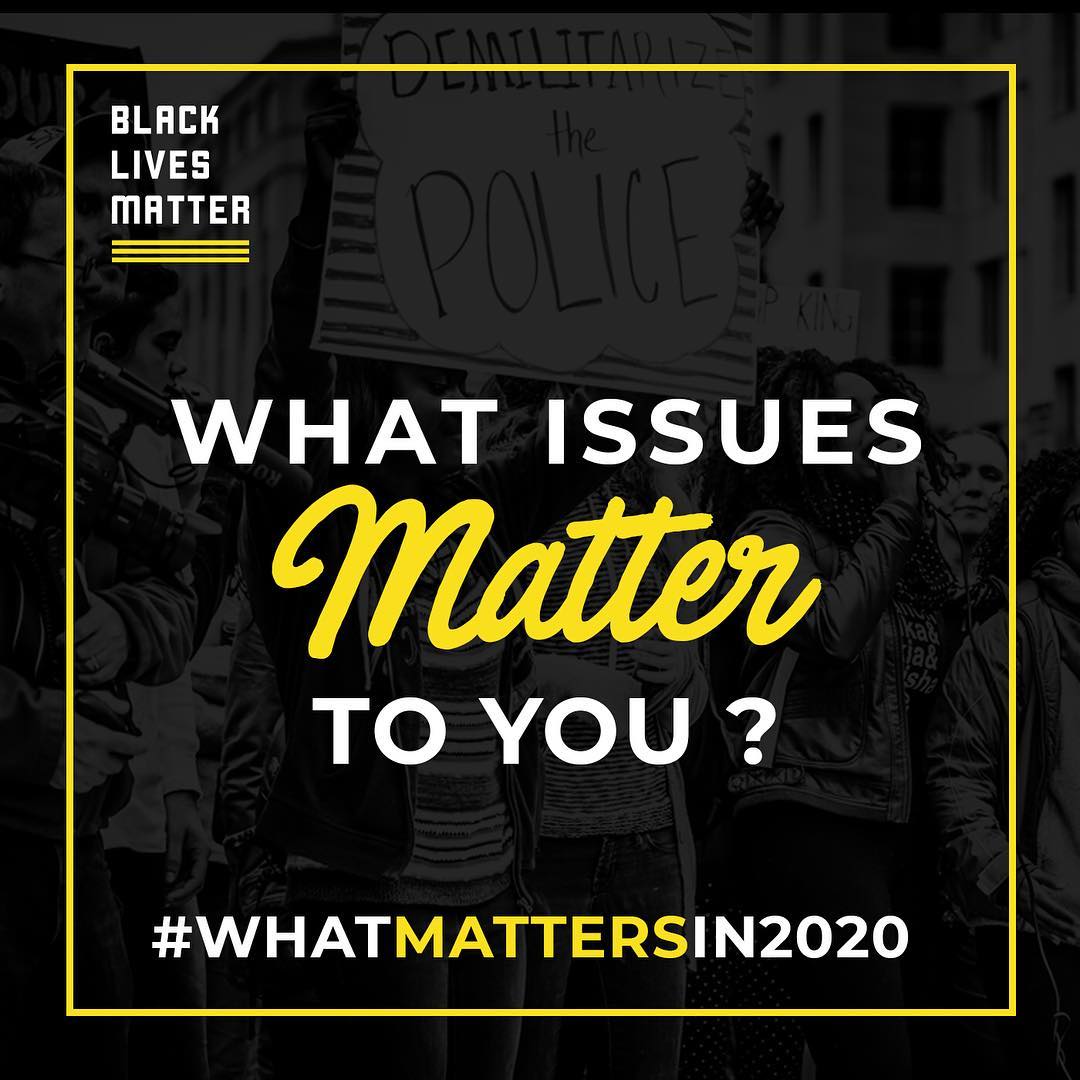}
    \includegraphics[width=3.4cm]{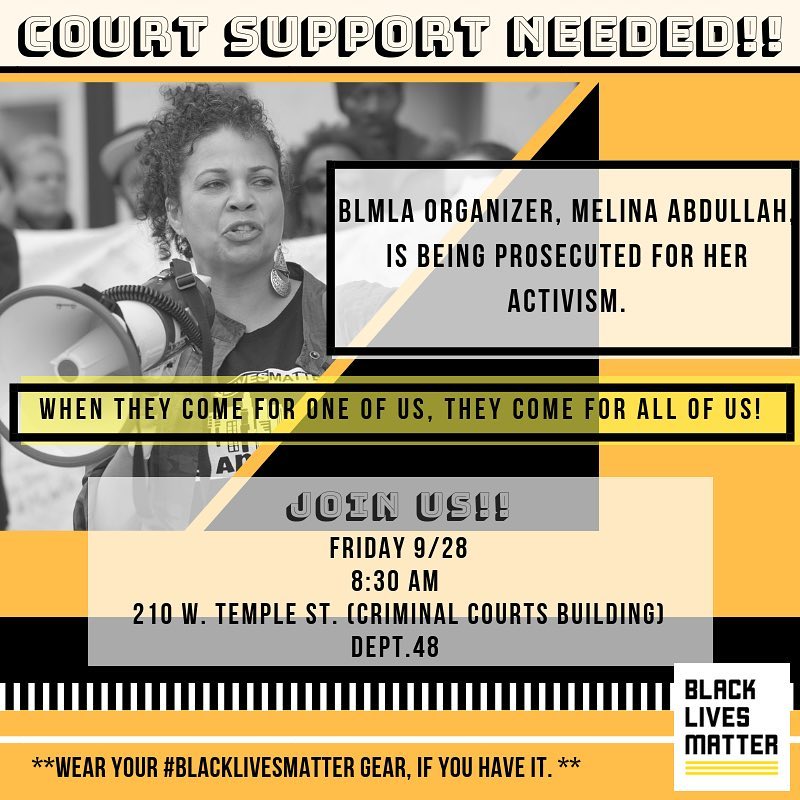}
    \includegraphics[width=3.4cm]{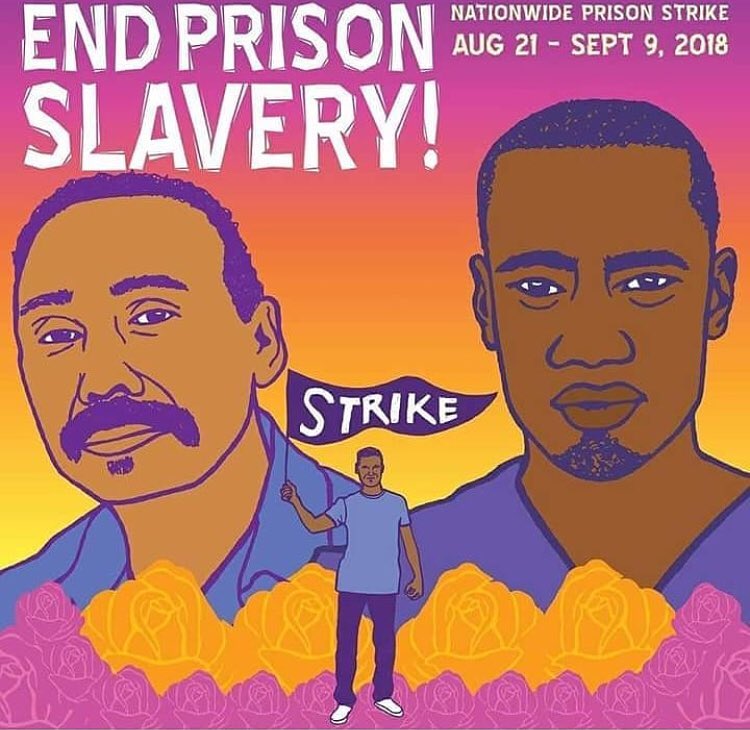}
    \includegraphics[width=3.4cm]{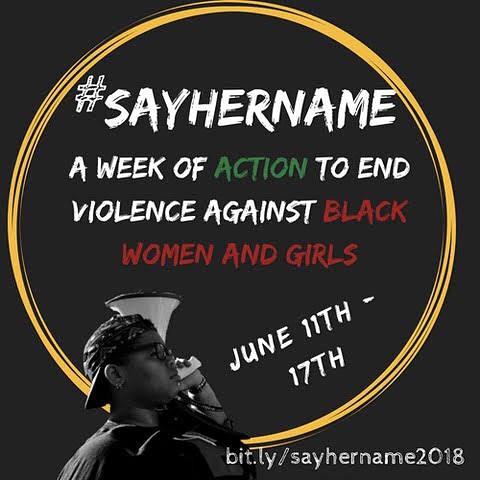}
    \includegraphics[width=3.4cm]{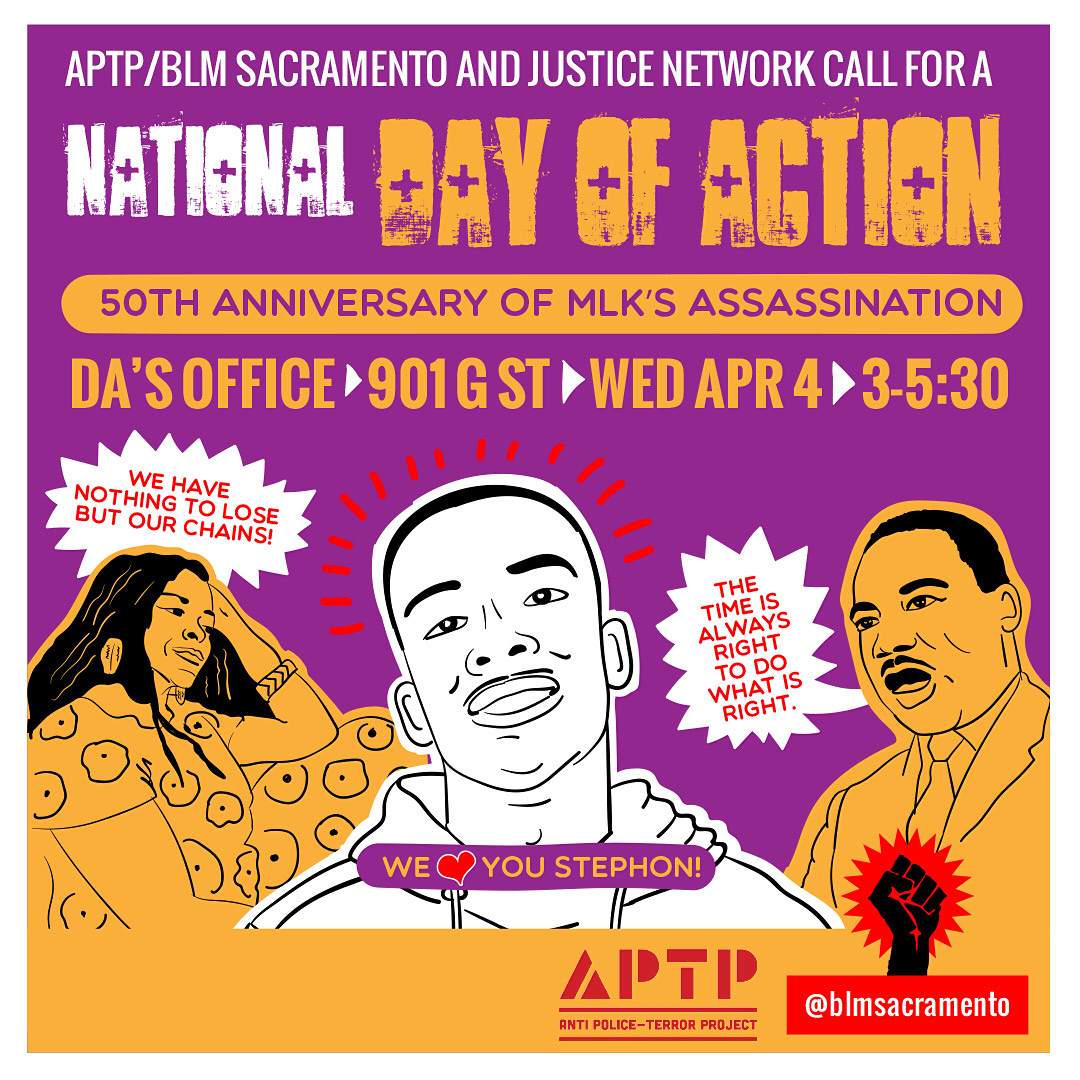}
    \includegraphics[width=3.4cm]{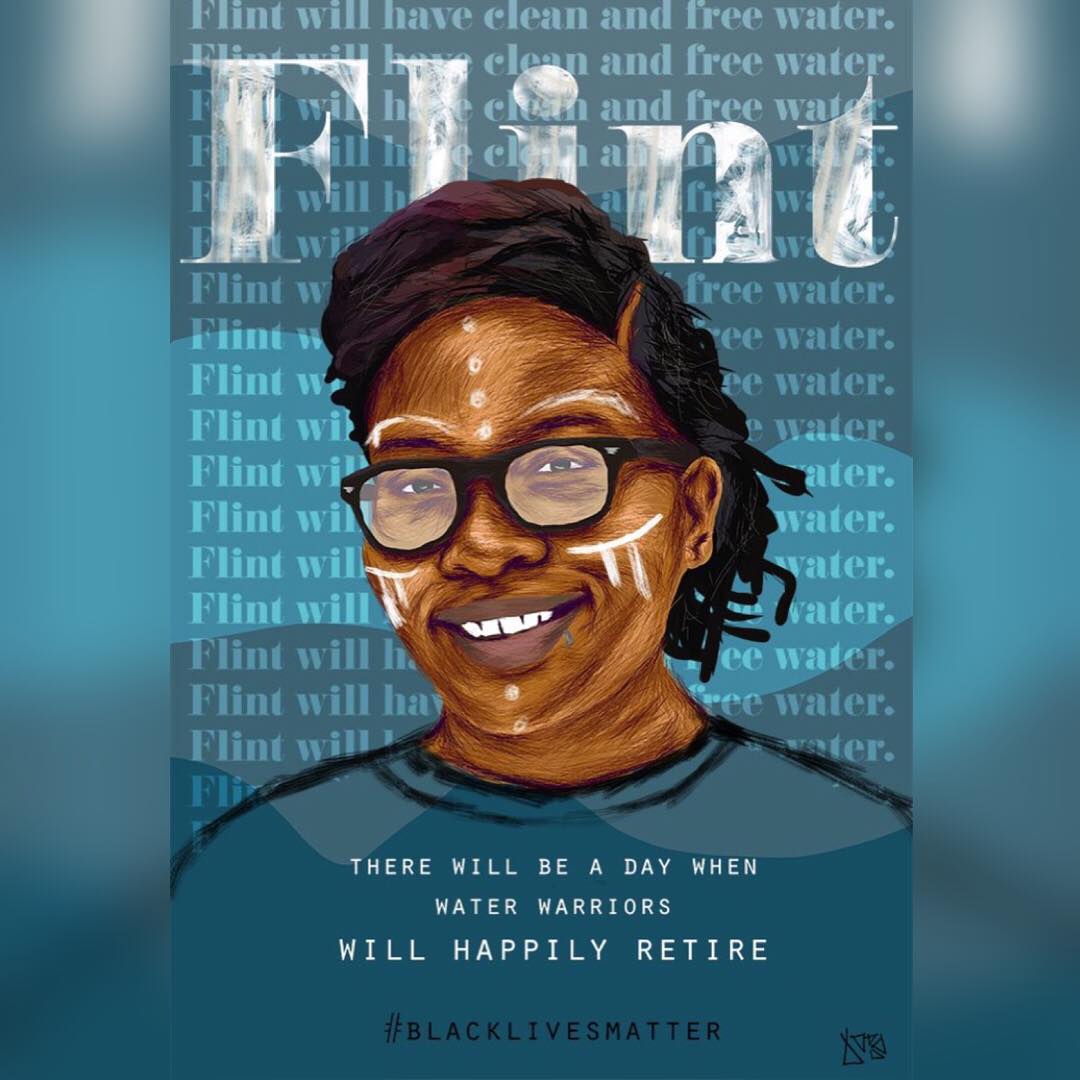}
    \includegraphics[width=3.4cm]{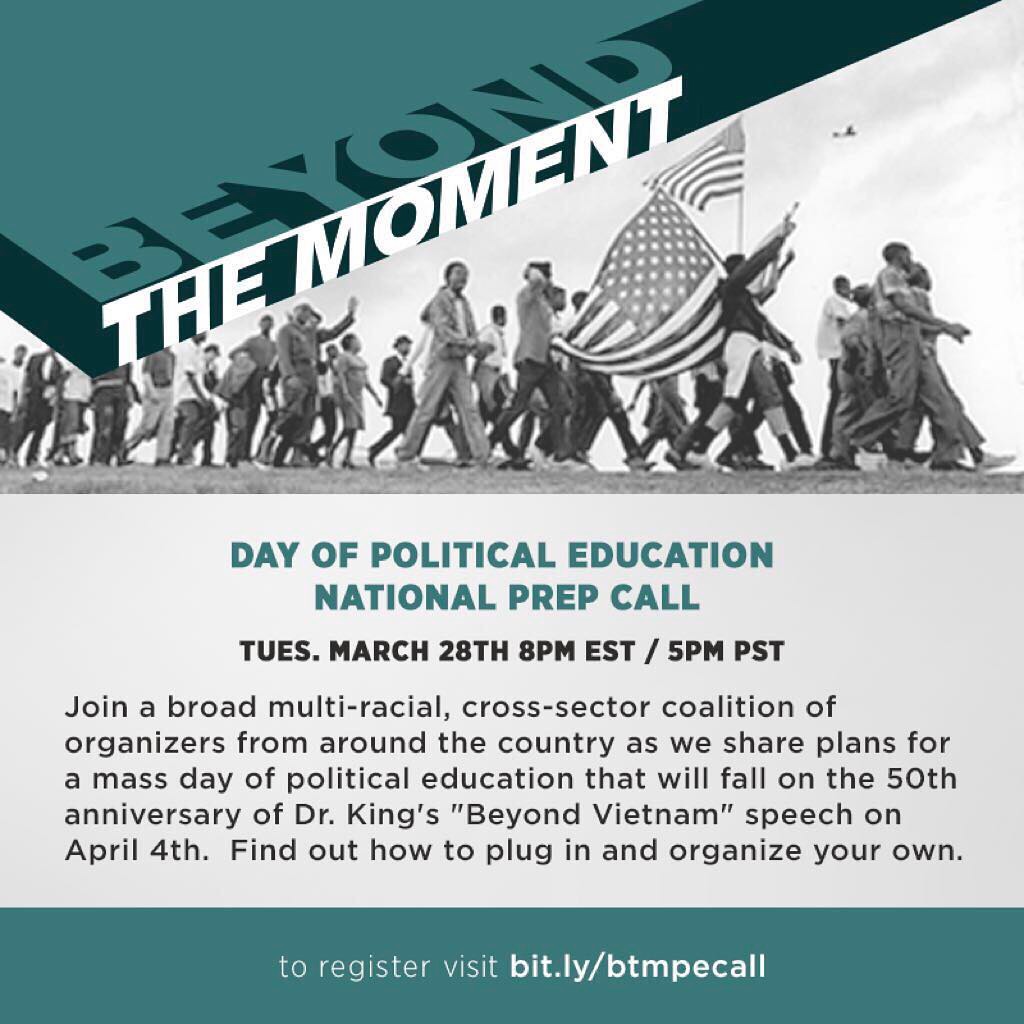}
    \includegraphics[width=3.4cm]{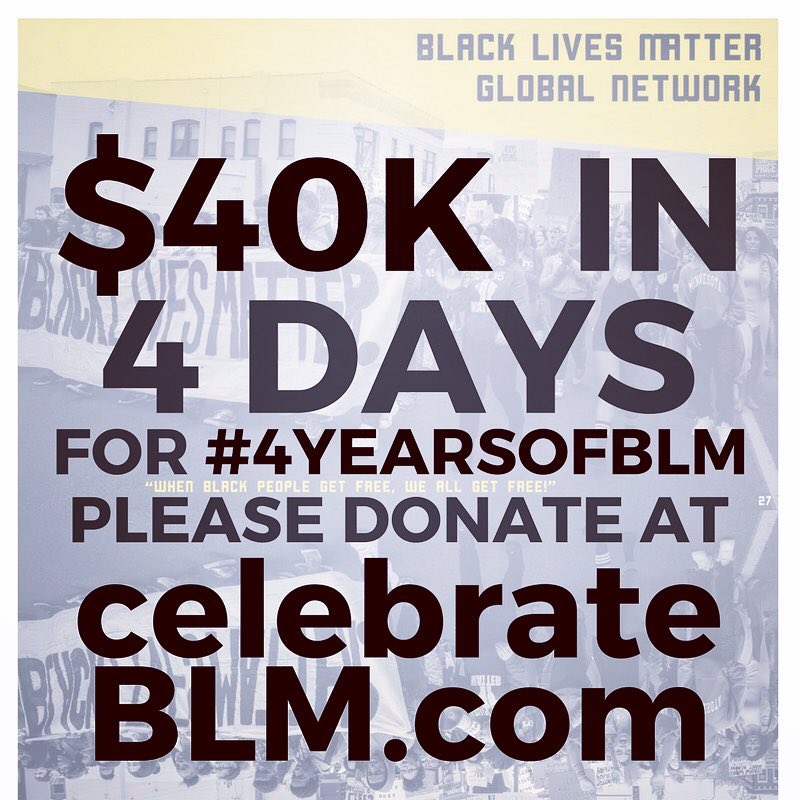}\\
\end{figure}
\footnotetext{Image content descriptions and citations are in A.1.}

\section{Introduction}

At the height of the COVID-19 pandemic, American activist Carmen Perez organized and trained 35,000 New York City-based Black Lives Matter protesters within just four days, all while located in California herself. Despite the protest’s immediacy, scale, and location, Perez’s team succeeded by leveraging virtual tools, most notably Instagram infographics. This was far from an isolated event: In the summer of 2020, spearheaded by the Black Lives Matter Movement\footnote{See 2.2.}, the social media platform Instagram erupted in infographics fighting for the protection of Black life through anti-racism resources and action items such as petitions, financial contribution routes, and protest information. While only 0.75\% of images posted on the Black Lives Matter Instagram account (@blklivesmatter) were infographics in 2014, this value soared and peaked at 62.25\% in 2020 (Table 3, Figure 4).

Historically, social movements have centered collective action: individuals unifying to contribute to a common cause, typically on the basis of a joint ideology and facilitated by organizational resources. Such ideologies manifest as action frames: the ``beliefs and meanings that inspire and legitimate'' action, or the guiding principles that frame movement processes and activities \cite{Benford00}. Facets of collective action include organization-wide coordination, social network oversight, unified movement action frames, inter-personal relationships, the reconciliation of internal differences, and high resource expenditure \cite{Bennett12}. A prime historical manifestation of collective action is ethnic movements: social movements that are organized by groups made distinct by their origin, culture, language, religion, territory, or phenotype \cite{Okamoto13}. Their objective is self-determination and a transformation of society’s viewpoint on cultural, political, and social issues. Often regarded as the Second Revolution following the Civil Rights Movement, the era from the 1960s through the 1980s was a breeding ground for U.S. ethnic movements such as the American Indian Movement, Black Power Movement, Asian American Movement, and Chicano American Movement \cite{SecondRevolution}. Second Revolution ethnic movements epitomize collective action by relying on organizational resources and presenting a unified front. 

In 2012, however, Bennett and Segerberg defined an alternate action formation known as connective action: sharing personalized content across social media networks and digital connections \cite{Bennett12}. Facets of connective action include a lack of coordination, personalized content and expression, individual action frames in movement communication, and minimal organizational operations. Connective and collective action were established as fundamentally different types of action formations \cite{Bennett12}. While collective action relies on group ties, organizational resources, common action frames, and a resolution of internal differences, connective action renders social good contributions as acts of personal expression. This framework suggests that connective action uses technological processes to yield action without the need for collective identity framing or organizational resources. 

Many modern-day ethnic movement activists have begun heavily utilizing digital media in their activism. In 2020, activists from across the world supported the Black Lives Matter movement by sharing Instagram posts in the form of an infographic: a ``visual presentation of information in the form of a chart, graph, or other image accompanied by minimal text, intended to give an easily understood overview, often of a complex subject'' \cite{Infographics-Def}. Figure 1 depicts several infographics that were posted by the Black Lives Matter Instagram account (@blklivesmatter). Propelled by the Black Lives Matter movement’s posts, ethnic movement activists across the platform began to centralize infographics in their virtual organizing \cite{Ables20, Dazed20}, resulting in the dispersion of an unprecedented number of viral activism-oriented infographics. 

Action formations---specifically connective and collective action---reveal the purpose, usage, needs, and motivation behind activism technology. As such, the explosion of Instagram infographics begs the question: \textit{Are Instagram infographics collective or connective action?} On the surface level, the answer is connective action—Instagram infographics digitally distribute individualized sociopolitical content through Instagram, a personal communication technology. In this paper, however, we argue that Instagram infographics in ethnic movements are a medium of deeply intertwined connective \textit{and} collective action, exemplifying that movement technologies can encompass the benefits and barriers of both. Infographics embody connective action by being distributed digitally, expediting fluid movement updates, and facilitating the curation of personalized content. Yet, infographics also embody collective action by cultivating shared movement identities, while carrying the same costs as traditional collective movement operations. In particular, Instagram infographic activism amplifies information dissemination for largescale education, necessitates the resolution of internal disputes for movement credibility, and leverages on-the-ground resourced efforts that drive forth transformative change.

Our study consists of a two-pronged mixed methods approach: (1) We conducted comparative interviews of Second Revolution ethnic movement activists and modern-day New Age ethnic movement activists to compare their experiences, perspectives, and organizing mechanisms. (2) We performed a data-mining case study on the Black Lives Matter Instagram page, revealing how infographics affect engagement and action-oriented community responses.

Our results indicate that the combination of connective and collective action is vital to modern-day ethnic movement infographic activism:

(1) Instagram infographics increase the scope of participation, educating movement participants to form a unified front. In particular, they supercharge information dissemination by circumventing physical, financial, and geographical constraints; they provide customizable key features and visual grabbers that draw in viewers; they serve as bite-sized entry points for new ethnic movement participants; they increase accessibility for those without formal education when movements integrate inclusive offline strategies. The expansive reach of infographics is exhibited by their communal traction on Instagram. While only 36.99\% of Black Lives Matter’s posted images from 2014-2020 were infographics, 50.44\% of likes and 54.19\% of comments were concentrated in infographics, suggesting that infographics increase engagement and therefore the scope of viewership. While an Instagram infographic’s content is initially personalized, the wide-scale distribution, expansive reach, and virality of the infographic establishes a unified movement front by broadening access to movement ideologies, education, and involvement. 

(2) Instagram infographics necessitate the reconciliation of movement disputes. While Instagram infographics force activists to cut corners to fit complex issues into confining slide-limits, activists have adopted connective methods to mitigate misinformation and establish movement legitimacy. The prioritization of reconciliation in pursuit of credibility exemplifies the need and desire to resolve movement differences in pursuit of a collective movement framing.

(3) Instagram infographics must be paired with high-resource efforts to yield transformative change. While infographics can cause passive, deceptive, and complacent movement involvement, they create tangible changes when paired with concrete non-virtual action. Indeed, infographics which call for action are more effective in creating action-based discourse. Only 23.85\% of total comments on the Black Lives Matter Instagram were action-oriented for non-action-oriented infographic posts, but 28.25\% of comments were action-oriented for action-oriented infographic posts, revealing that action-based infographics have sparked more action-based communal dialogue on the Black Lives Matter Instagram, a stepping stone to concrete action. The need to pair infographics with protests, financial contributions, and political work necessitates community resources, another feature of collective action. 

In addition to its connective flexibility, personalization, and digital nature, infographic activism embodies aspects of collective action. Our work provides an example of an online activism tool that centralizes connective and collective action, showcasing that information technologies are not restricted to a single action formation. In theory, Instagram infographics are connective; however, in practice, the ways in which they have been woven into movements yield collective output. Beyond introducing a rich research area at the intersection of connective and collective action, our insights can be used to help activists who must select movement tools to maximize movement benefits and minimize opportunity cost. By providing recommendations on how to expand scope and accessibility, mitigate misinformation, and optimize for legislative and societal transformation, we give ethnic movement activists a design framework to configure their infographics in a way that bridges connective and collective action. 

When designing technology in the social computing space, we must aim to meet the needs of both individuals and the collectives they form. To do so, researchers, designers, and scholars will have to consider and accept the marriage of connective and collective action when optimizing, implementing, and utilizing even the most personalizable digital media tools.

\section{Background: Yesterday and Today}
In this section, we outline the historical Second Revolution ethnic movements of the era between the 1960s and the 1980s and the outburst of infographics in the summer of 2020 in order to contextualize how ethnic movements originated, evolved, and relate to Instagram infographics.

\begin{figure}[h]
  \caption{Sample of Black Lives Matter Instagram Infographics Following the Murder of George Floyd (from @millenialblack @umaaraiynaas @dwardslife) \protect\footnotemark}
  \centering
  \includegraphics[width=3.4cm]{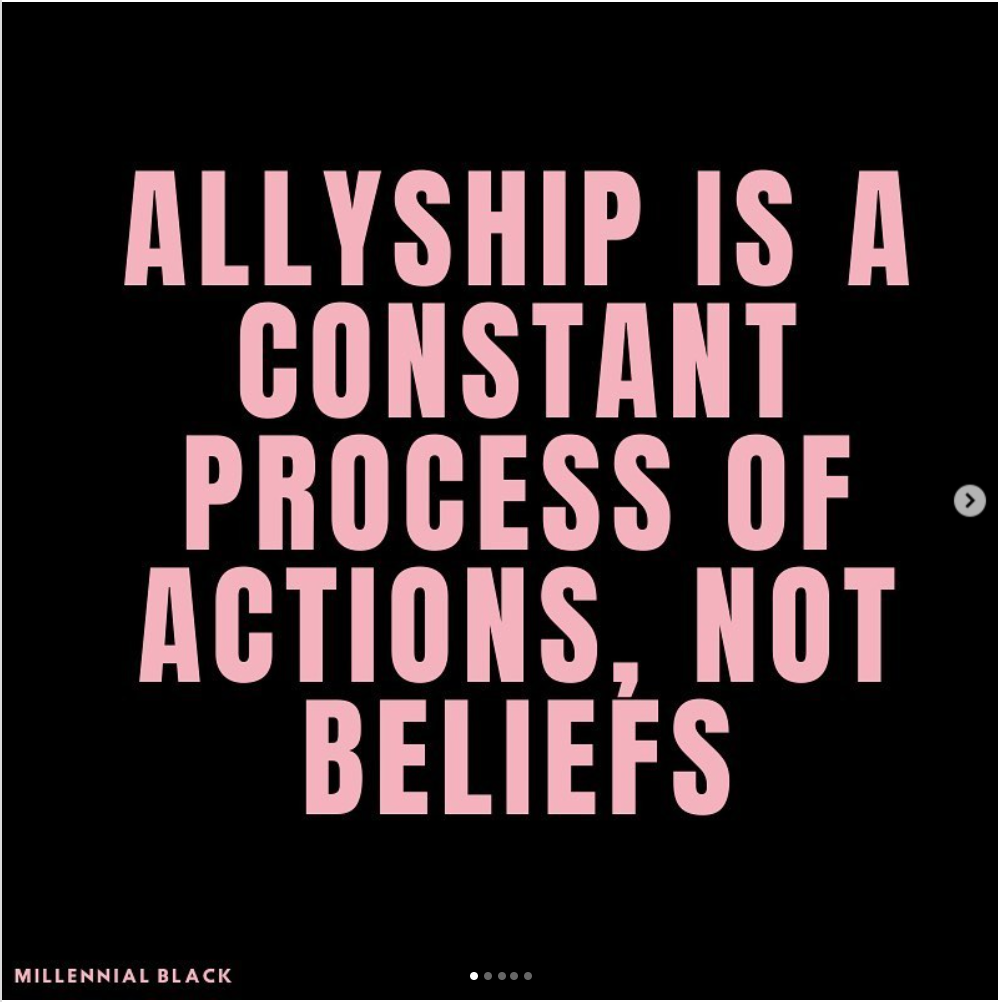}
  \includegraphics[width=3.4cm]{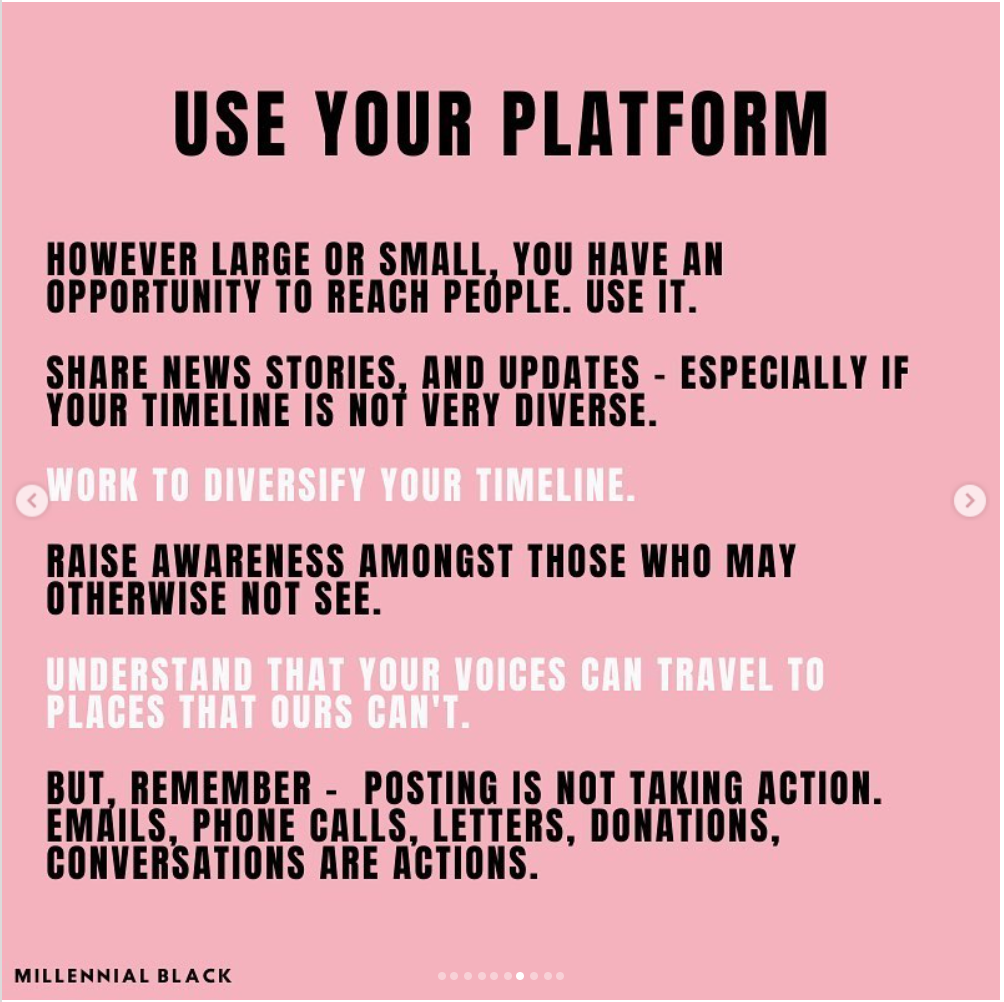}
  \includegraphics[width=3.4cm]{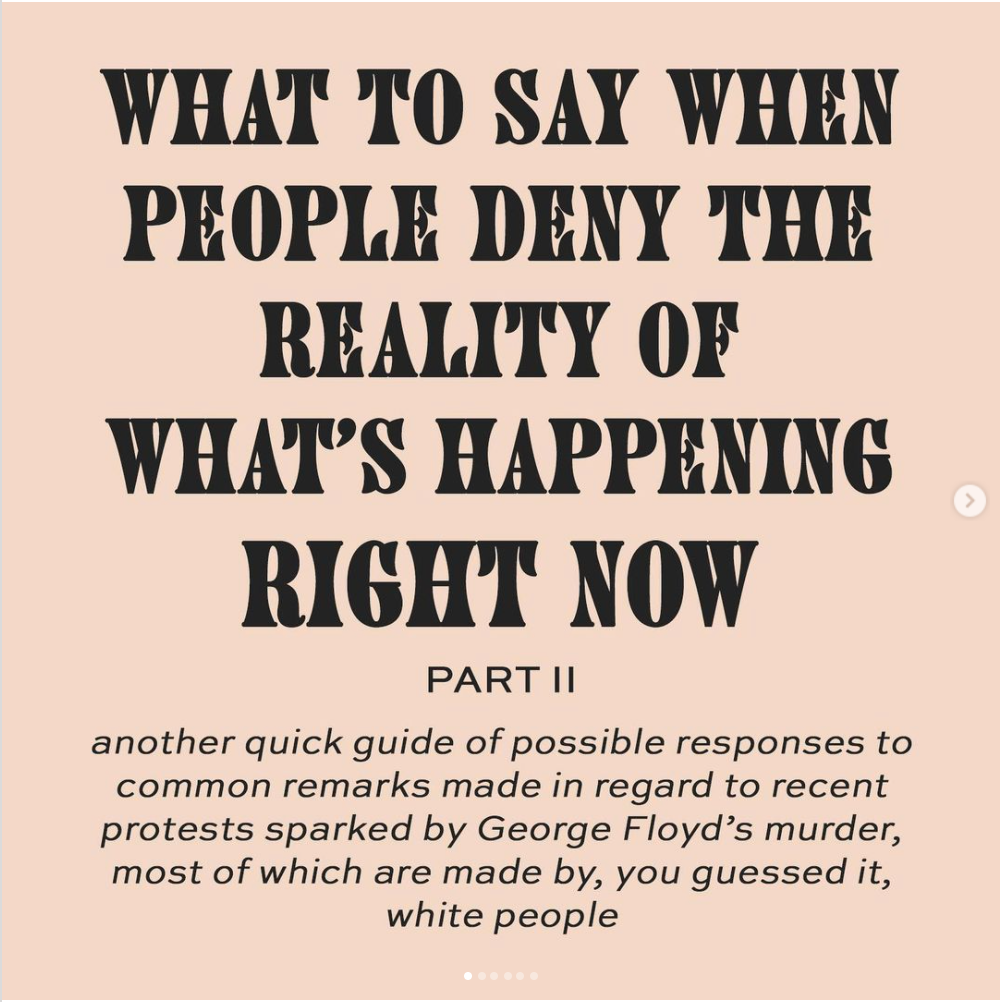}
  \includegraphics[width=3.4cm]{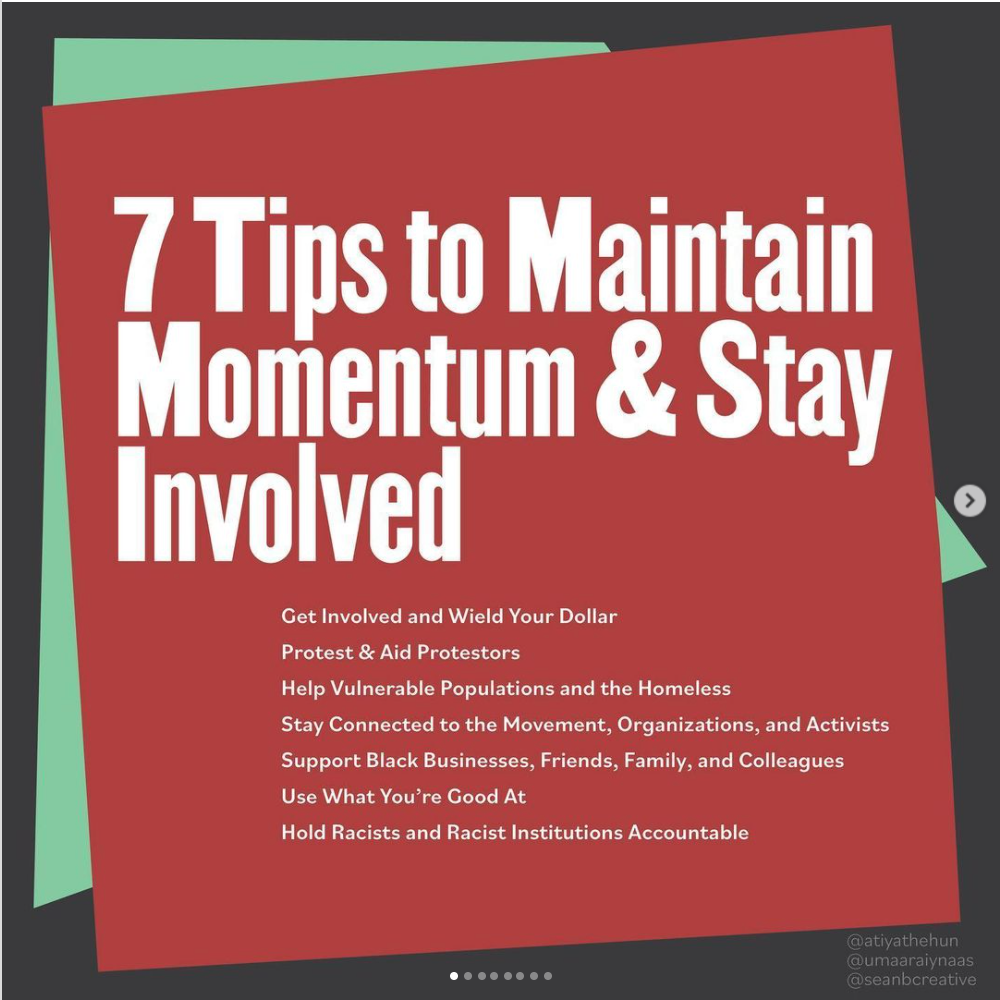}
\end{figure}
\footnotetext{Image content descriptions and citations are in A.2.}

\begin{figure}[h]
  \caption{A Sample of Infographics from asianawarenessproject on Instagram \protect\footnotemark}
  \centering
  \includegraphics[width=7cm]{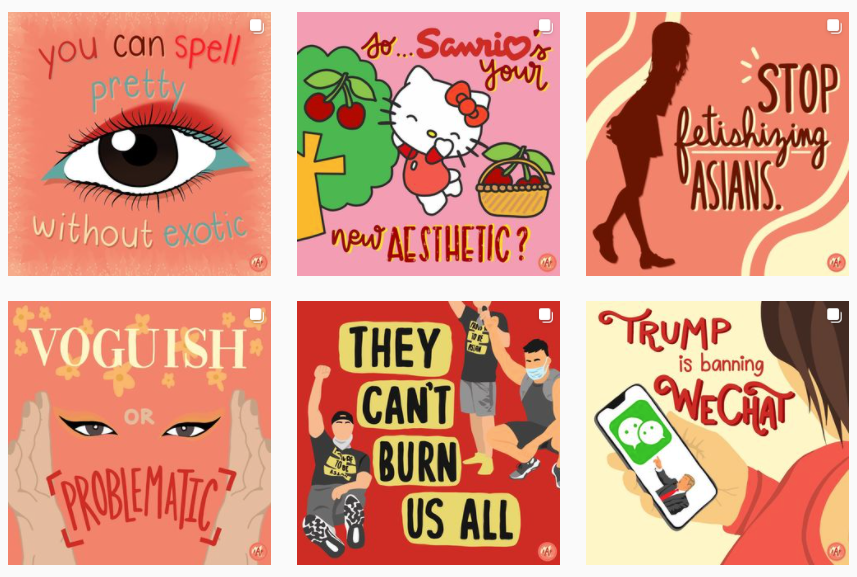}
\end{figure}
\footnotetext{Image content descriptions and citations are in A.3.}

\subsection{Ethnic Movements of the Second Revolution}
The Civil Rights Movement (1954-1964), often deemed the ``greatest mass movement in modern American history,'' fought for desegregating schools and national constitutional equality for Black Americans \cite{CivilRights}. In spite of the Civil Rights Movement's strides towards equity and justice, the ensuing era still left many communities disenfranchised, marginalized, and impoverished. Ethnicity-based social movements began to congregate in order to advocate for change, forming what we describe as the Second Revolution \cite{SecondRevolution}. In this section, we outline a handful of these movements to contextualize the experiences of the Second Revolution activists we interviewed. 

\subsubsection{American Indian Movement} The American Indian Movement (AIM) was founded in Minneapolis, Minnesota in 1968 with the objective of attaining sovereignty over illegally seized Native lands as well as economic and legal rights that the United States government has stripped from Indigenous communities for centuries \cite{Britannica}. Notable events include the occupation of San Francisco’s Alcatraz Island to demand sovereignty in 1969 and the 1973 Occupation of Wounded Knee in which 250 Sioux Indigenous Americans launched a 71-day occupation on the South Dakota Pine Ridge Reservation, calling global attention to the generational atrocities and unsafe living conditions inflicted upon the Indigenous American community \cite{Cooper15}.

\subsubsection{Black Power Movement} The Black Power Movement was spearheaded by the Black Panther Party, a revolutionary organization founded in October of 1966 in Oakland, California that strayed from the integrationist ideals of the Civil Rights Movement in an effort to procure Black self-determination through armed self-defense and socialist ideologies \cite{Archives16}. In addition to informing the public through a widely popular newspaper and political education classes, the Black Panther Party hosted several social services such as the Breakfast Program that served 20,000 school children daily, drug and alcohol awareness courses, early childhood education programs, and clothing and medical assistance programs \cite{Collier15}. 

\subsubsection{Asian American Movement} After the Asian American Political Alliance was founded by UC Berkeley graduate student Yuji Ichioka who coined the term “Asian American” \cite{YooAzuma16}, Chinese, Filipino, Korean, and Japanese Americans began to recognize the power in forming a communal consciousness to “bring together previously isolated and ineffective struggles against the oppression of the Asian communities into a coherent pan-Asian movement for social change” \cite{Wei93}. The Asian American Movement’s organizing included public protests against the Vietnam War, the fight for reparations for formerly interned Japanese Americans, and the movement for ethnic studies at San Francisco State and UC Berkeley \cite{Nittle19}. 

\subsubsection{Chicano American Movement} In poet Alurista’s 1969 Chicano manifesto, “El Plan Espiritual de Aztlán,” the land that ranges from what we now know as California and Texas were declared the Chicano homeland, giving Chicanos "something they could claim as their own" \cite{Aztlan}. By displacing Anglo-Americans from the perspectival center of North America, the concept of Aztlán was able to combat American xenophobia, while critiquing U.S. violations of land treaties like the 1848 Treaty of Guadalupe Hidalgo. In addition to protesting for jurisdiction over land, the Chicano American Movement fought for creating bilingual and bicultural programs and Chicano representation in educational and political roles as well as improvements in migrant farmworker socioeconomic conditions (led by César Chávez and Dolores Huerta who co-founded the National Farm Workers Association) \cite{Carrillo20}.

\subsubsection{West Asian American Movement} After Israel’s occupations of Palestine in 1967 and 1973, organizations fought to counter the Western narrative of the occupation. The Association of Arab American University Graduates (AAUG) sought to counter anti-Arab representation in scholarship and society at large. Years later, the Arab Community Center for Economic and Social Services (ACCESS) emphasized radical\footnote{The term "radical" (although carrying negative connotations in other contexts) was commonly used in Second Revolution ethnic movements to describe the pursuit of fundamental "radical" change in a sociocultural system \cite{Radical, NCID19}; in this paper we utilize this meaning for "radical," "radicalization," and other similar terms.} anti-imperialist thought and provided social services for Arab immigrants who fled war in Lebanon, Palestine, and Yemen \cite{Pennock17}.

\subsection{Summer 2020: Activist Infographics Take Off}

We define New Age ethnic movements to be ethnic movements that are ongoing and prominent today. Perhaps the most well-known and rapidly growing of these movements is Black Lives Matter (BLM), an organization founded in 2013 in response to the acquittal of the murderer of Trayvon Martin, a 17-year-old unarmed Black boy whose fatal, unsolicited murder sparked global outrage. BLM aims to “eradicate white supremacy and build local power to intervene in violence inflicted on Black communities by the state and vigilantes” \cite{BLM14}. The movement first spread globally when founders Patrisse Cullors, Alicia Garza, and Opal Tometi posted \#BlackLivesMatter on Twitter on July 13, 2013. The high-profile murders of several Black individuals then skyrocketed the usage of the phrase, as it developed into a mantra for the protection of Black life and an end to state-sanctioned violence \cite{Tillery19, Clark16, Garza14, Hockin18}. 

In the summer of 2020, the racially rooted murders of George Floyd, Breonna Taylor, Ahmaud Arbery, and others at the hands of police officers and white supremacists created waves in the Black Lives Matter movement. Infographics supporting BLM went viral on the social media platform Instagram, with the hashtag \#blacklivesmatter (Figure 2) accruing over 26 million posts and infographics on the Black Lives Matter Instagram page (Figure 1) rippling throughout the platform. These carefully-crafted infographics fought for the protection of Black life through anti-racism resources and action-items such as petitions and financial contribution routes \cite{Ables20, Dazed20}. 

Other ethnic movements and activist groups began to utilize infographics as well, as exemplified in Figure 3. Indigenous Americans have been sharing their culture through infographics that celebrate Indigenous People’s Day and Native Heritage Month, with \#indigenouspeoplesday home to over 217,000 posts. @dearasianyouth is an Instagram account that consists of over 100 Asian youth that share educational infographics about the Asian American community. Today, the Latinx community is focused on immigration rights and anti-deportation, with \#AbolishICE possessing over 193,000 posts, \#endfamilyseparation with over 28,000 posts, and \#closethecamps with over 63,000 posts. Palestinians and Palestinian Americans are fighting for the right to return to their homeland, with the hashtag \#freepalestine acquiring over 1.3 million posts as of 2020\footnote{Not all of these hashtags are accrued from U.S.-based activists or posts; we simply utilize these numbers in order to emphasize the online presence of the modern ethnic movements which have prominent U.S. roots and branches (although these movements also have prominent roots and branches elsewhere).}.

This paper not only verifies the unprecedented takeoff in ethnic movement infographics, but also investigates the significance of this phenomenon to digitally-mediated social movements.
\section{Related Work}

\subsection{Social Media Activism}

Technology and social media have long been utilized by social movements to meet community needs---often braided seamlessly (and sometimes ruggedly) into movement practices and strategies \cite{Tufekci17}. Bødker et al. found that volunteer-based communities curate functioning collectives through a close combination of circumstance, largescale communal strategies, and granular tactics involving technology \cite{Bodker16} and Liu discovered that activists strategically select and reject technologies on the basis of affordances \cite{Liu21}. In this section, we examine the preexisting research exploring the impact of digital activism modes on movement communication, credibility, and integration.

\subsubsection{Facilitating a Unified Movement Front} Technology---particularly social media---can help establish a borderless, unified movement front by documenting and fluidly disseminating central movement ideologies. 

\textsc{Sharing Core Movement Ideologies.} Ince et al. analyzed thousands of tweets with the hashtag \#BlackLivesMatter, discovering several core themes including movement solidarity, police violence, and movement tactics \cite{Ince17}. Twyman et al. discovered that Wikipedia allows us to virtually track the evolution of movement thought through the documentation and reappraisal of historical and contemporary events \cite{Twyman16}. Stanley studied an online community of fat, queer, and women hikers which united a diverse array of non-conforming bodies under a core ideology of queer mobility \cite{Stanley20}. Siddarth and Pal found that social media not only aided short-term movement goals, but also reframed bystander ideologies, values, and behaviors in the longterm, representing the gradual development of a unified ideological body \cite{Siddarth20}. Tillery discovered that Black Lives Matter organizations used Twitter to frame movement ideologies; in this case, by centering individual rights over broader identity framing \cite{Tillery19}. 

\textsc{Expanding the Movement Front.} Laer and Aelst discussed that social media has globalized movements by expanding them to the international scale \cite{Laer10} and Selander and Jarvenpaa found that social media broadens the engagement and interactions of social movement supporters \cite{Selander16}. Likewise, Enjolras et al. found that social media can expand reach by accessing different segments and demographics of movement populations \cite{Enjolras13}. 

\textsc{Accessibility of Information.} Crucially, disseminating unified movement information necessitates inclusion and accessibility. Ghoshal et al. found that the use of technology in grassroots movements can unfairly empower technically savvy leaders, resulting in inequitable participation \cite{Ghoshal20}. Siddarth and Pal labeled social media itself as a class-segregated entity \cite{Siddarth20}. Class struggle is a rift in social movement information dissemination that technolologically enabled movements must consider, as we explore when discussing the role of infographics in mass education.

Overall, past work indicates that activists have leveraged social media to widely disseminate movement frameworks. In our work, we assess the educational information dissemination capability of digital activism tools, particularly when connective and collective action are at play. 

\subsubsection{Challenging Credibility} Movements must not only spread information, but do so in a way that cultivates movement trust and resultant solidarity. As a result, Starbird et al. emphasized that disinformation is of critical concern to CSCW researchers since it threatens the integrity of the information space which it targets \cite{Starbird19}. This concern has expanded into the realm of social media, whose fluidity is incredibly susceptible to rapid-spreading rumors and misinformation \cite{AlrubaianAL-Qurishi19, Maddock15, Arif16, Liao13}. As a result, many researchers have studied information credibility on social media platforms \cite{Castillo11, Gupta14, Tavish19}. Mitra et al. found that transient spurts of collective attention on social media platforms are susceptible to a lack of credibility in the public eye \cite{Mitra18}. While social media can challenge credibility, Starbird and Palen determined that the use of Twitter in the 2011 Egyptian uprising succeeded in promoting crowd solidarity, increasing credibility and heightening situational awareness during events \cite{Starbird12}. Our work centralizes the activists in conversations around credibility---while social media activism challenges movement legitimacy, activists prioritize collective trust online.

\subsubsection{Supplementing Offline Movement Strategies}
Past work echoes that virtual engagement platforms complement physical action to enhance social movements. Donovan demonstrated that Networked Social Movements via the InterOccupy Movement provided crucial communication media when on-the-ground interaction did not suffice \cite{Donovan18}. Similarly, Massung et al. explained how virtual crowdsourcing for pro-environmental community activists increases the scalability of campaigns \cite{Massung13}. 

However, there are disagreements as to whether technology can reduce the need for offline tactics or whether overreliance can breed complacent movement involvement \cite{Rotman11}. Indeed, Earl and Kimport showed that virtual organizing sharply reduces the cost of protest and the need for physical engagements \cite{EarlKimport13}, yet Gerbaudo concluded that while social media serves as a supplementary activist tool for organizing, physical space is irreplaceable in social movements \cite{Gerbaudo12}. Indeed, Theocharis et al. studied several online activism campaigns, concluding that Twitter rarely facilitated coordination and protest organization \cite{Theocharis15}. Our work adds a layer of nuance; while we confirm that transformative change necessitates offline, action-oriented, and high-resource collective movement components, digital tools such as infographics can serve as a crucial pipeline.

\subsubsection{Instagram Activism} In the past decade, Instagram's centralization of visuals has captivated users \cite{Blystone20} and benefited activists. Alexander and Hahner found that the visual intimacy of the image-sharing platform allowed activists to revisualize and reframe, encouraging followers to adopt new understandings of Down syndrome \cite{Alexander17}. Parmelee and Roman found that Instagram has become a hub for political information and guidance \cite{Parmelee19}. Mendonça and Caetano found that Instagram's deployment of images can frame public understandings of political leaders \cite{Mendonca21}. Instagram centers imagery instead of text, which results in significant opportunity for creative and newfound modes of information dispersal such as Instagram infographics.

\subsection{Infographics and Activism}

\subsubsection{Infographics} HCI and visualization researchers have been increasingly exploring the use and effect of infographics. On a broad level, Lu et al. analyzed the use of Visual Information Flows to better understand how infographics relay information \cite{Lu20}. More specifically, several works underline the  \textit{benefits} of infographics in information delivery. Wang et al. found that graphical and textual annotations improve readability and accessibility \cite{Wang20}. Bateman et al. \cite{Bateman10} and Borgo et al. \cite{Borgo12} demonstrated that visual embellishments improve retention while Borkin et al. uncovered design strategies that create memorable infographics \cite{Borkin13}. Holistically, Haroz et al. revealed that simple pictographic representations improve memorability, speed of information discovery, and engagement \cite{Haroz15}. Our work builds on this literature by testing the theoretical benefits of infographics when integrated into social movements.

\subsubsection{Infographics and Activism} Limited research has unveiled key instances of virtual infographic use in \textit{activism}. Broadly, Amit-Danhi and Shifman stated that “digital political infographics” introduce a novel way of discussing, understanding, and relaying socio-political information \cite{Amit-DanhiShifman18}. For instance, quick-spreading digital media infographics expanded the initiative against land reclamation into the longest movement in Bali’s history \cite{Suwana20}. Likewise, Chicago climate activists used accessible infographics to reach community members that would be otherwise excluded from environmental planning discussions \cite{Bates20}. Similarly, Facebook infographics were a vital contributor to Ukraine’s EuroMaidan mobilization process by rapidly disseminating information \cite{Onuch15}.

While researchers have begun to uncover the rich area of intersection between activism and infographics, this realm is heavily untapped both in HCI and academia at large. Previous works have investigated either (a) the use of infographics as a medium of information dissemination or (b) the use of social media tools and platforms as they relate to activism. However, these topics have been analyzed in isolation. A space that has not been explored is the intersection of the two: the usage of social media infographics in advancing activist objectives. Analyzing the groundbreaking trend of ethnic movement infographics on Instagram provides insight into the potential of this untapped realm of information dissemination and its impact on technologically enabled activism.

\subsection{Connective and Collective Action} Collective action is primarily offline, expends organizational resources, and involves movement-wide ideological framing. Connective action is individualized, technologically facilitated, and lacks the need for organizational resources or internal dispute resolution \cite{Bennett12}. While Bennett and Segerberg acknowledge that connective and collective action may co-occur, researchers have outlined tensions between connective and collective action and debated the efficacy and value of the action formations.

Chu and Yeo discovered that inauthenticity of connective action movement participation incentivized young Hong Kong activists to pivot to embodied collective action participation \cite{Chu20}. Shahin and Ng claimed that connective action fails to lead to lasting change due to excess individualization, flexibility, and emotionless participation \cite{Shahin21}. Laer and Aelst argued that social media is unable to create inter-activist ties that yield sustained action \cite{Laer10}. Others dispute the value of connective action tools, claiming that digital technologies do not inherently transform movement dynamics \cite{Gonzalez16, Earl10}. Indeed, Pavan wrote that digital communication technologies are in fact components of collective action networks \cite{Pavan14}.

In contrast, Pond and Lewis found that connective action discourses may lack influence on the surface-level, but are able to create and propagated resultant action frames with broader impact \cite{Pond19}. Bang and Halupka similarly established connective action as a potential for self-governing and a departure from neoliberal reliance on unified political authorities \cite{Bang19}, while Castells similarly highlighted that connective action's self-configurability breeds decentralization \cite{Castells16}.

We problematize both approaches---connective and collective action each maintain their own faults and groundbreaking features. Analyzing the \textit{application-level} of activism technologies reveals rich information about the roles of technology in collective action \cite{Segerberg11}. For instance, Lee et al. outlined social media pipelines into both personalized and collective action outcomes in Hong Kong’s Umbrella Movement \cite{Lee17}. Ethnic movement activists' use of Instagram infographics evidences that digital media activism is not confined to just one action formation; the integration of digital media tools into social movements demands consideration of the affordances of both connective and collective action. CSCW considers and designs for collectives; recognizing the careful coexistence of connective and collective action is critical to effectively designing and evaluating technologies for movement collectives.

\section{Method}
Our granular research question is: \textit{How do Instagram infographics impact ethnic movements?} Our broader research question is: \textit{Are Instagram infographics collective or connective action?} To answer these questions, we used a two-pronged mixed methods approach consisting of an interview process and a data-scraping case study. 

\subsection{Comparative Activist Interviews}

Our study spotlights ethnic movement activists by comparing the insights of Second Revolution activists and modern-day Instagram activists. We focus on U.S.-based activists and primarily U.S.-based ethnic movements in order to draw sound parallels with Second Revolution movements. Introducing ethnic movement activist perspectives into academia is vital as activists are the most knowledgeable about movement intentions and core values. By going directly to the source, we can more accurately contextualize quantitative results. In contrast with a one-dimensional interview procedure, comparative interviews of two disjoint groups can reveal otherwise inaccessible conclusions about the unique changes that Instagram infographics have introduced. We effectively establish a "collective action baseline" by revealing the practices, challenges, and strategies of the offline Second Revolution ethnic movements. In turn, this allows us to distinguish collective action continuities and connective action mediums when turning our attention to New Age activists' use of Instagram infographics. 

To compare and contrast ethnic movements of modern day with their predecessors, we sought ethnic movement activists from our two eras of interest: Second Revolution (1960s-1980s) and New Age (modern-day). We used Instagram Direct Messaging and email to connect with activists and asked advisors and professors within the Ethnic Studies department at UC Berkeley for referrals to Second Revolution ethnic movement activists. We interviewed three Second Revolution activists and five New Age activists. Table 1 includes each activist's name, movement involvement, and bio.

\begin{table}[h]
\centering
\scriptsize
\caption{Interviewee Profiles}
\begin{tabularx}{\textwidth}{>{\raggedright}p{0.1\textwidth}|>{\raggedright}p{0.1\textwidth}|p{0.05\textwidth}|p{0.65\textwidth}}
\toprule
    Name & Movement(s) &Era& Description \\
    \midrule 
    Harvey Dong & Asian American Movement & Second Revolution & Professor of Ethnic Studies focusing on Asian American and Asian Diaspora Studies at UC Berkeley. In the 60s, was involved in the movement to establish UC Berkeley Ethnic Studies department, the Anti-War Movement, and the Asian American Political Alliance. Current research focuses on "Asian American and Third World social movement activism" \protect \footnotemark. \label{sec:Dong} \\
    \midrule
    Billy X. Jennings & Black Power Movement& Second Revolution & Former Black Panther with over 50 years of participation in the Black liberation struggle. Member of the Black Panther Party from 1968 to 1974. Founder of It’s About Time, a Black Panther Alumni committee which hosts events about the Black Panther Party’s legacy \protect \footnotemark.
    \label{sec:Jennings} \\
    \midrule
    Pablo Gonzalez & Chicano American Movement& Second Revolution & Professor of Chicano Studies at UC Berkeley. Facets of Chicano-American Movement ingrained in upbringing. Current research focuses on the housing crisis' implications on Black and Latinx communities. Manuscript discusses intersections of Chicano American Movement and Zapatista Indigenous Movement of Mexico \protect \footnotemark. \label{sec:Gonzalez} \\
    \midrule
    Carmen Perez & Latinx community, Black Lives Matter &New Age& Co-founder of the Women’s March on Washington \protect \footnotemark. President and CEO of The Gathering For Justice, a movement to combat racial inequities in the justice system \protect \footnotemark. Focuses on advocating for individuals impacted by police brutality, particularly Black and Brown communities.
    \label{sec:Perez} \\
    \midrule
    Zoha Raza & Muslim American Movement &New Age& Communications Coordinator at the Council on American-Islamic Relations, the largest Muslim American civil rights organization \protect \footnotemark. Runs CAIR’s social media accounts and routinely utilizes Instagram infographics to mobilize Muslim Americans and break stereotypes about the Muslim American community.
    \label{sec:Raza} \\
    \midrule
    Mary Celestin & Black Lives Matter &New Age& Founder of San Jose Strong, a grassroots organization to reinvent San Jose through local, community-level organizing \protect \footnotemark. Creates projects and resources to mobilize individuals for the Black Lives Matter Movement.
    \label{sec:Celestin} \\
    \midrule
    Hesham Jarmakani & Palestinian and Muslim American Movements &New Age& Political Director for Bears for Palestine, a UC Berkeley organization rooted in advocacy for Palestinian human rights. Has raised funds for Black-affiliated organizations and crises in Southwest Asia such as in Syria and Lebanon.
    \label{sec:Jarmakani} \\
    \midrule
    Keys of Palestine (Founder) & Palestinian American Movement &New Age& Organization which aims to shed light upon the silenced stories of elders in the Palestinian community through video submissions, photographs, and artwork in thematic infographics. Note: The founder of Keys of Palestine requested anonymity.
    \label{sec:KeysOfPalestine} \\
\bottomrule
\end{tabularx}
\end{table}
\footnotetext[7]{https://ethnicstudies.berkeley.edu/people/harvey-dong/}
\footnotetext[8]{http://www.itsabouttimebpp.com/home/home.html}
\footnotetext[9]{https://ethnicstudies.berkeley.edu/people/pablo-gonzalez-1/}
\footnotetext[10]{https://womensmarch.com/}
\footnotetext[11]{https://www.gatheringforjustice.org/}
\footnotetext[12]{https://www.cair.com/}
\footnotetext[13]{https://sanjose-strong.github.io/home/}

Each interview lasted between 30 minutes and 1 hour. All activists were asked to outline their experiences in their respective movement(s). Questions for Second Revolution activists were focused on the benefits and drawbacks of information dissemination mechanisms of the past and how the introduction of activism infographics has implicated the authenticity, togetherness, and effectiveness of ethnic movements. Questions for New Age activists were focused on strategies for Instagram infographic success, combating misinformation, and yielding tangible change.

Interviews were conducted via Zoom or telephone and recorded. We then executed an iterative procedure for our coding analysis:

\begin{enumerate}
    \item We first mapped out all striking quotes and sought to find patterns and connections. 
    \item We compartmentalized the quotes into their respective categories; these categories make up the skeleton of the subsections \textit{within} each of our main findings. 
    \item Next, we performed a second iteration to map these sub-themes into the three overarching themes that we will discuss in our findings.
    \item Finally, we performed a last iteration to examine how the three themes interacted with one another and what they revealed about digitally-facilitated social movements, leading us to draw our conclusions about the inseparability of connective and collective action.
\end{enumerate} 

\subsection{Instagram Data Scraping Case Study of Black Lives Matter}

We performed a case study on Black Lives Matter in order to explore the prominence and usage of Instagram infographics in one of the most pertinent movements of modern day. We chose Black Lives Matter because of its large following and consistent online presence since 2014, which makes it a reliable dataset. We sought to quantitatively determine (a) whether infographics have been increasingly used by the Black Lives Matter Instagram account (b) whether infographics affect movement scope or transformative impact. 

\subsubsection{Scraping Account Photos}

We utilized Instaloader, a tool that allows users to scrape pictures, videos, captions, comments, and other metadata from Instagram\footnote{https://instaloader.github.io/}. We extracted all images posted by the Black Lives Matter Instagram account (@blklivesmatter) as well as a JSON file of metadata associated with each image.

\subsubsection{Classifying Infographics}

To classify images as infographics or non-infographics, we identified the criteria of an infographic and determined whether or not images were infographics by hand. We drew upon the following definition:

\begin{quote}
    \textit{a visual presentation of information in the form of a chart, graph, or other image accompanied by minimal text, intended to give an easily understood overview, often of a complex subject.} \cite{Infographics-Def}
\end{quote}

Images which we manually classified as infographics met all three of the following criteria:
\begin{enumerate}
    \item The image is a visual presentation of information in the form of a chart, graph, or other image.
    \item The visual presentation is accompanied by minimal text.
    \item The image breaks down a complex subject, concept, or message into an easily understood overview.
\end{enumerate}

While we explored computer vision classifiers to bucket images as infographics and non-infographics, we decided against this as no classifiers were specific enough to make decisions that we deemed accurate in subsequent hand samples. 

\subsubsection{Defining Action Words}

We define Action Words to be terms indicative of calls to action beyond the infographic, particularly in the context of ethnic movements and Black Lives Matter. Our Action Keyword Bank was defined by browsing the action-oriented infographic images on Instagram, including all images we extracted from the Black Lives Matter Instagram page. Examining a post's Action Words is an objective way of gauging action-oriented language, messaging, and discourse both in the infographic and the comments. Action Words signal and inspire beyond-infographic action---reading, donating, protesting, etc. In this sense, Action Words serve as an identifier of the bridge between virtual organizing and physical, in-person interaction. 

\subsubsection{Analyzing Engagement and Action Words}

We wrote Python scripts\footnote{The scripts for this case study can be found at https://github.com/infographics-cscw/scripts.} in order to extract and analyze the following data:

\begin{enumerate}
    \item \textit{Engagement: } From each post's Instaloader JSON file, we extracted the number of likes and comments in order to determine whether posts with infographics reached or resonated with a broader audience.
    \item \textit{Action Words: } To extract image text, we used OpenCV, a library providing real-time computer vision functionality\footnote{https://opencv.org/}. In conjunction, we also used Python-tesseract, a wrapper for Google’s Tesseract-OCR Engine that can recognize text from images\footnote{https://pypi.org/project/pytesseract/}. The script iterates through each image filepath pulled from Instaloader, at which point OpenCV’s \texttt{imread} function reads each image, the \texttt{cvtColor} function converts each image’s colors to a more readable color combination, and \texttt{pytesseract} transforms the detected image text to a string. We then utilized this script to determine whether a post's images contained Action Words. We also downloaded a JSON file of the comments section of each Black Lives Matter post from 2014 through 2020 in order to determine whether Action Infographics (infographics with Action Words) drew a more action-oriented communal response than Non-Action Infographics (infographics without Action Words).
\end{enumerate}

\begin{table*}
  \caption{Action Keyword Bank \protect \footnotemark}
  \small
  \label{tab:commands}
  \begin{tabular}{ccccccccc}
    \toprule
    Petition & Call & Vote & March & Rise & Text & Action & Act & Demand \\
    Fight & Protest & Share & Come & Support & Help & Urge & Sign & Prosecute \\
    Donate & Defend & Change & Command & Shout & Cry & React & Now & Justice \\
    Resist & Read & Stop & Manifest & Join & Give & Lift & Assert & Protect \\ Stand & Take & Say & Liberate & Return & Heal & Raise & End & Register \\ Contact & Bring & Hear & Listen & Watch & Release & Strike & Preach & Abolish \\ Free & Defund & Duty & Email & Post\\
    \bottomrule
  \end{tabular}
\end{table*}
\footnotetext{Action Words are listed in no particular order.}

\subsection{Method Limitations}

    A few limitations exist: First, the selected interviewees depended on the ethnic movement activists we were familiar with or were referred to, meaning the sample of ethnic movement activists was largely localized. However, our activists were from a variety of movements and backgrounds and therefore shared vastly different experiences. Second, our hand-classification of infographics---despite strict adherence to the three-pronged definition---was subjective in some borderline images. However, the number ``unclear'' or ``borderline'' images was largely outnumbered, reducing the implications of such bias. Third, it is possible that a post is calling for action without including a word in our Action Keyword Bank or that a post includes a word in our Action Keyword Bank without calling for action. However, our method provides a key balance between objectivity and accuracy by defining a clearcut set of action-identifiers, removing subjectivity from the classification process.

\section{Findings}

Our core discovery is that Instagram infographic activism stems beyond connective action by epitomizing central facets of collective action. Perhaps the most convincing example of Instagram infographics' ability to leverage collective action is the Black Lives Matter movement's Instagram page. By any measure, the Black Lives Matter movement embodies collective action: they have a unified ideology of Black liberation and have utilized organizational resources in physical events including protests since their genesis in 2013. Crucially, infographics are more prominent on the Black Lives Matter Instagram page now than ever before. Figure 4 and Table 3 detail two trends in the Black Lives Matter Instagram account from 2014 through 2020: the absolute number of infographic images and the percentage of all images that were infographics.
\begin{table*}[h]
  \caption{Infographics in Black Lives Matter Instagram Posts Over Time}
  \small
  \label{tab:commands}
  \begin{tabular}{c|c|c} 
    \toprule
    Year &Number of Infographics &Percentage of Images That Were Infographics\\
    \midrule
    2014 & 1 & 0.75\% \\
    2015 & 13 & 12.75\% \\
    2016 & 27 & 40.91\% \\
    2017 & 41 & 35.04\% \\
    2018 & 62 & 44.29\% \\
    2019 & 50 & 44.25\% \\
    2020 & 155 & 62.25\% \\
    \bottomrule
  \end{tabular}
\end{table*}

\begin{figure}[h]
  \caption{Infographics in Black Lives Matter Instagram Posts Over Time}
  \centering
  \includegraphics[width=9cm]{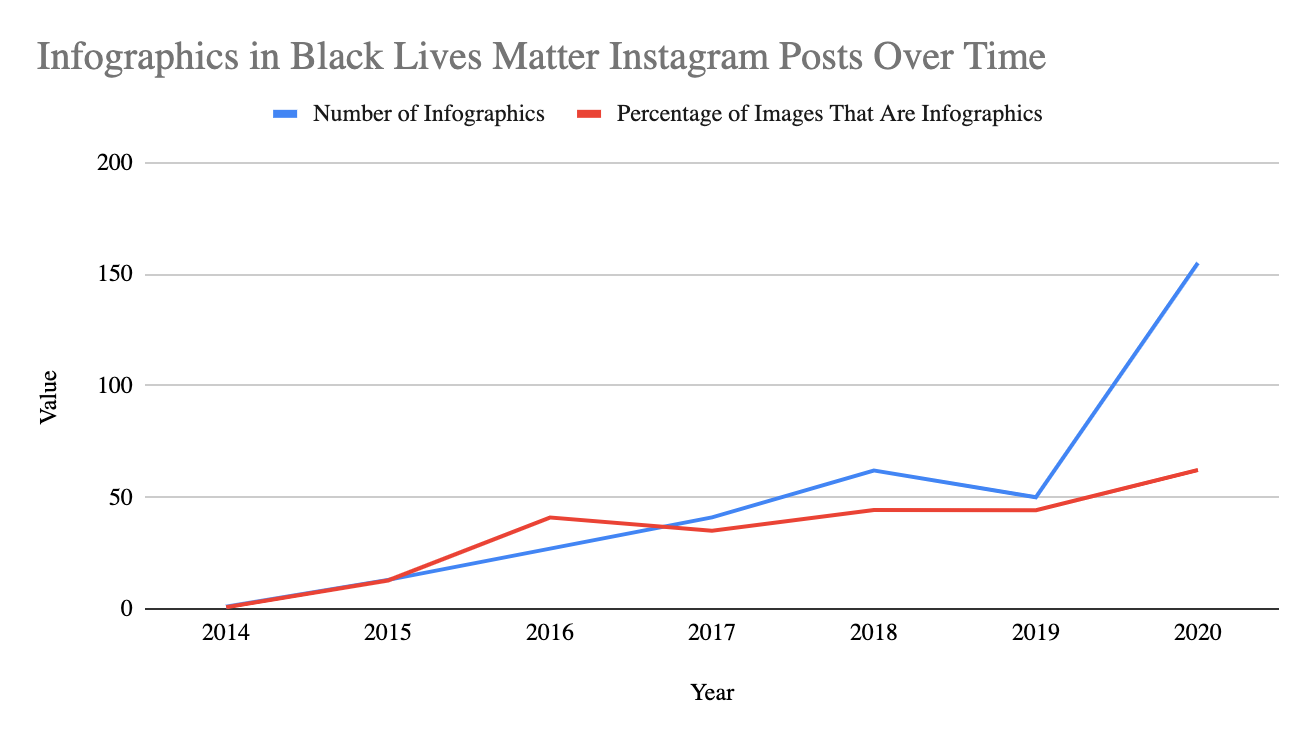}
\end{figure}

\begin{figure}[h]
  \caption{Juxtaposing Black Lives Matter (@blklivesmatter) Posts from 2014 (left) with 2020 (right) \protect\footnotemark}
  \centering
  \includegraphics[width=4cm]{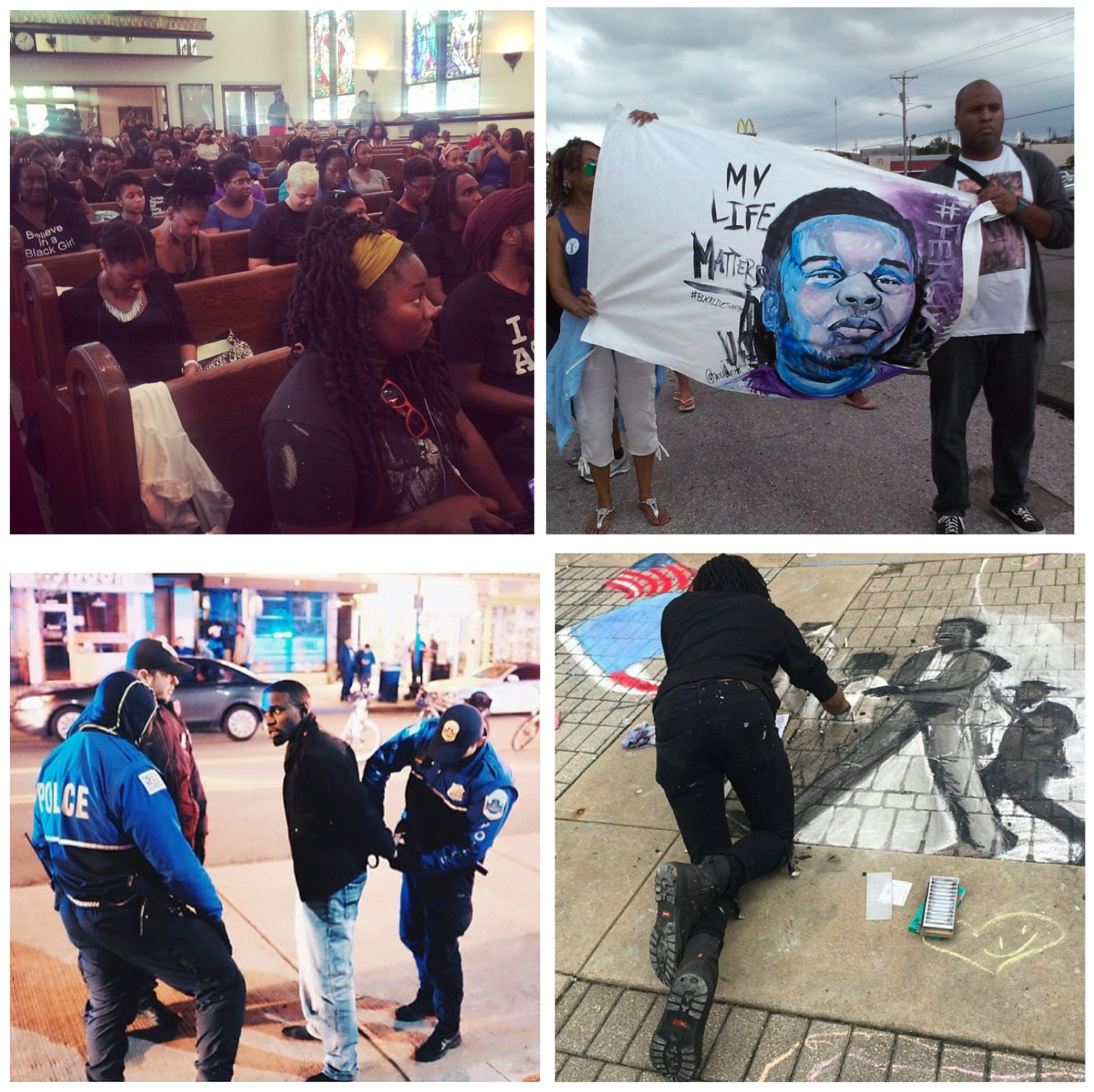}
  \hspace{1cm}
  \includegraphics[width=4cm]{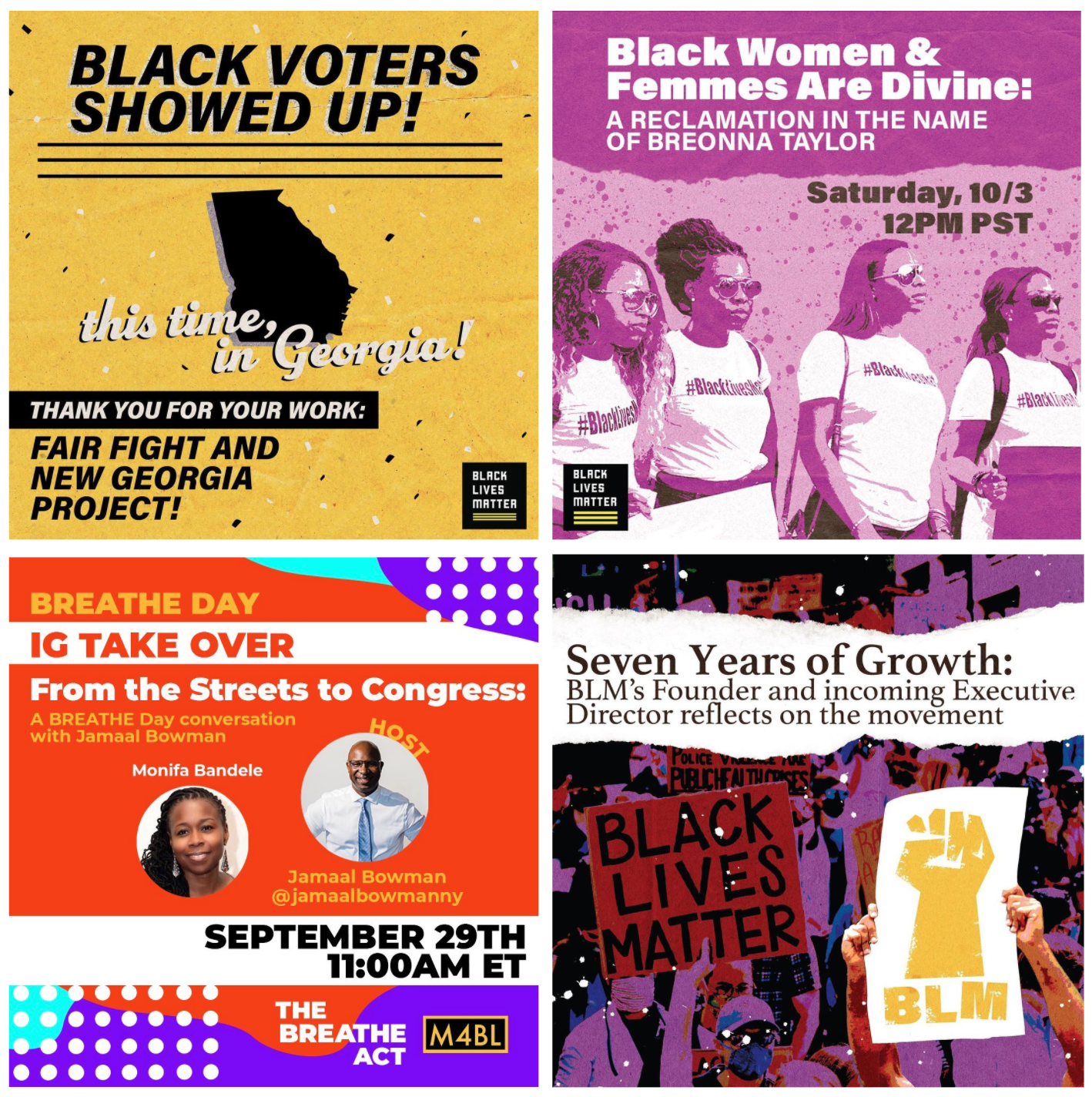}
\end{figure}
\footnotetext{Image content descriptions and citations are in A.4.}

As anecdotally depicted in Figure 5, while Black Lives Matter has been actively fighting against the loss of Black life and white supremacy for years, 2020 showcases a significant surge in infographics, with 155 posted infographics and 62.25\% of posted images being infographics. The increased usage of infographics demonstrates that a movement empirically rooted in collective action can leverage a personalizable digital media device traditionally categorized as a connective action tool. This complicates our understanding of connective and collective action: How can Instagram infographics just be another connective action device if a collective movement is utilizing them at the organizational level? Now, we will explore three manifestations of Instagram infographic activism as a bridge between connective and collective action: information dissemination for movement education, internal reconciliation for credibility, and high-resource efforts for transformative change.

\subsection{Increasing Scope for Collective Education} \label{sec:1}

Collective action requires that people unify about a common cause, the precursor to which is educating movement participants about the cause itself. Indeed, Second Revolution movements necessitated education of the masses. Today, infographics serve as powerful education tools by heightening ethnic movement information dissemination, allowing personalizable posts to attract numerous movement contributors who share a common objective, and curating a collective movement ideology.

\subsubsection{Information Sharing Methodologies of the Past} \label{sec:1.1}

In this section, we outline Second Revolution methods of obtaining and spreading ethnic movement information to establish a baseline for collective movement education methods. While television and radio were common-use technologies from the 1960s through the 1980s, activists predominantly relied on person-to-person contact because mainstream media outlets catered to a white audience. Primary mobilizing strategies included print (alternative movement newspapers, handbills, postcards), telephone, artwork (murals, print-art, graphics), and face-to-face events (festivals, protests, conferences).

\textsc{Physical Mobilization. } \hyperref[sec:Gonzalez]{Pablo Gonzalez} drew attention to the vitality of human-to-human contact:
\begin{quote}
    \textit{“Some of the initial ways that the Chicano Movement built its critical mass of folks to mobilize… [was] through a grassroots initiative that had to take place face-to-face. You had to be able to experience and meet across regions, across different communities, in order for you to really feel the power of the Chicano Movement.”}
\end{quote}

A prominent example of these physical interactions was the Black Panther Party's 10 10 10 Program, in which a Panther organizes ten people within a ten-block square, each of which agree to mobilize ten more people, producing a ripple effect that broadcasts movement information. Door-to-door operations like the 10 10 10 Program stemmed from the belief that direct contact—arriving at someone’s doorstep with a message and a flyer—was the best way to spread knowledge.

\textsc{Alternative Movement Print. } Movement newspapers were, as \hyperref[sec:Jennings]{Billy X. Jennings} called them, \textit{“the lifeblood”} of ethnic movements. The Black Panther Party extracted information from internal political education classes and dispersed this radical thought through 350,000 weekly newspapers, many of which are showcased in Figure 6.

\begin{figure}[h]
  \caption{Art in the Black Panther Party Newspaper \protect \footnotemark}
  \centering
  \includegraphics[width=3cm]{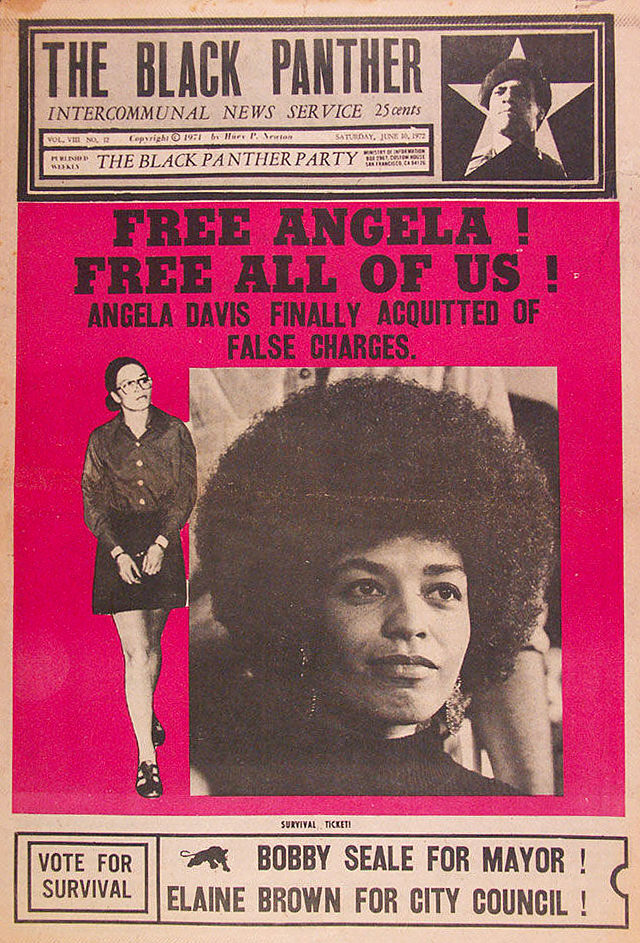}
  \includegraphics[width=3cm]{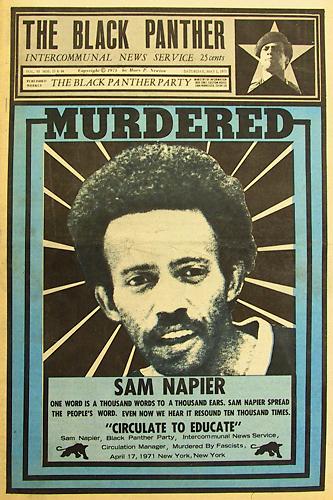}
  \includegraphics[width=3cm]{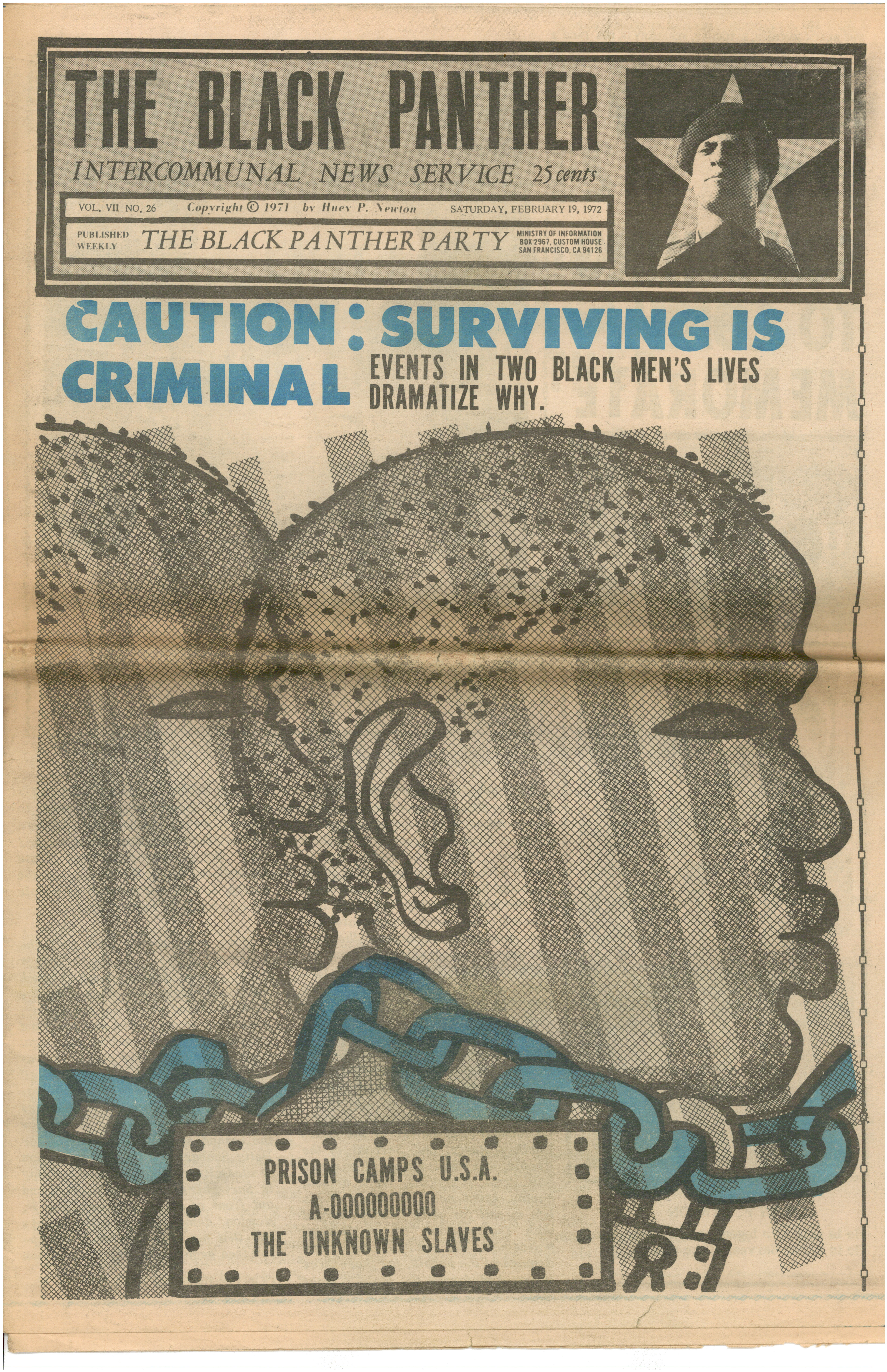}
  \includegraphics[width=3cm]{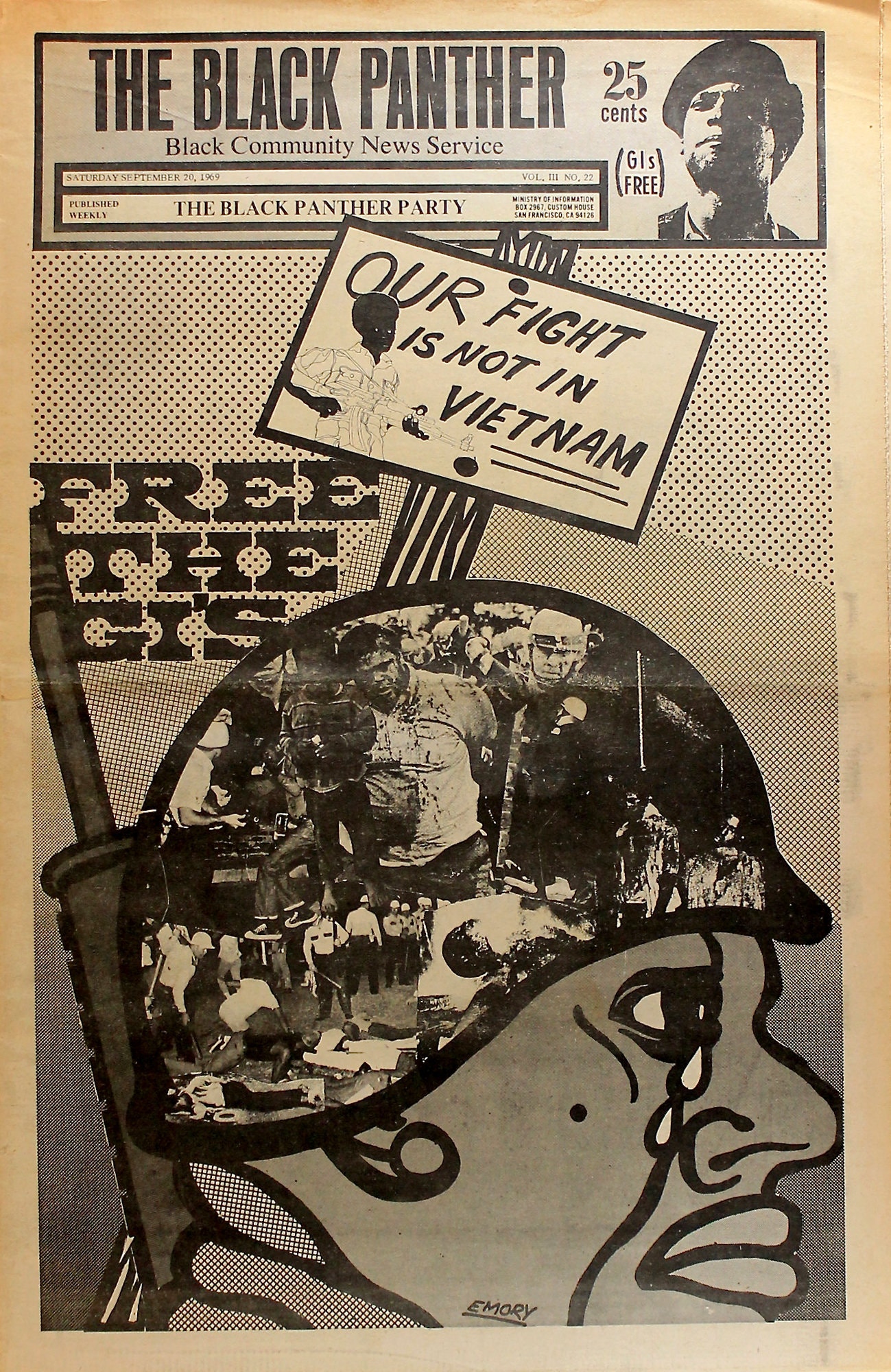}
  \includegraphics[width=3cm]{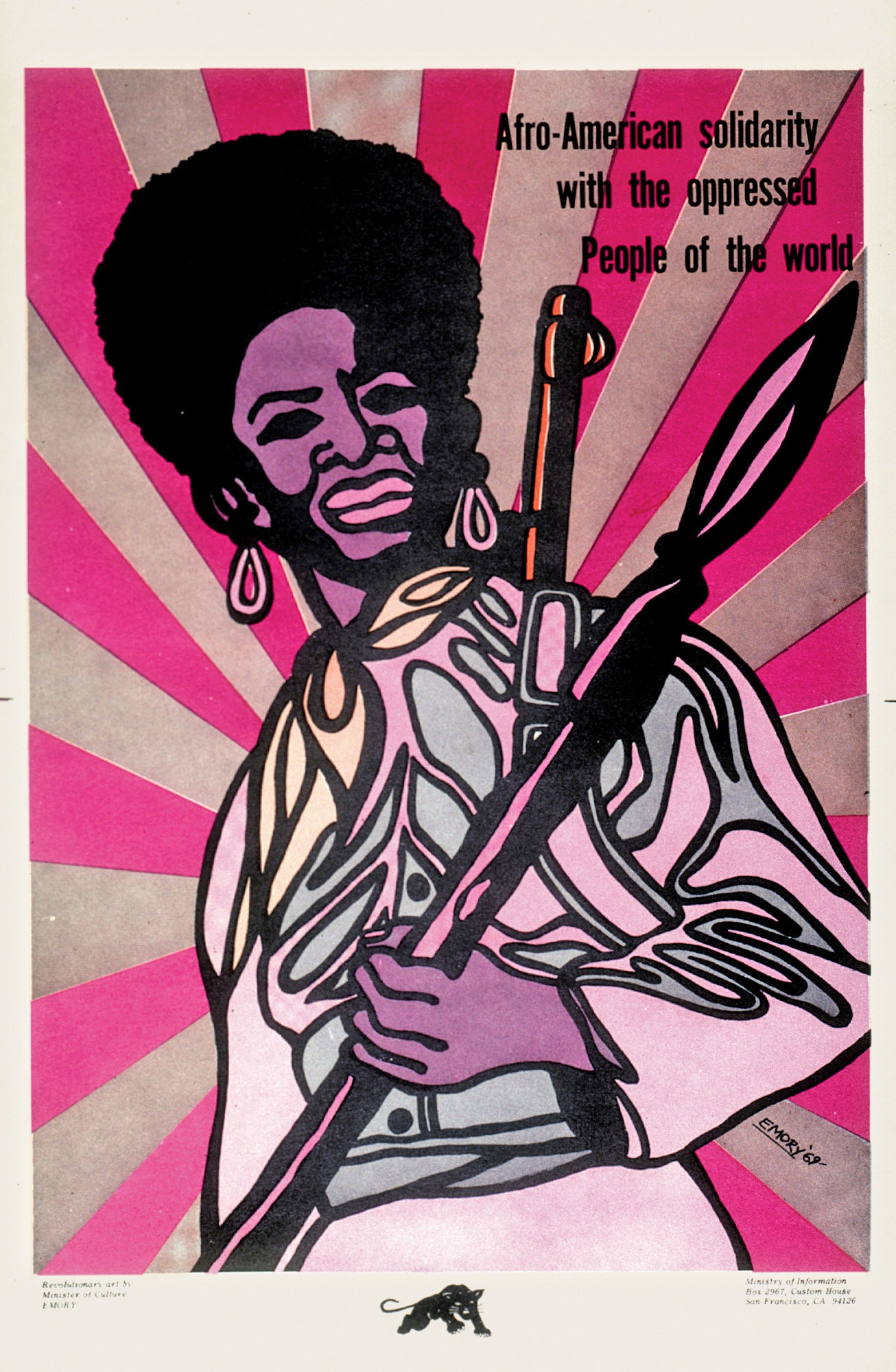}
  \includegraphics[width=9cm]{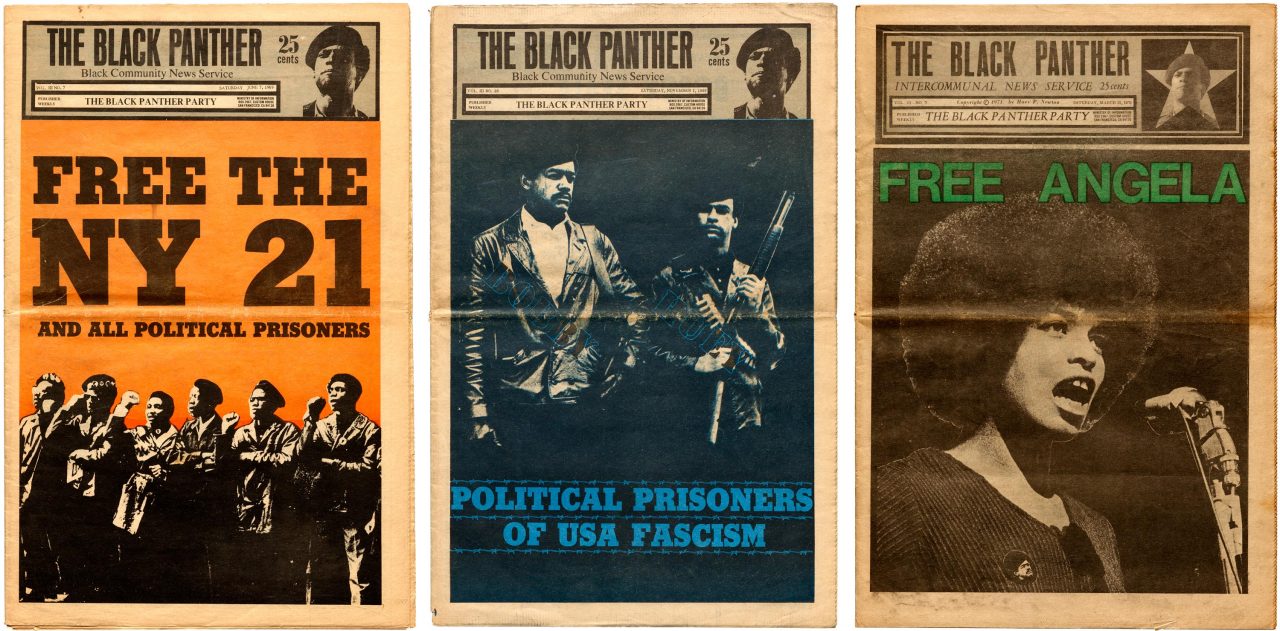}
\end{figure}
\footnotetext{Image content descriptions and citations are in A.5.}

By portraying a reality that the state and the mainstream media omitted, Second Revolution ethnic movements were able to maximize their reach through print, door-to-door operations, and physical events. The prominence of information-sharing mediums in Second Revolution movements certifies that information and education have historically been essential components of collective action.

\subsubsection{Overcoming Physical, Financial, and Geographical Obstacles} \label{sec:1.2} 

\begin{quote}
    \textit{“Information is the raw material for new ideas. Once we give people the information, they can come up with new ideas.”}  — \hyperref[sec:Jennings]{Billy X. Jennings}

\end{quote}
In this section, we examine how Instagram infographics expedite and expand information dissemination by reducing physical, financial, and geographical restrictions. As a result, the use of Instagram infographics can increase the reach of movement education in pursuit of establishing and spreading a collective movement ideology.

\textsc{Physical Labor Barrier. }\label{sec:1.2.1} Second Revolution activists depended on intricate and time-intensive tools and methods. \hyperref[sec:Jennings]{Billy X. Jennings} pressed the importance of knowing how to operate a memorigram machine as a Panther. Similarly, \hyperref[sec:Dong]{Harvey Dong} articulated the great measures that a San Francisco Asian American Movement community center took to spread their newsletter using a sister movement's Gestetner (Figure 7) machine: 
\begin{quote}
    \textit{“We knew that the Black Panther office in Oakland had a machine that could make a Gestetener sheet and that machine was expensive. So I remember driving over with our layout copy… and paying them 50 cents to have the master copy made and then taking it back and putting that into our Gestetner machine… and then you paste them up on lamposts.”}
\end{quote}

\begin{figure}[h]
  \caption{Gestetener Printing Machine \cite{Victorian}}
  \centering
  \includegraphics[width=5cm]{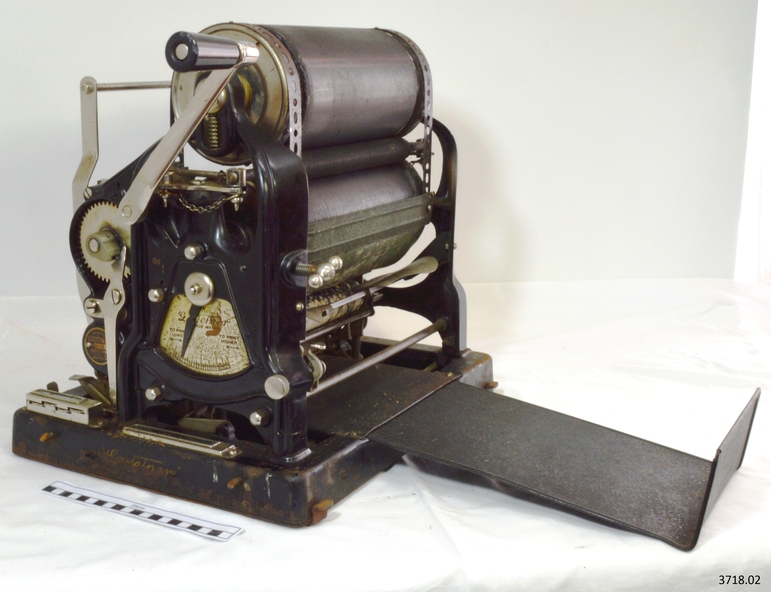}
\end{figure}

\hyperref[sec:Dong]{Harvey Dong} juxtaposed another onerous experience with the relative ease of posting an Instagram infographic today: 
\begin{quote}
    \textit{“I brought a stack of political newspapers [to Vancouver]. I would get stopped at the border and they’d ask, ‘Why are you bringing 50 copies of one newspaper?’… Today, it’s relatively easy. You type something, you research something, you proofread it, and you click a button and it goes out there.”} 
\end{quote}

Instagram infographics enable ethnic movement activists to replace physically exhausting operations with efficient tasks. A graphic through an online template can be crafted in minutes, while drafting a master copy of a newsletter takes hours and printing copies takes even longer. Likewise, the distribution of an Instagram infographic takes seconds, while passing out newsletters requires significant time and labor from multiple movement participants.

\textsc{Financial Limitations. } \label{sec:1.2.2} Second Revolution activists needed to practice resourcefulness to overcome government and institutional movement suppression, which added friction to the information-spreading process. \hyperref[sec:Dong]{Harvey Dong} continued that while many Bay Area ethnic movement groups had access to the Waller press to print newspapers for 500 dollars per issue, \textit{“you had to come up with the money.”} News outlets were reluctant to promote ethnic movement messages, forcing activists to pay substantial production costs for personal newsletters. In stark contrast, free Instagram infographic templates are widely available today on platforms such as Canva\footnote{https://www.canva.com/instagram-posts/templates/}, eliminating the financial restrictions that once established an upper bound on the information-sharing capabilities of activist groups.

\textsc{Geographical Constraints. } \label{sec:1.2.3} All three Second Revolution activists noted that their movement entrance was largely attributable to luck: being in the right place at the right time. For example, UC Berkeley's Free Speech Movement of the 1960s---which involved sit-ins and mass protests against the Vietnam War---is credited with the first revolt of the era to bring "mass civil disobedience tactics in the civil rights movement" to college campuses \cite{FSM}. \hyperref[sec:Dong]{Harvey Dong} believes that a significant reason behind their radicalization was attending UC Berkeley since the work of UC Berkeley’s Free Speech Movement helped curate \textit{"a tremendous amount of information"} in an atmosphere that was conducive to socio-political dialogue.

Likewise, \hyperref[sec:Jennings]{Billy X. Jennings} emphasized the significance of attending college in the same city as the Black Panther headquarters:

\begin{quote}
    \textit{“Prior to 1968 in America, there was no Black history, no Black education course on a public basis… Fortunately, for us, the national headquarters was in Oakland, so we had the benefit of leadership knowledge as well as… political education classes."} 
\end{quote}

Likewise, \hyperref[sec:Gonzalez]{Pablo Gonzalez} highlighted that being raised in Berkeley meant being the \textit{“product of movements.”} Thus, Second Revolution radicalization environments were largely reliant on the location of activists' upbringings and college campuses.

\textsc{Bypassing Borders. } In contrast, modern-day technological advancements circumvent traditional geographical constraints. \hyperref[sec:Celestin]{Mary Celestin} articulated:
\begin{quote}
    \textit{“Technology has an immense amount of power to actually bring [together] a lot of movements that would be otherwise fragmented.”}
\end{quote}

While physical tools such as newspapers can effectively disperse information to generate a movement ideology in a local community, infographics generate a unified front by curating fluid global educational channels. Indeed, posting an infographic has the ability to bridge people on opposite ends of a country. During a pandemic, \hyperref[sec:Perez]{Carmen Perez} utilized virtual methods including Instagram infographics to assemble and prepare 35,000 New York City-based Black Lives Matter protesters. Not only was the timespan of training only four days, but \hyperref[sec:Perez]{Perez} was also located remotely in California. \hyperref[sec:Perez]{Perez} added: 
\begin{quote}
    \textit{“It allows us to stretch our work beyond our immediate physical space.”} 
\end{quote}

\hyperref[sec:Raza]{Zoha Raza} confirmed that while in-person events are confined to local residents, virtual events are reached by all Instagram followers and \hyperref[sec:Dong]{Harvey Dong} explained that the Black Lives Matter movement has used virtual tools to spread into suburbs and isolated communities. 

\hyperref[sec:Jennings]{Billy X. Jennings} claimed that if the Black Panthers had Instagram infographics, they would \textit{"cross all racial and class lines"}:

\begin{quote}\textit{“The Black Panther Party would not be just confined to a college campus or a Black community. It would be read by the masses of America.”} \end{quote}

The Black Panther Party hosted several social services such as the Breakfast Program that served 20,000 school children daily \cite{Collier15}. While the Breakfast Program was incredibly expansive and impactful, perhaps if the Black Panthers had Instagram infographics, Black folks living nowhere near Oakland could have engaged in radical thought and created branches of Panther social service programs to feed the hungry and care for the ill. This alternative reality could have affected thousands by operating at a much larger scale. By expanding access to movement-wide ideologies and information, Instagram infographics are able to reach, educate, and galvanize otherwise unreachable community members. As a result, Instagram infographics can play a role in mass education, establishing collective movement underpinnings that were \textit{less} widespread and unified in the past. 

\textsc{Echo Chambers as a Counterargument. } \label{sec:1.2.4} Instagram infographics do not permeate all location restrictions due to information bubbles which are created either by online echo chambers or the exclusion of offline communities. \hyperref[sec:Jarmakani]{Hesham Jarmakani} explained: 

\begin{quote}
    \textit{“So while we may be able to get some of the voters from industrialized metropolitan cities… where a lot of younger people use social media, we also have to factor in the opportunity cost of focusing on social media alone resulting in us not being able to focus on other demographics, especially older people, which, whether we like it or not, tend to show up consistently to the [election] polls.”}
\end{quote}

Instagram infographics cater to urban areas and younger audiences, an unfortunate similarity with Second Revolution activism. People tend to "follow" accounts with similar belief systems, so information is likely to remain within concentrated bubbles. While Instagram infographics chip away at many physical, financial, and geographical obstacles, they are limited by echo chambers, an age-old continuity.

\subsubsection{A Unique Combination of Features} \label{sec:1.3}

Instagram infographics are able to \textit{uniquely} maximize ethnic movement scope by allowing activists to leverage features that alternative social media platforms lack.

\textsc{Room For Creativity. } \label{sec:1.3.1} \hyperref[sec:Perez]{Carmen Perez} suggested that while Facebook limits private users to 5,000 friends and Twitter limits characters per post, Instagram allows movement leaders to include more content with increased flexibility, allowing these customizable posts to reach a broader audience. Personalizability and flexibility are components of connective action, yet in the case of Instagram infographics, they facilitate collective education.

\textsc{“Story” Feature. } \label{sec:1.3.2} \hyperref[sec:Celestin]{Mary Celestin} spotlighted a scope-enhancing Instagram feature:

\begin{quote}
    \textit{“They have a really cool approach to permanency and impermanence. Their work with the stories was really, really smart… When it comes to digestible, bite-sized information that can be consumed and then shared and then commented on, Instagram does a really good job.”}
\end{quote}
By allowing activists to create permanent posts on their feed, while giving movement participants the opportunity to share these permanent posts to their stories temporarily and in seconds, Instagram’s “Story” feature facilitates the fluid spread of infographics. 

\textsc{A Visual Gallery. } \label{sec:1.3.3} Instagram was, from its birth in 2010, designed to be a photo-sharing application \cite{Blystone20}. This is precisely why \hyperref[sec:Jarmakani]{Hesham Jarmakani} accredits Instagram with being ideal for infographic activism: 
\begin{quote}
    \textit{“Instagram is very accessible because it is a platform solely devoted for imagery… We’ve been in the age of visual appeal. I mean, when we look at musicians, all of our musicians have to be attractive, even though we care more about their music than how they look because we want to have something that’s visually enticing as shallow as that is… that’s the way the world operates right now… I feel like Instagram is the most ideal platform because it is a platform solely dedicated to that. When you click on Instagram, you know what you’re going to see—just picture after picture after picture.”} 
\end{quote}

Visuals provide a “hook” into a post’s content; a post without an infographic may never draw in a movement participant, but placing that same thought within a visual representation might entice viewers to engage. \hyperref[sec:Raza]{Zoha Raza} agreed, emphasizing that infographics have eye-catching depictions that draw audiences in: 
\begin{quote}
    \textit{“It’s all about pictures. The pictures capture the person and then they look at the caption.”}
\end{quote}

Thus, Instagram infographics attract participants and facilitate rapid information spread through eye-catching visual appeal and features that are versatile for both the activist and the information spreader. As a result, Instagram infographic activism has become repost-oriented rather than solely individualistic---most participants post movement infographics to their Stories more than they create infographics. So while an infographic is purely connective when originally designed and posted, viral infographics that ripple through Instagram accounts document collective movement ideologies and values shared by posters.

\subsubsection{A Bite-Sized Entry Point} \label{sec:1.5}

By trimming down hefty issues into digestible entry points, Instagram infographics are able to introduce a wider audience into movements. \hyperref[sec:Perez]{Carmen Perez} defined an infographic entry point as \textit{"an opportunity for learning and online dialogue"} which eventually develops into offline discourse and movement education beyond the scope of the infographic. Infographics are ideal for individuals who lack time, training, and knowledge of where to start. \hyperref[sec:Raza]{Zoha Raza} provided a key example:
\begin{quote}
    \textit{“It’s possible that people don’t have time to read through ‘Who is a Muslim, what does a Muslim stand for.’ [With Instagram infographics], someone is scrolling through Instagram and they see ‘Swipe to learn more about the American Muslim Movement’ and they find new information that they didn’t know before and it’s right there, it’s in their face.”}
\end{quote}

The founder of \hyperref[sec:KeysOfPalestine]{Keys of Palestine} noted that Instagram functions as their organization's entry point by drawing attention to their primary content and enabling users to travel to their website or podcast for more concentrated information. In this way, a movement participant's educational experience is not limited to the content they see on Instagram; rather, it extends to the plethora of resources which the infographics introduce. Instagram infographics provide digestible introductions to movement issues and avenues for subsequently pursuing one’s own research, increasing the scope of movement participation to encompass otherwise busy or initially uninterested individuals. 

\subsubsection{Strategies and Metrics for Success} \label{sec:1.4}

Beyond the virality of the infographic itself, New Age activists have curated several strategies and metrics to optimize their infographic impact and maximize their audiences. 

\textsc{Strategies}
\label{sec:1.4.1} \begin{itemize}
    \item Themes: Both \hyperref[sec:KeysOfPalestine]{Keys of Palestine} and \hyperref[sec:Raza]{Zoha Raza} utilize color schemes for their posts in order to garner attention. \hyperref[sec:KeysOfPalestine]{Keys of Palestine} uses keys to represent the right of return to one’s homeland, as well as tea and seeds to culturally connect to Palestinians. Unified symbols, themes, and colors allow movement participants to gain familiarity with the visual structure of a movement's posts, allowing recognizable movement posts to stand out in a viewer's feed.
    \item Infographics as a Highlighter: Since \hyperref[sec:KeysOfPalestine]{Keys of Palestine} is focused on posting videos of Palestinain elders’ stories, they utilize infographics to draw attention to lines or quotes of significance from the videos. Highlighters draw in movement participants by providing eye-catching hooks and understandable takeaways.
    \item Partners: \hyperref[sec:Raza]{Raza}, \hyperref[sec:Perez]{Perez}, and \hyperref[sec:Jarmakani]{Jarmakani} all emphasized the significance of tagging partners in their posts and having their partners repost their content in their feeds to break into their social media bubbles. \hyperref[sec:Jarmakani]{Jarmakani} highlighted the importance of partner engagement:
    \begin{quote}
        \textit{“One thing that definitely helps is having people shoutout whatever posts we make, but something even more important is having people within different solidarity groups also share… We do have 1,000 followers, but 800 to 900 of them probably know each other or are within the same area or the same demographic. So it’s important to have more solidarity orgs share these so they can reach different individuals, different communities, different demographics.”}
    \end{quote}
    
    Social media echo chambers can hinder information dissemination by siloing radical thought into bubbles that are inaccessible to other groups. When partners re-post infographics, activists break into their partners' information bubbles, exceeding the limitations of their Instagram following. In doing so, movements are able to share a collective ideology across distinct information bubbles and communities.
\end{itemize}

\textsc{Metrics. } \label{sec:1.4.2} Some activists analyze specific metrics to optimize their viewership. \hyperref[sec:Raza]{Zoha Raza} looks to likes, new followers, and shares to gauge success, while \hyperref[sec:Jarmakani]{Hesham Jarmakani} evaluates story shoutouts and event attendance. \hyperref[sec:Perez]{Carmen Perez} explained the value of tracking metrics:
\begin{quote}
    \textit{“I did a post people were really invested in, so that made me want to post more examples of the way in which Latinos have shown up for the African American community… That has me have an internal conversation about what it looks like to actually post more of that versus maybe something that did not do so well.”} 
\end{quote}

On the contrary, \hyperref[sec:Celestin]{Mary Celestin} noted that posts that are more broad and aesthetically pleasing do best, but that San Jose Strong does not seek to adapt their strategies based upon which infographics do better: 
\begin{quote}
    \textit{“I think the approach of building community online on an Instagram platform is very different from being an influencer trying to get as many likes… as possible… They’re a different marketing style.”}
\end{quote}

While there are commonalities between the strategies and metrics of ethnic movement activists who utilize Instagram, some take into account those metrics in decision-making for future posts, while others maintain that ethnic movement activism should be less about what the public would find more “likeable” or “sharable” and instead simply about the content that should be broadcasted. This represents a unique tension between educational \textit{content} and educational \textit{reach}---when infographics hold the weight of collective thought, activists must prioritize the expansiveness of their message or the message itself. In this sense, activists can opt for connective customizability, but also have the option to prioritize broader collective messaging.

\subsubsection{Accessibility of Education} \label{sec:1.6}

Instagram infographics pose both accessibility improvements and concerns when it comes to dispersing movement information.

\textsc{Decreasing Educational Barriers. } \label{sec:1.6.1} 
\hyperref[sec:Perez]{Carmen Perez} highlighted that infographics improve accessibility for those without a college education:
\begin{quote}
    \textit{“Information in the past that did not use infographics was only accessible to those that had more resources versus [today, you can get involved] if you have a phone or a computer… If you went to college, then you understood what the Chicano Movement was about. Now, if you are online, you can be informed of the origin of the Chicano Movement.”} 
\end{quote}

Indeed, perhaps if \hyperref[sec:Jennings]{Billy X. Jennings} or \hyperref[sec:Dong]{Harvey Dong} had been raised in the age of the infographic, they would have been able to access radical thought prior to their college educations. \hyperref[sec:Jarmakani]{Hesham Jarmakani} noted that the digestible format of infographics can make information more accessible for those without familiarity with dense academic texts:
\begin{quote}
    \textit{“As a person who grew up speaking Arabic and English and never fully developing or fully grasping one language as other people do, I tend to find a lot of progressive texts very inaccessible where most of the time, it tends to be a lot of French philosophers trying very, very hard to out-jargon one another. I have a copy of Foucault’s Discipline and Punish that I’ve never been able to get through because it’s just very dense. I feel like Instagram infographics are able to break down very complex ideas into very bite-size materials.”}
\end{quote}

Reducing the necessity of formal education is crucial, as school funding and resources remain segregated based on race and socioeconomic status: In the United States, white students out-graduate students of color by a 10 point margin \cite{Rates}. Instagram infographics ameliorate the flaws of an unfair schooling system; a college education is no longer a precursor to movement engagement, improving the accessibility of movement introductions.

\textsc{Offline Movement Components for Accessibility. } Instagram infographic activism necessitates supplementary offline educational means to include individuals who lack access to technology; otherwise, the use of Instagram infographics in isolation poses an accessibility concern. \hyperref[sec:Gonzalez]{Pablo Gonzalez} explained:
\begin{quote}
    \textit{“Our communities don’t always have access to social media… When we’re talking about making policy that will impact communities of color, but also poor communities, in social media, we reproduce the very notion of a middle class family but never do I see conversation about poor working class families, so that already tells you that it over-reproduces a discourse that already happens in society that has no consideration about a growing population in the United States that's poor, under the poverty wage, [increasingly] houseless that don't have resources.”}
\end{quote}
Indeed, people who make infographics likely have a phone and internet, but an unhoused person could learn about, partake in, or even lead an on-the-ground protest, as evidenced by the houseless movement spearheaded by Black women in Oakland \cite{Paulas20}. This concept is summarized by \hyperref[sec:Jarmakani]{Hesham Jarmakani}:
\begin{quote}
    \textit{“It takes privilege having to use Instagram, let alone having a phone and I feel like that’s the kind of social capital that we don’t really discuss in a lot of organizing where some people might not have access to these things.”}
\end{quote}
In a world where people of color are often constrained to the bottom of the economic pyramid \cite{APA17}, it is crucial that ethnic movements are inclusive of those who cannot access technology, something that a movement entirely reliant on Instagram infographics does not offer. Ethnic movements often seek justice for individuals who are suppressed by oppressive institutions; therefore, online movements can fall short of meeting movement objectives by keeping vital information from those most impacted by movement topics.

\textsc{A Continuity: Art as a Method of Increased Accessibility. } \label{sec:1.6.3} Instagram infographics and ethnic movements of the past both use visuals as a mechanism of increased accessibility. Just as \hyperref[sec:Jarmakani]{Jarmakani} noted that those with language barriers are able to process infographics more clearly, \hyperref[sec:Jennings]{Billy X. Jennings} explained that art in the Black Panther newspaper was vital to accessibility:
\begin{quote}
   \textit{“The Black community is not a reading community during the 60s and 70s. We had a lot of illiteracy… so the artwork in the Panther paper was very important because people looked at those images… People could relate to that artwork. They might not read the article… but that artwork would give them the basis of what the article is about.”} 
\end{quote}

Long before the age of the Instagram infographic, ethnic movement activists leveraged visuals to empower their communities and foster change. The Black Panther Party newspaper depicted criticisms of institutionalized racism, police brutality, poverty, and imperialism through thought-provoking imagery (Figure 6) that was able to "cut through the high illiteracy rates in poor communities," heightening movement accessibility and scope \cite{Sudbanthad08}.

Likewise, the Chicano Movement is reputed for their printmaking, graphics, and murals; in the Los Angeles area alone, more than a thousand Chicano murals have been created since 1965, like those in Figure 8. Chicanx activists "rooted their muralism in support for the struggles of the poorest, most exploited members of their communities," centralizing themes such as Aztlàn and labor exploitation \cite{Reed05, Mejorado20}.

\begin{figure}[h]
  \caption{Chicano Murals \protect\footnotemark}
  \centering
  \includegraphics[width=5cm]{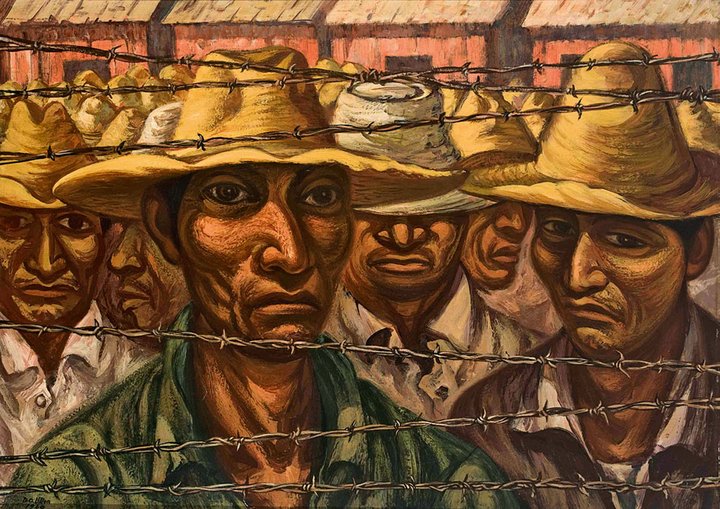}
  \includegraphics[width=5cm]{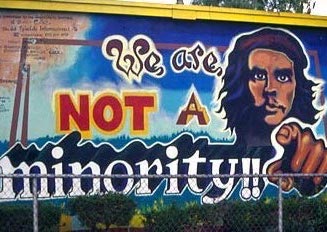}
\end{figure}
\footnotetext{Image content descriptions and citations are in A.6.}

Ethnic movement art catalyzed change by circumventing educational obstacles, curating movement culture, and evoking thought and emotions that image-less words could not encompass. Illiteracy should not be a roadblock to movement participation. By presenting content through visuals, infographics (just like their newspaper-art predecessors) convey messages to those with lower levels of literacy, reflecting an evolved movement strategy whose roots have existed for decades.

Overall, while Instagram infographics increase the accessibility of information through visual representations and a lack of a necessity for higher education, they are limited in accessibility when it comes to bringing offline communities to the table. When offline educational means exist, Instagram infographic activism can disperse movement education to otherwise excluded movement participants, further expanding the movement unified front. 

\subsubsection{Examining Scope Through Black Lives Matter Engagement}

Since more traction on a post indicates that the post either reached or resonated with more people, increased engagement suggests increased scope of movement participation. As such, we sought to explore whether infographics cultivate more engagement (measured by likes and comments) on the Black Lives Matter Instagram page.

Figure 9 and Table 4 denote the percentages of likes and comments belonging to infographic images aggregated across 2014-2020 juxtaposed with the percentage of posted images which were infographics.

\begin{table*}[h]
  \caption{Comments and Likes in Comparison to Infographics}
  \small
  \label{tab:commands}
  \begin{tabular}{c|c|c} 
    \toprule
    Infographic Percentage&Likes From Infographics&Comments From Infographics\\
    \midrule
    36.99\% & 50.44\% & 54.19\%\\
    \bottomrule
  \end{tabular}
\end{table*}

\begin{figure}[h]
  \caption{Overall Comments and Likes in Comparison to Infographics}
  \centering
  \includegraphics[width=9cm]{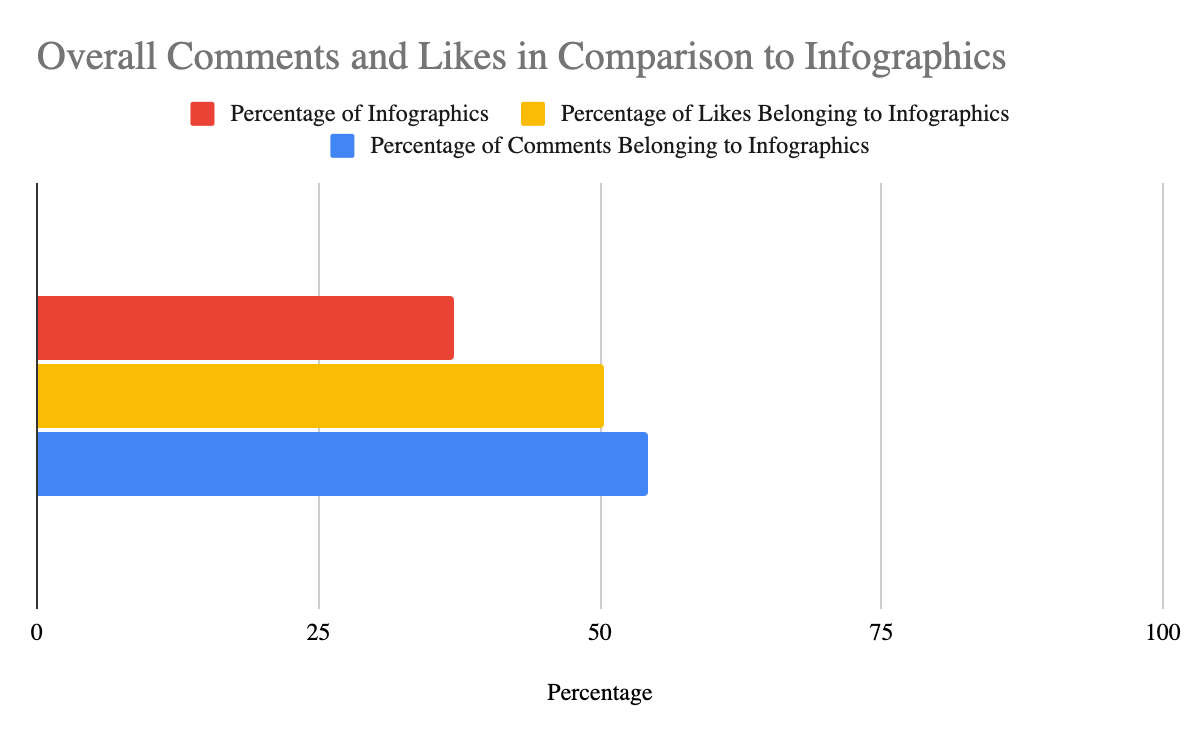}
\end{figure}

Overall, while only 36.99\% of posted images were infographics, 50.44\% of likes and 54.19\% of comments belonged to infographics from 2014 through 2020. This suggests that the Black Lives Matter Instagram's engagement is concentrated within their infographic posts. We further analyzed the distribution of likes and comments to confirm the correspondence between more engagement and more infographics. Figure 10 and Table 5 describe the infographic post percentage in the 0th-20th, 20th-40th, 40th-60th, 60th-80th, and 80th-100th percentile ranges for both post like count and comment count. The dotted lines in Figure 10 mark 36.99\%, the overall infographic post percentage.

\begin{table*}[h]
  \caption{Black Lives Matter Instagram Infographic Post Percentage By Like and Comment Percentiles}
  \small
  \label{tab:commands}
  \begin{tabular}{c|c|c|c|c} 
    \toprule
    \thead{Percentile\\Range}&\thead{Raw Like\\Count Range}&\thead{Likes Percentile\\Infographic Percentage}&\thead{Raw Comment\\Count Range}&\thead{Comments Percentile\\Infographic Percentage} \\
    \midrule
    0th-20th &65-113&1.86\% &0-2& 2.04\%\\
    20th-40th &113-1,335&34.36\% &2-38& 27.43\%\\
    40th-60th &1,335-5,792&49.38\% &38-129& 51.22\%\\
    60th-80th &5,792-28,844&43.21\% &129-312& 44.10\%\\
    80th-100th &28,844-598,222&56.17\% &312-3,465& 57.67\%\\
    \bottomrule
  \end{tabular}
\end{table*}

\begin{figure}[h]
  \caption{Black Lives Matter Instagram Infographic Post Percentage By Like and Comment Percentile}
  \centering
  \includegraphics[width=9cm]{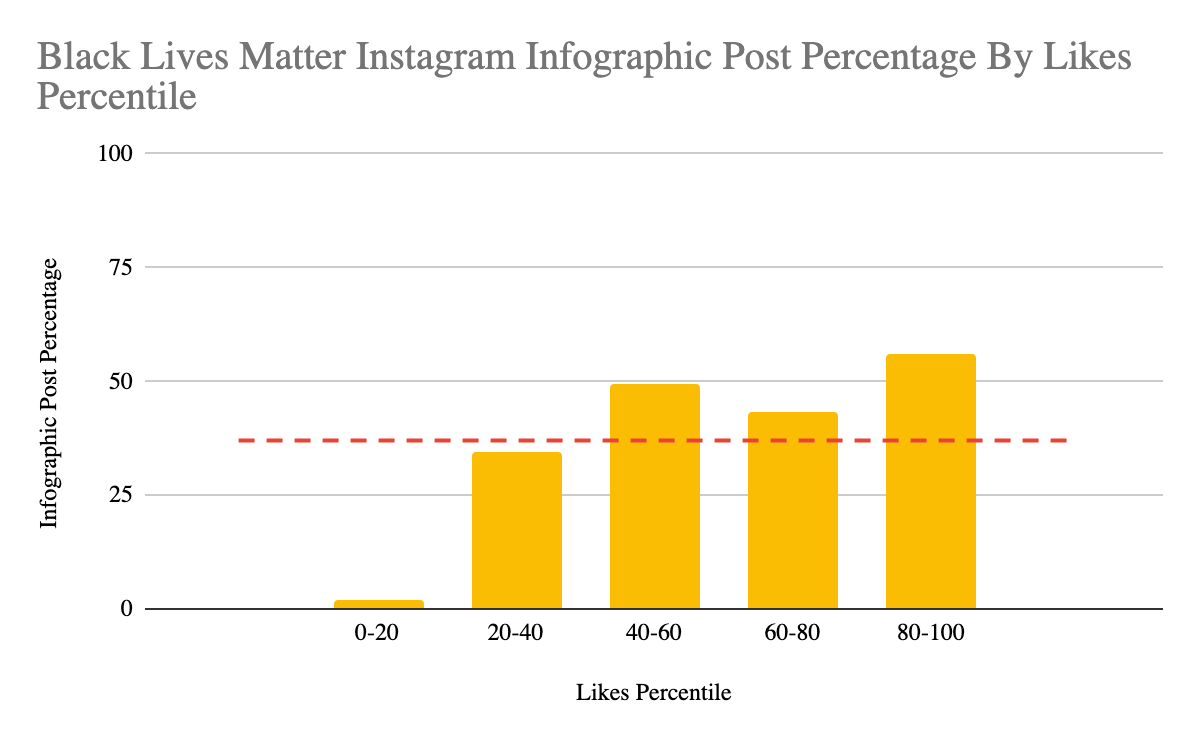}
  \includegraphics[width=9cm]{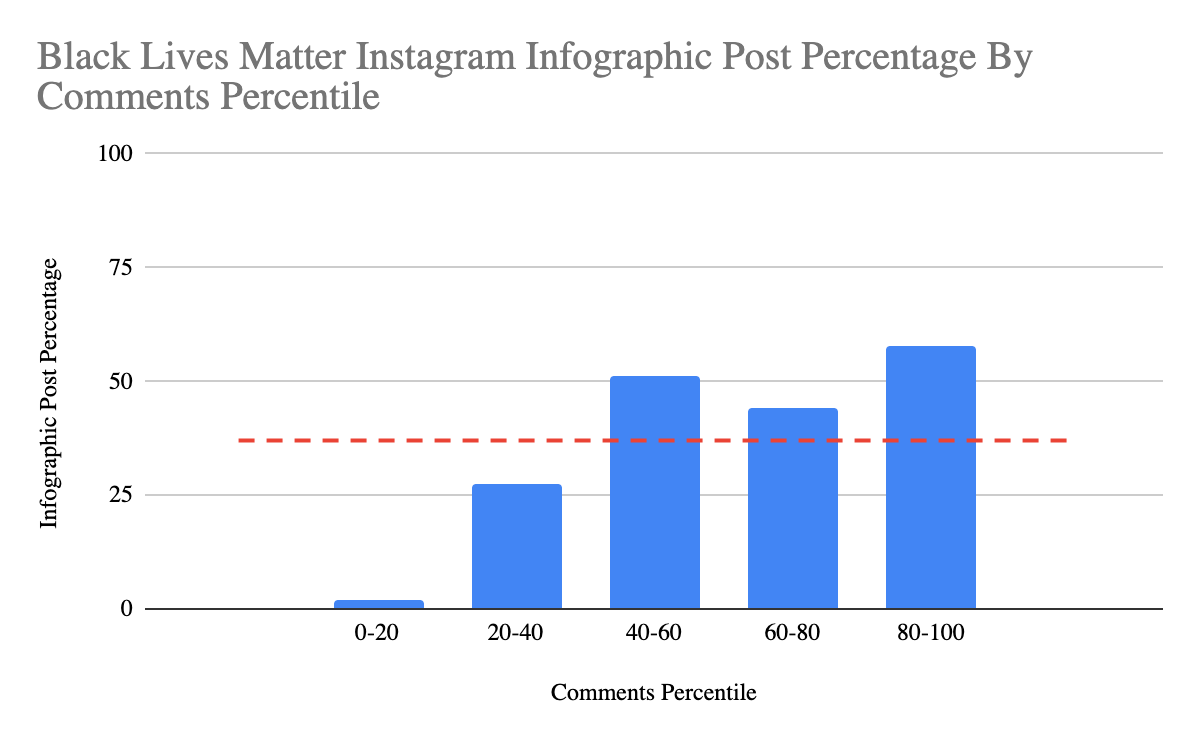}
\end{figure}

For both likes and comments, the 40th-60th, 60th-80th, and 80th-100th percentile ranges all exceed the overall infographic post percentage of 36.99\%, showing that posts on the upper half of the like and comment distributions are more likely to be infographics than posts with less traction. Furthermore, the 80th-100th like and comment percentile ranges are comprised of 56.17\% and 57.67\% infographic posts, respectively; the top 20\% most liked and 20\% most commented posts contain the highest portion of infographics. Our examination of engagement percentiles supports our aggregated statistic by indicating that more engaged posts disproportionately contain infographics.

When posts receive significantly more engagement, it is likely that they have reached or resonated with more people to accumulate those likes and comments. Therefore, our case study supports our qualitative finding that Instagram infographics increase the scope of information dissemination. While the efficacy of the infographic may depend on the content of the comments or the action participants take after engaging with the post, our findings show that infographics are widely distributed and viewed. This scope facilitates the education of the masses, an invaluable prerequisite to collective action. 

\subsubsection{Scope: Connective and Collective Action} \label{sec:1.7}

Collective action necessitates that a movement educates and informs its participants with movement-wide ideologies, values, and objectives. Despite the fact that Instagram infographics are subject to information echo chambers and urban information bubbles, this is outweighed by New Age activists’ careful crafting of their infographic strategies, the visual appeal and quick-spreading features of Instagram, as well as infographics’ ability to significantly reduce geographical, physical, financial, and schooling constraints. However, Instagram infographics do necessitate offline educational means to ensure economic accessibility.

When it comes to heightening the scope of information dissemination, infographic activism is connective in that infographics are digitally mediated, information begins as personalized content, and dispersed information is not always managed by one centralized organization. Infographic activism has also become collective by establishing movement-wide priorities and ideologies at an unprecedented scale. Pure connective action is individualized and focuses on personalized expression; however, while infographics may be personalized at the point of creation, their rapid and largescale circulation makes content more communal than individual, developing collective action frames. What was once newspapers and flyers is now infographics---values and action-items are documented and spread across geographical and socioeconomic lines to generate movement-wide ideologies. 

\subsection{Practicing Reconciliation for Collective Credibility} \label{sec:2}

Collective action involves the curation of a unified movement front. As such, movement organizers must forge strategies to reconcile disagreements to preserve movement-wide resilience \cite{Bennett12}. Connective action lacks this need: since content is digitally mediated and personalized, two distinct actors discussing one topic can share diametrically opposed ideologies with little movement-wide side-effects. This is far from the case in Instagram infographic activism, where movement participants across the platform use seemingly connective features to reframe ideologies and resolve differences in order to sustain consistent movement messaging.

\subsubsection{Cutting Corners} \label{sec:2.2}

Instagram infographics are particularly susceptible to misinformation. By over-summarizing and omitting critical context, Instagram infographics can serve as a form of movement erasure. A post's credibility is compromised when complex issues are underexplained due to Instagram’s 10-slide limit. \hyperref[sec:Jarmakani]{Hesham Jarmakani} explained why some issues are better saved for more “roomy” information dissemination mechanisms:
\begin{quote}
    \textit{“It’s very, very hard to try and fit a very, very complex issue into just six or seven slides and I personally find it very demoralizing [when infographics say things like] ‘here’s everything that’s happened in Syria in six or seven slides.' I feel like it takes away from the severity of the situation… It’s very easy, if not necessary, to have to cut corners to try and fit it into that infographic.”}
\end{quote}

When reducing largescale issues that organizers have fought for for years, infographics can be a disservice to the movements themselves.  \hyperref[sec:Raza]{Zoha Raza} confirmed:
\begin{quote}
    \textit{“You are trying to fit all of the information about a whole subject on a small square… Two sentences cannot sum up an entire movement…”}
\end{quote}

Therefore, infographics that leave out crucial information in an attempt to compartmentalize rigorous topics can spread misinformation and undermine communal trust. This is a counterpoint to the bite-sized entry point feature of Instagram infographics; some exceedingly complex topics are impossible to pack into simple, bite-sized formats. The affinity for cutting corners with Instagram infographics creates possibilities for misinformation, compromising movement credibility and creating internal rifts and ideological disconnects.

\subsubsection{Preserving Credibility} \label{sec:2.3}

To combat the aforementioned misinformation concerns, activists use several strategies to build collective trust and legitimacy, bridging differences to preserve unified, consistent movement messaging. And yet, they use personalizable, online features to do so.

\textsc{Resources. } \label{sec:2.3.1}
\hyperref[sec:Perez]{Carmen Perez} mentioned adhering to a specific infographic structure: a title slide, a definitions slide, arguments slide(s), and \textit{always} a sources slide.
\hyperref[sec:Celestin]{Mary Celestin} similarly put a particular emphasis on sharing resources, especially with those who question one’s credibility. In response to people criticizing an effort to remove police from schools, \hyperref[sec:Celestin]{Celestin} explained:
\begin{quote}
    \textit{“We responded by saying, ‘Hey, everyone, we saw some comments of folks concerned about school shooters via our demands to remove cops from schools. Fear of a school shooting is a valid concern. It is important to understand the full scope of this petition. This is not a call for lawlessness in our community. This is a call to create new safety measures formed by all members of our community. For those who are genuinely curious to learn more about police in schools, please check out the following resources.’ And then we linked six different sources.”} 
\end{quote}

By including resources, activists ensure that critics cannot undermine their credibility, prompting participants to trust, follow, and help promote the ethnic movement. Crucially, this reconciliation tactic relies on the connective nature of infographics---activists are able to include resources and establish collective trust \textit{due} to the connective customizability of the infographic.

\textsc{Taking Updates. } \hyperref[sec:Celestin]{Celestin} revealed another crucial facet of building legitimacy:
\begin{quote}
    \textit{“In terms of building trust, it comes from engaging and also being willing to take updates… One of my first [Instagram infographic] guides involved defunding the [police] budget in San Jose and that involves a lot of fact checking and a lot of touching base with people. People would have comments… Sometimes we’ve taken it and been like ‘Oh, that was right, that was a mess-up on our end.’"} 
\end{quote}

The ability to update an infographic yet again relies on its connective personalizability and flexibility---no organization or largescale entity must be consulted by an activist before updating a post. By being willing to improve one’s stance when presented with new information, ethnic movement leaders are able to showcase their advocacy is not only based in passion for radical change, but also a genuine effort to spread truth. In this way, the adaptation of movement messaging in response to communal pushback exemplifies infographic activists' desire to build collective trust using a connective feature. 

\textsc{The Power in the Comments Section. } \label{sec:2.3.2} \hyperref[sec:Raza]{Zoha Raza} emphasized that the Instagram comments section acts as an inherent fact-checker. With a misinformed newspaper article, prospective ethnic movement participants cannot critically analyze or respond to the information presented. The comments section of an Instagram infographic, however, is likely to contain backlash and critical commentary in the case of misinformation, provoking viewers not to take the the faulty graphic at face value. Movement messaging thereby remains consistent since misinformed infographics are more likely to generate responses articulating the faults of an infographic, informing viewers that the graphic disrupts a collective value, ideology, or objective. Individual commenters rely on individualized decision-making, but the impact of their aggregated discourse is profoundly collective.

\textsc{Infographic Inception. } \label{sec:2.3.3} \hyperref[sec:Jarmakani]{Hesham Jarmakani} highlighted a phenomenon which we label as infographic inception: When ethnic movement activists view a misinformed infographic, they will often make their own infographics to fact-check and clarify the original post. Activists often incorporate images of the original graphic within their own, using fonts, colors, and shapes to highlight the points they are responding to. In this way, the activists use the personalizable nature of the infographic to hold the community accountable and signal to movement participants when an internal dialogue is counterproductive and disagrees with core movement values.

\subsubsection{Credibility: Connective and Collective Action} \label{sec:2.4}

Collective action necessitates the reconciliation of internal disputes in pursuit of a unified movement front. Instagram infographics threaten ethnic movement credibility by forcing activists to cut critical information out of their narrative, undercutting communal trust. However, many New Age ethnic movement activists have leveraged the customizability of the Instagram infographic to preserve credibility by taking updates, using comments as a fact-checker, and responding to faulty infographics with their own. 

If infographic activism was purely connective, individual activists would not make an effort to establish communal trust by exhausting resources to reconcile their differences---personalized content does not necessitate movement-wide acknowledgement and acceptance in order to meet its objectives. Efforts to counter faulty information, signal legitimacy, and remain accountable suggests that Instagram infographic activism is also collective: activists exhaust resources in order to bridge differences and preserve unified movement ideologies. So while Instagram infographics are connective in that no single organization is mandating the resolution of disputes and reconciliation tactics utilize connective features, movement participants themselves exercise collective action by prioritizing movement-wide trust.

\subsection{Expending Organizational Resources for Collective Transformation} \label{sec:3}

Collective action involves expending organizational resources and coordinating movement-wide action. Purely connective action does not necessitate coordination on an organizational scale and therefore lacks the intensive expenditure of movement-wide resources. Crucially, transformative legislative and societal change rarely stops at the infographic, but rather operates in inseparable conjunction with high-resource on-the-ground collective efforts.

\subsubsection{Isolated Infographic Passivity} \label{sec:3.2}

The isolated use of Instagram infographics can yield passive movement involvement. While the physical nature of Second Revolution movements preserved authenticity among participants, the effortlessness of infographic engagement can reduce virtual movement efficacy when movement-wide resources are not expended in on-the-ground efforts.

\textsc{Physical Movement as a Prerequisite to Material Change. } \label{sec:3.2.1} \hyperref[sec:Jennings]{Billy X. Jennings} outlined the effort discrepancy between virtual and physical movements:
\begin{quote}
    \textit{“You are deliberately doing something… It’s different from just flipping a switch or just sharing something. It doesn’t take much consciousness to do that… You can be a so-called revolutionary in your house, but we need activists. We need people putting theory in practice in the real time… That stuff has to be based on struggle. It has to be based on direct activity in the streets…”}
\end{quote}

More specifically, \hyperref[sec:Jennings]{Jennings} emphasized that the Black Panther Party required members to read two hours a day and attend political education classes twice a week. In this way, the Panther became consciously involved in the movement and thereby active in its growth:
\begin{quote}
    \textit{“You are the spark. You have to make the flyers. All that direct, hands-on contact was what really grew the party legacy.”}
\end{quote}

Those who invest time, energy, and effort through disobeying authority in the streets and physically calling for change feel a sense of ownership in movements. This sense of ownership, in turn, drives longterm commitment, spirit, and a desire to produce concrete outcomes. In contrast, \hyperref[sec:Gonzalez]{Gonzalez} noted that when it comes to Instagram infographic activism:
\begin{quote}
    \textit{“I can do this in my chair at home, but not have to leave my home, and that in and of itself does not lead to any particular material change.”}
\end{quote}

The lack of a physical prerequisite within pure virtual movements undercuts efficacy by eliminating the emotional connection, consciousness, and resultant action involved in traditional ethnic movements.

\textsc{A Vehicle for Complacency. } \label{sec:3.2.2} The danger of Instagram infographics is that it allows those who engage exclusively with the infographic to develop complacency in their activism, reducing efficacy. 
\hyperref[sec:Jarmakani]{Hesham Jarmakani} explained:
\begin{quote}
    \textit{“It gives people the excuse to say, ‘Well, if you just read an infographic or you just make an infographic, that’s enough work for the day’ and I feel like it lowers the bar… You think the activism starts at the infographic and that you don’t have to vote yes on Prop 16, you don’t have to go out into the protest and protest, you don't have to donate to Black Lives Matter or other organizations when you have the wallet.”}
\end{quote}

A like button does not indicate true movement involvement or dedication, but it makes the liker feel as if they have made a valuable contribution, procuring an infinite cycle of actionless likes and shares. In fact, an Instagram infographic could be shared hundreds of times without a single real-world implication. The establishment can only be swayed by concrete actions: A corrupt politician will remain eternally ignorant to a post trapped within a social sphere of complacent likers and sharers.

\textsc{The Like Button as a Deceptor. } \label{sec:3.2.3} \hyperref[sec:Gonzalez]{Pablo Gonzalez} warned of the threat posed by an infographic's like button by drawing an analogy to white America in the Second Revolution era:
\begin{quote}
    \textit{“You have white Americans in the 70s and 80s, who might have said in person, ‘I'm against racial violence’ but voting, organizing differently. They decided we needed to be jailed… they decided we needed to be bombed… they elected Richard Nixon and they furthered and promoted the genocides in Southeast Asia and in Central America. And so [white America] became a silent majority.”}
\end{quote}

Just as white America was complicit in the growth of violence, deindustrialization, and dismantling of the welfare state, a white person could very well like an infographic about putting an end to racial violence and proceed to vote "no" on a key Proposition that uplifts marginalized communities. In contrast, physical activism circumvents deception by revealing authentic intentions.

\hyperref[sec:Gonzalez]{Gonzalez} elaborated:
\begin{quote}
    \textit{“Social media can portray one thing and and actually be another because where do you find out the complexity of a person, if not through day to day in-person interaction when you're faced with making decisions and having conversations… [That’s when we] express how we see the world.”}
\end{quote}

Overall, by introducing a sense of complacency and acting as a deceptor for movement participation, Instagram infographics can fall victim to passive movement involvement, undermining concrete change. Thus, while some New Age activists identified likes as a metric of infographic success, it is possible for disingenuous likers to artificially inflate the number of individuals who took action after engaging with a post. This, to some degree, decreases the validity of infographic engagement metrics in determining movement impact. Further, it evidences that connective action in isolation can fall short of yielding transformative movement output---when the onus is on isolated individuals to act, they have full jurisdiction over the authenticity of their virtual movement participation, eliminating the accountability that exists in physical movements. As a result, infographics that lack tangible, physical components can fail to achieve movement objectives.

\subsubsection{The Necessity for Both Information and Action} \label{sec:3.3}

Ethnic movement activism necessitates a close conjunction of information \textit{and} action. Information-sharing educates the community and action allows the community to take that newfound perspective and apply it to improving society. \hyperref[sec:Jennings]{Billy X. Jennings} best described this relationship:
\begin{quote}
    \textit{“It’s a combination of both. It’s like dissecting a human being. You got a lot of parts in there and if you mess with one part, the other part ain’t gonna work. Everything is interconnected with something else… Malcolm had this theory and a lot of information he was trying to get to the people. Okay, Instagram can illuminate some basic facts with graphics. There’s nothing wrong with that. But those graphics have to be backed up with rallying people to some action…”}
\end{quote}

If Malcolm X had infographics at his disposal, he might have frictionlessly spread his message, but for the movement to be just as meaningful, his participants would need to demonstrate the same initiative as an exclusively physical setting. \hyperref[sec:Gonzalez]{Pablo Gonzalez} highlighted the fatal flaw of disrupting the information-action dynamic in favor of social media:
\begin{quote}
    \textit{“Social media defines social movements as solely based on information. Social movements are not based on information. Social movements are based on an attempt to try to change social and power relations in society… I think [Instagram infographics and social media] has been, to some degree, effective to spread awareness, but movements aren’t just based on awareness. There has to be action.”}
\end{quote}

Indeed, \hyperref[sec:Jennings]{Billy X. Jennings} explained that the Breakfast Program \cite{Collier15} emerged from fundamental socialist theories shared by the Black Panther Party, a prime example of how information is \textit{only} valuable when acted upon. \hyperref[sec:Jennings]{Jennings} explained the outcome of the Panther's information-action balance:
\begin{quote}
    \textit{“The concrete things that we learned in the party, we were able to put into practice and start institutions in the community that exist to this day.”}
\end{quote}

Ultimately, it is necessary to pair the information dissemination of Instagram infographics with direct action. Reach only matters when those who are being reached are applying shared knowledge. Otherwise, that reach serves to be futile. 

\subsubsection{Transformative Output} \label{sec:3.1}

There are promising examples of Instagram infographics helping ethnic movement activists attain tangible legislative changes, suggesting that Instagram infographics can play a role in transformative movement output when paired with on-the-ground efforts.
\hyperref[sec:Perez]{Carmen Perez} exemplified two instances: (1) \hyperref[sec:Perez]{Perez} and fellow organizers were able to raise the age of criminal responsibility in New York by creating infographics with opposing legislators' images and contact information, encouraging movement participants to demand change. (2) \hyperref[sec:Perez]{Perez} and fellow organizers helped the Monte Rosa sisters, whose brother was killed at the hands of police in Vallejo, to demand justice by crafting infographics that gave movement participants resources to sign petitions and call the governor and attorney general. \hyperref[sec:Perez]{Perez} concluded: 
\begin{quote}
    \textit{"The opportunity here is the fact that we’re able to make legislative changes by putting pressure on those elected officials we know are not on our side and bringing awareness to a broader community versus maybe just having a small meeting about it that isn’t available visibly to others…"} 
\end{quote}

In both instances, \hyperref[sec:Perez]{Perez} utilized Instagram infographics to spread crucial information and mobilize a larger group of people to fight for Black and Brown communities. Those who were unable to protest in person could still help by signing petitions, calling legislators, or sharing the infographics themselves. \hyperref[sec:Celestin]{Mary Celestin} similarly used Instagram infographics about how to contact one’s city council members and join city council meetings to help eliminate San José Mayor Liccardo’s “Strong Mayor Grab,” which would have granted him additional powers and extended his term by two years \cite{WipfLaPlante20}. 

Broadly, Zoha Raza, Carmen Perez, and Keys of Palestine all often ensure that their infographics include an action item, such as a shortened URL or a link to donate. Therefore, Instagram infographics can directly facilitate crucial legislative strides by bringing more voices to the table and equipping those voices with avenues of action.

However, each of these movement achievements necessitated that numerous movement participants utilized the information in the infographic to take movement-wide collective action. Traditional resource-intensive movement efforts---bird-dogging, protesting, calling, petitioning---were consistently utilized. Transformative Instagram infographic activism leverages the infographic as a personalizable information spreader, but also relies on tangible high-resource collective initiatives. 

\subsubsection{Examining Transformation Through Black Lives Matter Action-Oriented Language} Our qualitative findings suggest that action-oriented posting undercuts complacency by encouraging movement participants to make concrete, high-resource movement contributions. We sought to determine whether action-oriented infographics generate a more action-oriented response from movement participants in comparison to infographics with other focuses. To do so, we analyzed the comments posted on the Black Lives Matter Instagram page from 2014-2020. We isolated our sample to comments posted no later than 2020 to remain consistent with our sample of posts---i.e. we are ensuring our dataset spans 2014-2020. We also restricted our sample to posts which contained infographics (of which there were 298 with comments sections).

We define Action Words to be words from our Action Keyword Bank and Non-Action Words to be words which are not in our Action Keyword Bank. Action Comments are comments containing Action Words. Similarly, Action Infographics are infographic-containing posts which contain Action Words, while Non-Action Infographics are infographic-containing posts which do not contain Action Words.

\begin{table*}[h]
  \small
  \caption{Action Words in Black Lives Matter Instagram Comments for Action vs. Non-Action Infographics}
  \label{tab:commands}
  \begin{tabular}{c|c|c} 
    \toprule
    &Action Infographics&Non-Action Infographics\\
    \midrule
    Number of Commented Action Words & 5,026 & 6,828  \\
    Number of Total Commented Words & 79,921 & 144,129  \\
    \midrule
    \textbf{Action Word Percentage}&\textbf{6.29\%}&\textbf{4.74\%} \\
    \bottomrule
    \toprule
    Number of Action Comments & 2,968 & 4,230  \\
    Number of Total Comments & 10,508 & 17,739  \\
    \midrule
    \textbf{Action Comment Percentage}&\textbf{28.25\%}&\textbf{23.85\%} \\
  \end{tabular}
\end{table*}

\begin{figure}[h]
  \caption{Action Words in Black Lives Matter Instagram Comments for Action vs. Non-Action Infographics}
  \centering
  \includegraphics[width=9cm]{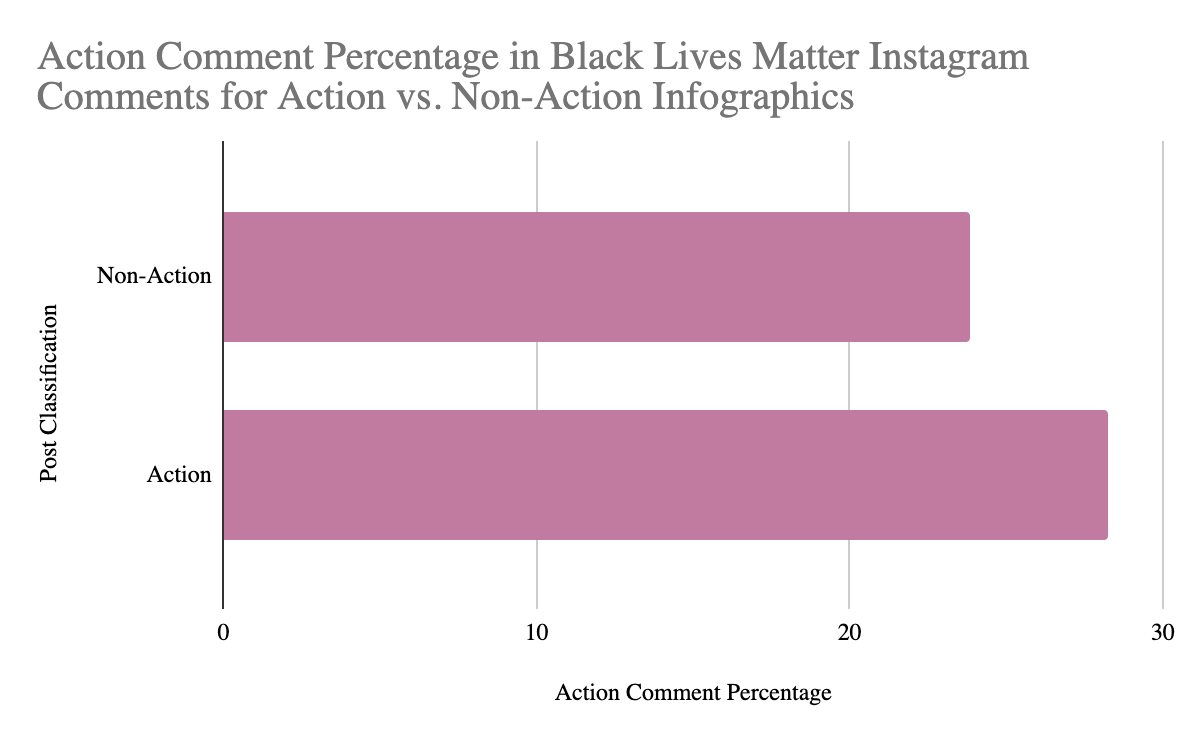}
  \includegraphics[width=9cm]{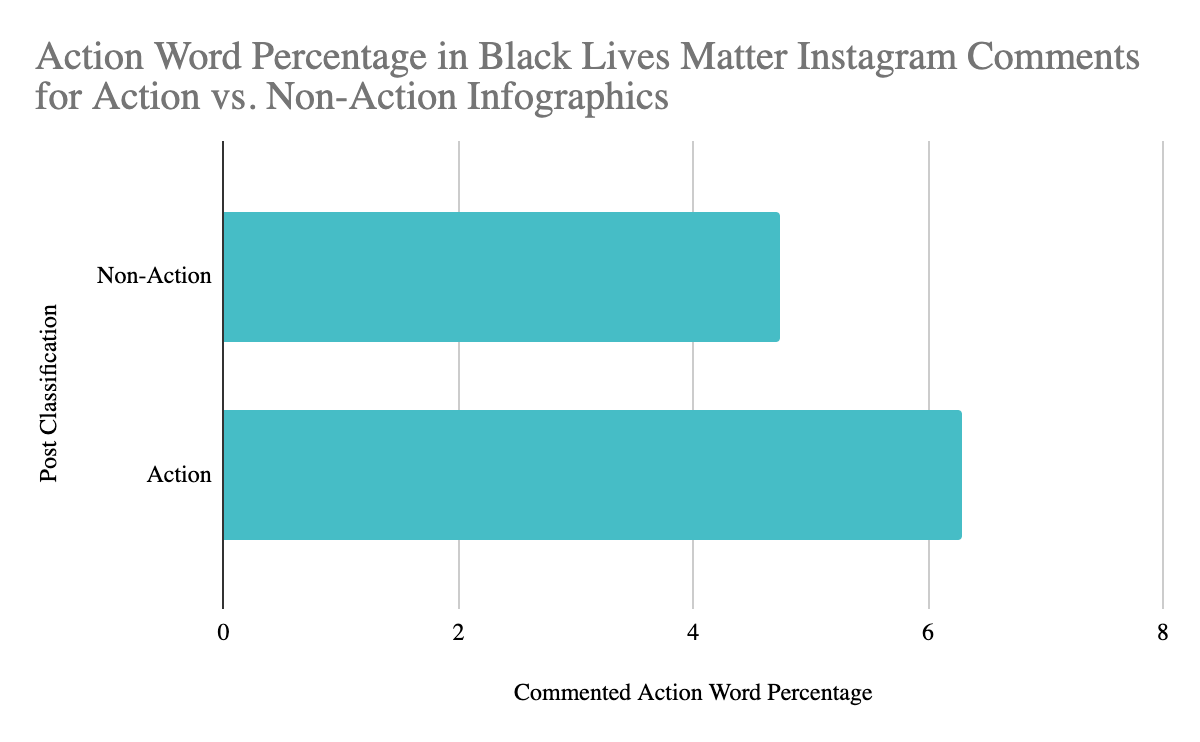}
\end{figure}

\begin{table*}[h]
  \small
  \caption{Action Words in Black Lives Matter Instagram Comments By Infographic Action Word Count \protect\footnotemark}
  \label{tab:commands}
  \begin{tabular}{c|c|c|c|c|c|c} 
    \toprule
    \thead{Post Action\\Words}&\thead{Action\\Comments}&\thead{Total\\Comments}&\thead{Action Comment\\ Percentage}&\thead{Action\\Words}&\thead{Total\\Words}&\thead{Action Word\\Percentage}\\
    \midrule
    0&4,230&17,739&23.85\%&6,828&144,129&4.74\%\\
    1&1,018&4,007&25.41\%&1,660&29,956&5.54\%\\
    2&251&857&29.29\%&399&7,135&5.59\%\\
    3&1,054&3,528&29.88\%&1,870&25,420&7.36\%\\
    4&195&655&29.77\%&310&5,558&5.58\%\\
    5&176&509&34.58\%&335&4,660&7.19\%\\
    6&43&142&30.28\%&66&1,262&5.23\%\\
    7&33&114&28.95\%&71&1,198&5.93\%\\
    9&15&60&25.00\%&22&499&4.41\%\\
    10&47&136&34.56\%&91&1,145&7.95\%\\
    11&51&172&29.65\%&80&1,451&5.51\%\\
    14&9&32&28.13\%&13&188&6.91\%\\
    19&55&239&23.01\%&83&994&8.35\%\\
  \end{tabular}
\end{table*}
\footnotetext{Rows with 0 posts or 0 comments are omitted.}

\begin{figure}[h]
  \caption{Action Words in Black Lives Matter Instagram Comments By Infographic Action Word Count}
  \centering
  \includegraphics[width=9cm]{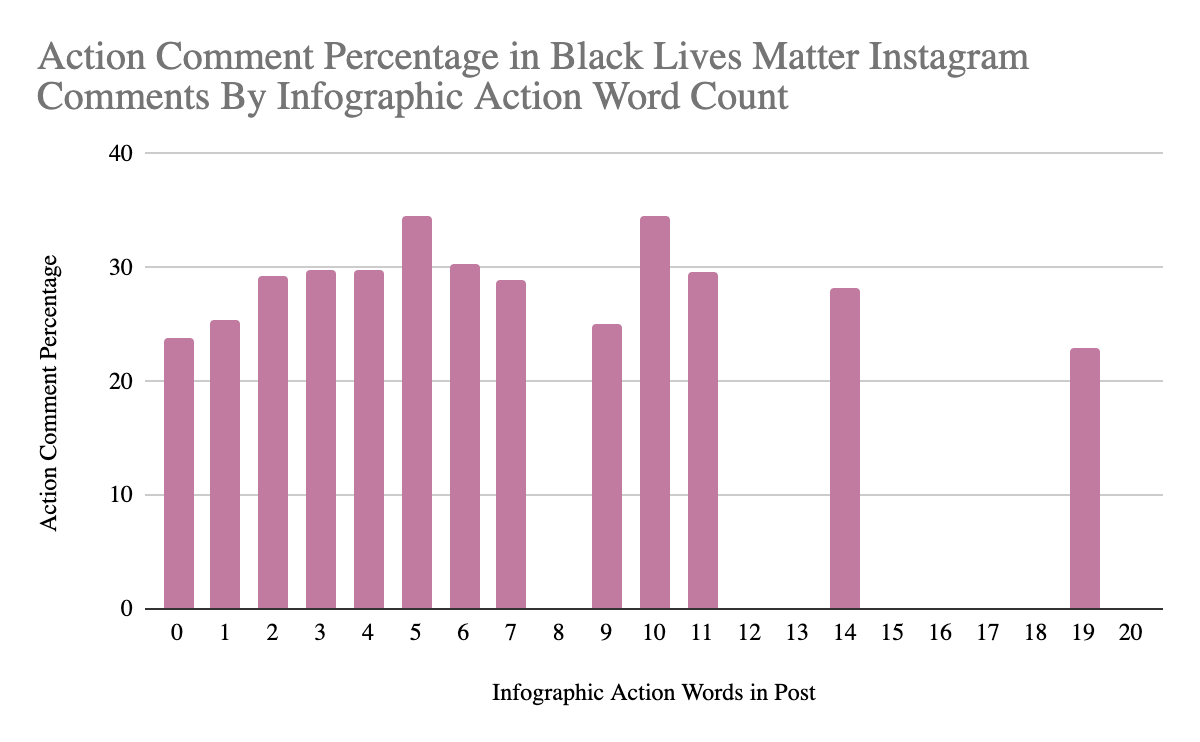}
  \includegraphics[width=9cm]{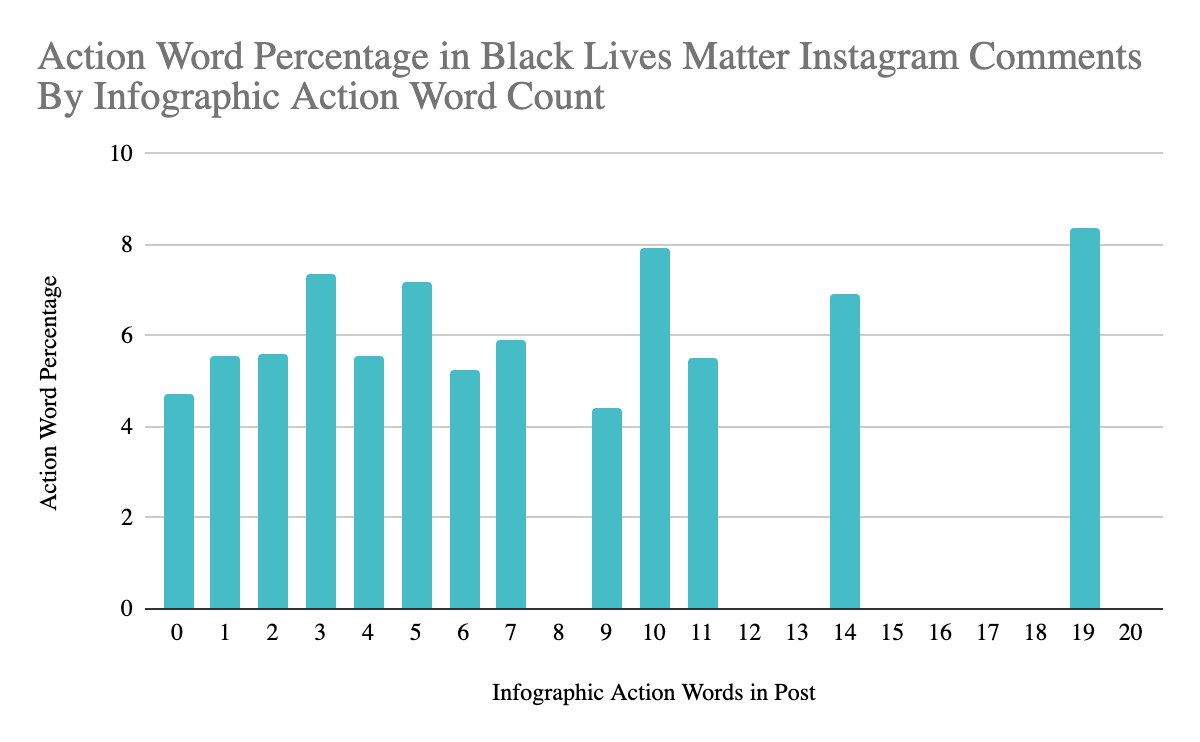}
\end{figure}

Table 6 and Figure 11 showcase the total number of comments and commented words as well as the number of Action Comments and commented Action Words for Action Infographics vs. Non-Action Infographics. While only 23.85\% of total comments were Action Comments for Non-Action Infographics, 28.25\% of comments were Action Comments for Action Infographics. In order to account for the fact that some Action Comments may contain a higher concentration of Action Words than others, we also calculated these statistics for commented words. While only 4.74\% of total commented words were Action Words for Non-Action Infographics, 6.29\% of commented words were Action Words for Action Infographics. Action Word percentages aligned with Action Comment percentages throughout the case study. Overall, our aggregated statistics reveal that Action Infographics accrue a greater portion of Action Words in the comments in comparison to Non-Action Infographics. 

Additionally, we examined the distribution of these values across the varying number of Action Words in a post. Table 7 and Figure 12 denote the total number of comments and commented words as well as the the total number of Action Comments and commented Action Words, broken up by number of Action Words present in the infographics of the post. Both Action Words and Action Comments gradually climbed from lower percentages to higher percentages until about three to five infographic Action Words, at which point the Action Comment Percentage and Action Word Percentage fluctuate mildly. We speculate that a potential reason for the three-to-five word cutoff is that calls to action exhibit diminishing additional value per word: Surpassing a certain threshold of Action Words could potentially confuse the infographic's message, intent, and next steps, trading off with the added value of more action-oriented language. When dealing with a small number of infographic Action Words, however, value is added per word. Notably, our distribution analysis revealed no significant high-magnitude outliers in percentage for words or comments, validating the soundness of our aggregated Action vs. Non-Action statistic.

When infographics contain action-oriented language, they spark a more action-oriented verbal response from the community. Infographics that call for transformative change by centering focused action items provoke the community to act upon, consider, or at the very least discuss playing a role in said change. So while an individually designed and posted Black Lives Matter Action Infographic may not necessitate movement-wide resources, the action-oriented language of such posts elicits high-resource verbal responses to a greater degree than non-action-oriented posts. Be it discussing petitions, protests, or donations, community members engage in dialogue that has the potential to emit output. This output, in turn, expends collective resources and yields concrete grassroots change. In this way, Action Infographics play a role in the pipeline from information to action---they inform with the intent of cultivating transformative conversations and initiatives.

\subsubsection{Transformation: Connective and Collective Action} \label{sec:3.4}

If infographics (a) are not paired with physical movements and (b) do not inform with the intention of explicit material changes in shifting societal power structures or meeting short-term legislative objectives, they produce complacency and deception. However, there have been significant instances of tangible change through Instagram infographics when avenues of direct action are centralized. As evidenced by Black Lives Matter, action-oriented infographics can catalyze action-oriented communal responses.

Instagram infographic activism can be futile through a purely connective lens: individualized action eliminates the authenticity, robustness, and unity of physical movements. The consistent use of traditional Second Revolution movement strategies in infographic activism suggests that transformative Instagram infographic activism relies upon adjacent movement-wide operations and boots on the ground which definitionally expend organizational resources. By depending on traditional physical tactics and consuming movement resources, transformative Instagram infographic activism shares the costs of collective action. In this way, the connective aspects of Instagram infographic activism are means to an end---serving as features and perks that bridge information and action.

\section{Discussion and Future Work}

Our investigation of Instagram infographics in ethnic movements can bridge movement technology action frames, serve as a framework for ethnic movement activists, and cultivate rich future work.

\subsection{Bridging Action Frames}

Our examination of Instagram infographics in ethnic movements showcases that the \textit{integration} of digital activism tools into a movement adds complexity and nuance to action formation dynamics. To perceive and evaluate the action formations encompassing a tool, features cannot be observed and analyzed in isolation---they must instead be considered at the application level. In this paper, we have examined three such application-level Instagram infographic phenomena: scope for education, reconciliation for credibility, and resource expenditure for transformative change. Each of these areas reveal important considerations for technology design.

\subsubsection{Scope: Maximizing Educational Reach}

While Instagram infographics have finite scope due to echo chambers and internet access constraints, activists have  optimized their virtual organizing using Instagram’s features, borderless nature, and reduction in physical, schooling, and financial limitations. Therefore, Black Lives Matter’s infographics have more engagement, traction, and scope. Years ago, Second Revolution activists centered political education to establish and share core movement ideologies, objectives, and values. Today, digital movement tools have facilitated the dispersal of collective movement information \cite{Ince17, Twyman16, Stanley20, Siddarth20, Tillery19} at an unprecedented scale \cite{Laer10, Selander16, Enjolras13}. Instagram infographics have taken this further by becoming a New Age documentation of movement education, rippling throughout wider communities with a fraction of the Second Revolution friction. By patching together otherwise fragmented communities for under a unified message, Instagram infographic activism uses connective customizability to breed collective action.

While digital activism tools with educational objectives for activism can be connective in their personalizability---an individual ultimately has the jurisdiction over the design, content, and post---they can also become collective as Instagram infographics have when integrated into a movement. The subsequent responsibility to deliver core movement messages amplifies the significance of addressing the tensions between coexisting facets of connective and collective action. For example, it is vital to reach and educate all community members by creating accessible information dissemination mediums. In the case of Instagram infographics, the collective uses of the tool exacerbate the need to combat the exclusion of offline communities and continue designing for lower literacy levels. Another area that challenges infographics is including those with visual impairments whose screen-readers cannot pick up text from images, particularly when the personalizability of infographics prioritizes aesthetics over readability \cite{Arum17}. When impressed with the responsibility of educating community members and delivering political education, digital activism tools cannot be designed merely to optimize for individualized, low-barrier information dispersal. They must be designed for the need at the movement integration level: educating community members.

\subsubsection{Credibility: Reconciling for Trust}

Because they reduce the barrier to entry of content creation and distribution, Instagram infographics allow overconsolidated content and misinformation to detract from movement trust. In response, New Age ethnic movement activists have curated mechanisms to preserve credibility and maximize the scope-level benefits of infographics while instilling trust in movement participants, including sharing resources, taking community updates, engaging in the comments section, and creating response infographics. Connective action centers the individual---its beauty is in its personalizability. Individuals post what they believe, advocate for what they need to, and do not rely on movement-wide compatibility with their belief systems. The resultant misinformation \cite{AlrubaianAL-Qurishi19, Maddock15, Arif16, Liao13} and desire for information credibility \cite{Castillo11, Gupta14, Tavish19, Mitra18, Starbird12} has been of keen interest. When tools are integrated into movements, however, we see a need for the reconciliation of clashing movement information. New Age activists utilizing Instagram infographics have demonstrated a movement-wide need to resolve internal differences and build trust, a far from individualistic phenomenon. In establishing an implicit system of checks and balances using connective features, activists using Instagram infographics prioritize the development and preservation of a consistent, trusted collective message.

The prioritization of credibility invites us to consider activist-level and community-level reconciliation in digital activism design. In the case of Instagram infographics, ethnic movement activists presented a need to engage with one another and their communities in order to establish movement-wide legitimacy. This discussion inspires the exploration of ways in which virtual activists can congregate, discuss, debate, and collaborate effortlessly. In addition, it raises the importance of creating features for community members to not only respond to activist content, but for their concerns to \textit{reach} activists in a way that serves community interests without overwhelming recipients. The use of connective features such as comments and post-update flexibility urge us to preserve customizability and low-barrier information sharing when creating reconciliation tools. Finally, it invites us to deeply consider ways in which activists can garner communal trust and build credibility with community members. Instagram infographic activists have presented a desire to reconcile activist differences to establish unity, warranting exploration of technology design that leverages personalizability to brew collective movement messaging and a trusted unified front.

\subsubsection{Transformation: Facilitating High-Resource Change}

Infographics that spread information without the underlying objective of action can yield performative activism, complacency, and disingenuous participation, undermining movement efficacy. Fortunately, when activists pack their infographics with action-items, they elicit a more action-based response, as showcased by the Black Lives Matter Instagram account. Ethnic movement activists that centralize tangible action items are able to undercut complacency by arming their audience with the tools they need to spark change \textit{beyond} the infographic. Indeed, transformative output is garnered by the close combination of information and action and researches have evaluated and debated the role of technology in physical movement efforts \cite{Donovan18, Rotman11, Massung13, EarlKimport13, Gerbaudo12, Theocharis15}. Activists can inform and inspire community members via Instagram infographics, but tangible output is the result of beyond-the-graphic organizing, protesting, and mobilization---all exhausting movement-wide resources. While connective action does not expend organizational resources, transformative Instagram infographic activism has been paired with definitionally collective resource-intensive efforts.

While Instagram infographics and other digital activism tools can theoretically remain fully virtual and resourceless, the necessity of acting beyond the infographic is introduced when considering how infographics integrate into preexisting movements. This invites us to design tools that bridge the gap between connective information-sharing capabilities of social media activism and the ensuing high-resource efforts that happen on the ground. We must explore how we can fortify the pipeline between receiving information about a movement on social media and taking high-resource transformative action. How can we ensure that users do not stop at the action-oriented commenting and follow through beyond the infographic? Further, we should explore whether technology can directly facilitate the high-resource, transformative aspects of movements. Information dissemination and discourse have been facilitated by social media tools, but can (and \textit{should}) technology play a role in output-level engagements such as physical events and protests? This collective need for resource-intensive efforts in social movements encourages us to explore innovative methods of reducing movement cost in the best interest of activists and served communities without compromising activist values. The collective nature of high-resource efforts in Instagram infographic activism urges us to consider the bounds, limits, and non-information-based roles that technology can play in movements to benefit high-resource efforts.

Overall, virtual activism tools must be evaluated at the integration level. In the case of Instagram infographic activism, the marriage of connective and collective action raises rich design implications for activism-focused technology design. 

\subsection{Design Implications for Ethnic Movement Activists}

There are also infographic-level initiatives that activists can pursue to account for the connective and collective nature of transformative infographic activism. In this section, we discuss our findings to derive design implications for ethnic movement activists who seek to balance connective customizability and collective momentum.

\subsubsection{Designing for Scope} We provide the following scope-enhancing design recommendations: (1) Activists can include visual attention-grabbers such as color schemes across posts, cartoon graphics, and vibrant palettes. (2) Activists can tag partner organizations, influencers, and fellow activists and should consider sending them the infographic to post to their own feeds. Building partnerships facilitates breaking into alternate social media bubbles to establish collective thought. (3) Activists should determine whether they prioritize objective success metrics and reach, or strict adherence to values and personal messaging. Only then should the activist adapt content based upon likes, comments, event attendance, or story reposts. (4) Activists should use infographics as an entry point to the core topics of their movement and provide valuable introductions with avenues to learn more and increase involvement. (5) Activists should formulate supplementary non-infographic strategies to counteract the echo chamber phenomenon and overcome accessibility issues, such as outreach to older age groups and houseless communities that lack internet access.

\subsubsection{Designing for Credibility} We provide the following credibility-enhancing design recommendations: (1) Activists should be cautious when summarizing movement information so as to avoid cutting corners and excluding valuable information. Specific topics are better for infographics, while more hefty topics are better to introduce in the infographic, but guide movement participants to further resources, rather than oversummarizing. (2) Activists should clearly include resources in the infographic and caption. (3) Activists should monitor the comments section and be willing to take updates from the community should there be an infographic error and clarify these updates in the caption. (4) Activists can optionally engage in infographic inception when another infographic is spreading egregious misinformation about the movement.

\subsubsection{Designing for Transformation} We provide the following transformation-enhancing design recommendations: (1) Activists should include action-oriented elements such as petitions, phone numbers, protest information, etc. Links (preferably with shortened URLs) should be included whenever possible to make taking action seem accessible and efficient, yet impactful. (2) Activists should strive for a balance between informative infographics and action-oriented infographics to maintain the information-action equilibrium. (3) Activists should centralize offline strategies, such as protests, bird-dogging, and social work to counteract complacency and deception, and further curate tangible changes.

\subsection{Future Work}

In addition to the technology design implications in 6.1, there are several avenues that can allow us to more deeply explore the impact of Instagram infographics on ethnic movements. Text analysis of the comments section can be used to gauge whether infographics are more prone to misinformation-indicative responses, revealing the objective indicators of whether an infographic is likely to be faulty. Our case study could be expanded to a larger group of ethnic movement Instagram accounts as accessing a broader dataset would allow us to extract quantitative results that apply to ethnic movement activism as a whole. Computer vision platforms that can identify shapes, images, color schemes, text, or graphics would increase the objectivity of infographic detection. A related design-oriented study expanding our design implications could more intricately examine specific infographic strategies and their success levels through user interviews and engagement experiments, providing a succinct, instruction-oriented guide for infographic production. A study of content-level engagement could delve into the "slides" feature of Instagram---determining whether participants explore content all slides or depart after a catchy header or title. In order to evaluate the efficacy of action-oriented posts, strategies to track and measure action taken by movement participants beyond the comments section via petitions, phone calls, protest attendance, etc. could be explored. Finally, while infographics have most definitely been used by ethnic movement activists, they are also accessible to ethnic movement oppressors. Future work could explore infographic content moderation by considering how we can avoid malicious uses of Instagram infographics.

\section{Conclusion}

Our comparative interviews of ethnic movement activists and data analysis of the Black Lives Matter Instagram underscore three ways in which Instagram infographics embody a union of connective and collective action: increased scope for education, reconciliation for credibility, and high-resource efforts for transformative change. While personalizable and low-resource in isolation, the connective features of Instagram infographics have been leveraged to establish a movement-wide front: Yesterday’s newspaper education and discourse-rich conferences are today’s Instagram infographics. As such, activism technologies need not embody singular action formations: by evaluating Instagram infographics at the movement integration level, our work inspires the design of infographics and activism technology that centralize the marriage of connective and collective action. When connective digital activism tools are used collectively, they must be designed as such: to serve collectives.

\begin{acks}
Thank you to the ethnic movement activists who gave us invaluable insights and made this work possible. Special thanks to Emily Pedersen, Wesley Deng, and Ishan Balakrishnan for their continued mentorship, advice, and encouragement.
\end{acks}

\bibliographystyle{ACM-Reference-Format}
\bibliography{sample-base}


\begin{thebibliography}{93}


\ifx \showCODEN    \undefined \def \showCODEN     #1{\unskip}     \fi
\ifx \showDOI      \undefined \def \showDOI       #1{#1}\fi
\ifx \showISBNx    \undefined \def \showISBNx     #1{\unskip}     \fi
\ifx \showISBNxiii \undefined \def \showISBNxiii  #1{\unskip}     \fi
\ifx \showISSN     \undefined \def \showISSN      #1{\unskip}     \fi
\ifx \showLCCN     \undefined \def \showLCCN      #1{\unskip}     \fi
\ifx \shownote     \undefined \def \shownote      #1{#1}          \fi
\ifx \showarticletitle \undefined \def \showarticletitle #1{#1}   \fi
\ifx \showURL      \undefined \def \showURL       {\relax}        \fi
\providecommand\bibfield[2]{#2}
\providecommand\bibinfo[2]{#2}
\providecommand\natexlab[1]{#1}
\providecommand\showeprint[2][]{arXiv:#2}

\bibitem[\protect\citeauthoryear{??}{Pra}{[n.d.]}]%
        {PrattLibrary}
 \bibinfo{year}{[n.d.]}\natexlab{}.
\newblock \bibinfo{title}{The Black Panther Party}.
\newblock
\newblock
\urldef\tempurl%
\url{https://www.prattlibrary.org/research/guides/the-black-panther-party}
\showURL{%
\tempurl}


\bibitem[\protect\citeauthoryear{??}{Vic}{[n.d.]}]%
        {Victorian}
 \bibinfo{year}{[n.d.]}\natexlab{}.
\newblock \bibinfo{title}{Gestetner Machine, c. 1922 - 1929}.
\newblock
\newblock
\urldef\tempurl%
\url{https://victoriancollections.net.au/items/58b3b81fd0ce260f2cdfe2b0}
\showURL{%
\tempurl}


\bibitem[\protect\citeauthoryear{??}{Rad}{[n.d.]}]%
        {Radical}
 \bibinfo{year}{[n.d.]}\natexlab{}.
\newblock \bibinfo{booktitle}{\emph{Radical movement}}.
\newblock
\urldef\tempurl%
\url{https://the-definition.com/term/radical-movement}
\showURL{%
\tempurl}


\bibitem[\protect\citeauthoryear{??}{FSM}{[n.d.]}]%
        {FSM}
 \bibinfo{year}{[n.d.]}\natexlab{}.
\newblock \bibinfo{title}{Visual History: Free Speech Movement, 1964}.
\newblock
\newblock
\urldef\tempurl%
\url{https://fsm.berkeley.edu/free-speech-movement-timeline/}
\showURL{%
\tempurl}


\bibitem[\protect\citeauthoryear{??}{Bri}{1998}]%
        {Britannica}
 \bibinfo{year}{1998}\natexlab{}.
\newblock \bibinfo{booktitle}{\emph{American Indian Movement}}.
\newblock
\urldef\tempurl%
\url{https://www.britannica.com/topic/American-Indian-Movement}
\showURL{%
Retrieved November 21, 2020 from \tempurl}


\bibitem[\protect\citeauthoryear{??}{Lib}{2012}]%
        {Libcom}
 \bibinfo{year}{2012}\natexlab{}.
\newblock \bibinfo{title}{The Black Panther: newspaper of the Black Panther
  Party}.
\newblock
\newblock
\urldef\tempurl%
\url{https://libcom.org/history/black-panther-newspaper-black-panther-party}
\showURL{%
\tempurl}


\bibitem[\protect\citeauthoryear{??}{BLM}{2014}]%
        {BLM14}
 \bibinfo{year}{2014}\natexlab{}.
\newblock \bibinfo{booktitle}{\emph{About}}.
\newblock
\urldef\tempurl%
\url{https://blacklivesmatter.com/about/}
\showURL{%
Retrieved November 21, 2020 from \tempurl}


\bibitem[\protect\citeauthoryear{??}{Arc}{2016}]%
        {Archives16}
 \bibinfo{year}{2016}\natexlab{}.
\newblock \bibinfo{booktitle}{\emph{The Black Panther Party}}.
\newblock
\urldef\tempurl%
\url{https://www.archives.gov/research/african-americans/black-power/black-panthers#bpintro}
\showURL{%
Retrieved November 21, 2020 from \tempurl}


\bibitem[\protect\citeauthoryear{??}{Civ}{2016}]%
        {CivilRights}
 \bibinfo{year}{2016}\natexlab{}.
\newblock \bibinfo{booktitle}{\emph{The Modern Civil Rights Movement,
  1954-1964}}.
\newblock
\urldef\tempurl%
\url{https://www.nps.gov/subjects/civilrights/modern-civil-rights-movement.htm}
\showURL{%
Retrieved July 05, 2021 from \tempurl}


\bibitem[\protect\citeauthoryear{??}{Sec}{2016}]%
        {SecondRevolution}
 \bibinfo{year}{2016}\natexlab{}.
\newblock \bibinfo{booktitle}{\emph{The Second Revolution, 1965-1980}}.
\newblock
\urldef\tempurl%
\url{https://www.nps.gov/subjects/civilrights/secondrevolution.htm}
\showURL{%
Retrieved December 29, 2020 from \tempurl}


\bibitem[\protect\citeauthoryear{??}{Sum}{2017}]%
        {SummerLove17}
 \bibinfo{year}{2017}\natexlab{}.
\newblock \bibinfo{title}{Remembering the Black Panther Party Newspaper}.
\newblock
\newblock
\urldef\tempurl%
\url{https://summerof.love/remembering-the-black-panther-party-newspaper/}
\showURL{%
\tempurl}


\bibitem[\protect\citeauthoryear{??}{Rat}{2020}]%
        {Rates}
 \bibinfo{year}{2020}\natexlab{}.
\newblock \bibinfo{booktitle}{\emph{Public High School Graduation Rates}}.
\newblock
\urldef\tempurl%
\url{https://nces.ed.gov/programs/coe/indicator_coi.asp}
\showURL{%
Retrieved November 21, 2020 from \tempurl}


\bibitem[\protect\citeauthoryear{??}{Daz}{2020}]%
        {Dazed20}
 \bibinfo{year}{2020}\natexlab{}.
\newblock \bibinfo{booktitle}{\emph{Speaking to the people behind the viral
  anti-racist graphics on Instagram}}.
\newblock
\urldef\tempurl%
\url{https://www.dazeddigital.com/life-culture/article/49516/1/people-behind-viral-anti-racist-graphics-instagram-black-lives-matter-resource}
\showURL{%
Retrieved November 21, 2020 from \tempurl}


\bibitem[\protect\citeauthoryear{Ables}{Ables}{2020}]%
        {Ables20}
\bibfield{author}{\bibinfo{person}{Kelsey Ables}.}
  \bibinfo{year}{2020}\natexlab{}.
\newblock \bibinfo{booktitle}{\emph{Selfies and sunsets be gone: The latest
  Instagram trend is PowerPoint-style presentations}}.
\newblock
\urldef\tempurl%
\url{https://www.washingtonpost.com/technology/2020/08/15/instagram-race-activism-slideshow-graphics/}
\showURL{%
Retrieved November 21, 2020 from \tempurl}


\bibitem[\protect\citeauthoryear{Acosta}{Acosta}{[n.d.]}]%
        {Acosta}
\bibfield{author}{\bibinfo{person}{Teresa~Palomo Acosta}.}
  \bibinfo{year}{[n.d.]}\natexlab{}.
\newblock \bibinfo{title}{Chicano Mural Movement}.
\newblock
\newblock
\urldef\tempurl%
\url{https://www.tshaonline.org/handbook/entries/chicano-mural-movement}
\showURL{%
\tempurl}


\bibitem[\protect\citeauthoryear{Alexander and Hahner}{Alexander and
  Hahner}{2017}]%
        {Alexander17}
\bibfield{author}{\bibinfo{person}{Kara~P. Alexander} {and}
  \bibinfo{person}{Leslie~A. Hahner}.} \bibinfo{year}{2017}\natexlab{}.
\newblock \showarticletitle{Chapter 12. The Intimate Screen: Revisualizing
  Understandings of Down Syndrome through Digital Activism on Instagram}.
\newblock \bibinfo{journal}{\emph{Social Writing/Social Media: Publics,
  Presentations, and Pedagogies}} (\bibinfo{year}{2017}),
  \bibinfo{pages}{225–243}.
\newblock
\urldef\tempurl%
\url{https://doi.org/10.37514/per-b.2017.0063.2.12}
\showDOI{\tempurl}


\bibitem[\protect\citeauthoryear{Alrubaian and AL-Qurishi}{Alrubaian and
  AL-Qurishi}{2019}]%
        {AlrubaianAL-Qurishi19}
\bibfield{author}{\bibinfo{person}{Majed Alrubaian} {and}
  \bibinfo{person}{Muhammad AL-Qurishi}.} \bibinfo{year}{2019}\natexlab{}.
\newblock \showarticletitle{Credibility in Online Social Networks: A Survey}.
\newblock \bibinfo{journal}{\emph{IEEE Access}}  \bibinfo{volume}{7}
  (\bibinfo{date}{Dec.} \bibinfo{year}{2019}), \bibinfo{pages}{2828--2855}.
\newblock
\urldef\tempurl%
\url{https://doi.org/10.1109/ACCESS.2018.2886314}
\showDOI{\tempurl}


\bibitem[\protect\citeauthoryear{Amit-Danhi and Shifman}{Amit-Danhi and
  Shifman}{2018}]%
        {Amit-DanhiShifman18}
\bibfield{author}{\bibinfo{person}{Eedan~R Amit-Danhi} {and}
  \bibinfo{person}{Limor Shifman}.} \bibinfo{year}{2018}\natexlab{}.
\newblock \showarticletitle{Digital political infographics: A rhetorical
  palette of an emergent genre}.
\newblock \bibinfo{journal}{\emph{New Media \& Society}} \bibinfo{volume}{20},
  \bibinfo{number}{10} (\bibinfo{year}{2018}), \bibinfo{pages}{3540--3559}.
\newblock
\showISSN{1461-4448}
\urldef\tempurl%
\url{https://doi.org/10.1177/1461444817750565}
\showDOI{\tempurl}


\bibitem[\protect\citeauthoryear{Arif, Shanahan, Chou, Dosouto, Starbird, and
  Spiro}{Arif et~al\mbox{.}}{2016}]%
        {Arif16}
\bibfield{author}{\bibinfo{person}{Ahmer Arif}, \bibinfo{person}{Kelley
  Shanahan}, \bibinfo{person}{Fang-Ju Chou}, \bibinfo{person}{Yoanna Dosouto},
  \bibinfo{person}{Kate Starbird}, {and} \bibinfo{person}{Emma~S. Spiro}.}
  \bibinfo{year}{2016}\natexlab{}.
\newblock \showarticletitle{How Information Snowballs: Exploring the Role of
  Exposure in Online Rumor Propagation}. In
  \bibinfo{booktitle}{\emph{Proceedings of the 19th ACM Conference on
  Computer-Supported Cooperative Work \& Social Computing}} (San Francisco,
  California, USA) \emph{(\bibinfo{series}{CSCW '16})}.
  \bibinfo{publisher}{Association for Computing Machinery},
  \bibinfo{address}{New York, NY, USA}, \bibinfo{pages}{466–477}.
\newblock
\showISBNx{9781450335928}
\urldef\tempurl%
\url{https://doi.org/10.1145/2818048.2819964}
\showDOI{\tempurl}


\bibitem[\protect\citeauthoryear{Arum}{Arum}{2017}]%
        {Arum17}
\bibfield{author}{\bibinfo{person}{Nenden~Sekar Arum}.}
  \bibinfo{year}{2017}\natexlab{}.
\newblock \showarticletitle{Infographic: Not Just a Beautiful Visualisation}.
\newblock \bibinfo{journal}{\emph{Obtenido de https://www. academia.
  edu/31903865/Infographic\_Not\_Just\_a\_Beautiful\_Visualisation}}
  (\bibinfo{year}{2017}).
\newblock


\bibitem[\protect\citeauthoryear{Association}{Association}{2017}]%
        {APA17}
\bibfield{author}{\bibinfo{person}{American~Psychological Association}.}
  \bibinfo{year}{2017}\natexlab{}.
\newblock \bibinfo{booktitle}{\emph{Ethnic and Racial Minorities \&
  Socioeconomic Status}}.
\newblock
\urldef\tempurl%
\url{https://www.apa.org/pi/ses/resources/publications/minorities}
\showURL{%
Retrieved January 11, 2021 from \tempurl}


\bibitem[\protect\citeauthoryear{Bang and Halupka}{Bang and Halupka}{2019}]%
        {Bang19}
\bibfield{author}{\bibinfo{person}{Henrik Bang} {and} \bibinfo{person}{Max
  Halupka}.} \bibinfo{year}{2019}\natexlab{}.
\newblock \showarticletitle{Contentious connective action: a new kind of
  life-political association for problematizing how expert systems operate}.
\newblock \bibinfo{journal}{\emph{Information, Communication \& Society}}
  \bibinfo{volume}{22}, \bibinfo{number}{1} (\bibinfo{year}{2019}),
  \bibinfo{pages}{89--104}.
\newblock
\urldef\tempurl%
\url{https://doi.org/10.1080/1369118X.2017.1355402}
\showDOI{\tempurl}
\showeprint{https://doi.org/10.1080/1369118X.2017.1355402}


\bibitem[\protect\citeauthoryear{Bateman, Mandryk, Gutwin, Genest, McDine, and
  Brooks}{Bateman et~al\mbox{.}}{2010}]%
        {Bateman10}
\bibfield{author}{\bibinfo{person}{Scott Bateman}, \bibinfo{person}{Regan~L.
  Mandryk}, \bibinfo{person}{Carl Gutwin}, \bibinfo{person}{Aaron Genest},
  \bibinfo{person}{David McDine}, {and} \bibinfo{person}{Christopher Brooks}.}
  \bibinfo{year}{2010}\natexlab{}.
\newblock \showarticletitle{Useful junk?: the effects of visual embellishment
  on comprehension and memorability of charts}. In
  \bibinfo{booktitle}{\emph{Proceedings of the SIGCHI Conference on Human
  Factors in Computing Systems}} \emph{(\bibinfo{series}{CHI '10})}.
  \bibinfo{publisher}{ACM Press}, \bibinfo{address}{New York, NY},
  \bibinfo{pages}{2573--2582}.
\newblock
\urldef\tempurl%
\url{https://doi.org/10.1145/1753326.1753716}
\showDOI{\tempurl}


\bibitem[\protect\citeauthoryear{Bates}{Bates}{2020}]%
        {Bates20}
\bibfield{author}{\bibinfo{person}{Julie~Collins Bates}.}
  \bibinfo{year}{2020}\natexlab{}.
\newblock \bibinfo{booktitle}{\emph{Local Expertise, Global Effects: Amplifying
  Activist Arguments for Climate Change Action}}.
\newblock
\urldef\tempurl%
\url{http://enculturation.net/Local_Expertise_Global_Effects}
\showURL{%
Retrieved January 5, 2021 from \tempurl}


\bibitem[\protect\citeauthoryear{Benford and Snow}{Benford and Snow}{2000}]%
        {Benford00}
\bibfield{author}{\bibinfo{person}{Robert~D. Benford} {and}
  \bibinfo{person}{David~A. Snow}.} \bibinfo{year}{2000}\natexlab{}.
\newblock \showarticletitle{Framing Processes and Social Movements: An Overview
  and Assessment}.
\newblock \bibinfo{journal}{\emph{Annual Review of Sociology}}
  \bibinfo{volume}{26}, \bibinfo{number}{1} (\bibinfo{year}{2000}),
  \bibinfo{pages}{611–639}.
\newblock
\urldef\tempurl%
\url{https://doi.org/10.1146/annurev.soc.26.1.611}
\showDOI{\tempurl}


\bibitem[\protect\citeauthoryear{Bennett and Segerberg}{Bennett and
  Segerberg}{2012}]%
        {Bennett12}
\bibfield{author}{\bibinfo{person}{W.~Lance Bennett} {and}
  \bibinfo{person}{Alexandra Segerberg}.} \bibinfo{year}{2012}\natexlab{}.
\newblock \showarticletitle{THE LOGIC OF CONNECTIVE ACTION}.
\newblock \bibinfo{journal}{\emph{Information, Communication \& Society}}
  \bibinfo{volume}{15}, \bibinfo{number}{5} (\bibinfo{year}{2012}),
  \bibinfo{pages}{739--768}.
\newblock
\urldef\tempurl%
\url{https://doi.org/10.1080/1369118X.2012.670661}
\showDOI{\tempurl}
\showeprint{https://doi.org/10.1080/1369118X.2012.670661}


\bibitem[\protect\citeauthoryear{Blystone}{Blystone}{2020}]%
        {Blystone20}
\bibfield{author}{\bibinfo{person}{Dan Blystone}.}
  \bibinfo{year}{2020}\natexlab{}.
\newblock \bibinfo{booktitle}{\emph{The Story of Instagram: The Rise of the \#1
  Photo-Sharing Application}}.
\newblock
\urldef\tempurl%
\url{https://www.investopedia.com/articles/investing/102615/story-instagram-rise-1-photo0sharing-app.asp}
\showURL{%
Retrieved January 9, 2021 from \tempurl}


\bibitem[\protect\citeauthoryear{Borgo, Abdul-Rahman, Mohamed, Grant, Reppa,
  Floridi, and Chen}{Borgo et~al\mbox{.}}{2012}]%
        {Borgo12}
\bibfield{author}{\bibinfo{person}{Rita Borgo}, \bibinfo{person}{Alfie
  Abdul-Rahman}, \bibinfo{person}{Farhan Mohamed}, \bibinfo{person}{Philip~W.
  Grant}, \bibinfo{person}{Irene Reppa}, \bibinfo{person}{Luciano Floridi},
  {and} \bibinfo{person}{Min Chen}.} \bibinfo{year}{2012}\natexlab{}.
\newblock \showarticletitle{An Empirical Study on Using Visual Embellishments
  in Visualization}.
\newblock \bibinfo{journal}{\emph{IEEE Transactions on Visualization and
  Computer Graphics}} \bibinfo{volume}{18}, \bibinfo{number}{12}
  (\bibinfo{year}{2012}), \bibinfo{pages}{2759--2768}.
\newblock
\showISSN{1077-2626}
\urldef\tempurl%
\url{https://doi.org/10.1109/TVCG.2012.197}
\showDOI{\tempurl}


\bibitem[\protect\citeauthoryear{Borkin, Vo, Bylinskii, Isola, Sunkavalli,
  Oliva, and Pfister}{Borkin et~al\mbox{.}}{2013}]%
        {Borkin13}
\bibfield{author}{\bibinfo{person}{Michelle~A Borkin},
  \bibinfo{person}{Azalea~A Vo}, \bibinfo{person}{Zoya Bylinskii},
  \bibinfo{person}{Phillip Isola}, \bibinfo{person}{Shashank Sunkavalli},
  \bibinfo{person}{Aude Oliva}, {and} \bibinfo{person}{Hanspeter Pfister}.}
  \bibinfo{year}{2013}\natexlab{}.
\newblock \showarticletitle{What makes a visualization memorable?}
\newblock \bibinfo{journal}{\emph{IEEE Transactions on Visualization and
  Computer Graphics}} \bibinfo{volume}{19}, \bibinfo{number}{12}
  (\bibinfo{year}{2013}), \bibinfo{pages}{2306 -- 2315}.
\newblock
\showISSN{1077-2626}
\urldef\tempurl%
\url{https://doi.org/10.1109/TVCG.2013.234}
\showDOI{\tempurl}


\bibitem[\protect\citeauthoryear{Bødker, Korsgaard, Lyle, and
  Saad-Sulonen}{Bødker et~al\mbox{.}}{2016}]%
        {Bodker16}
\bibfield{author}{\bibinfo{person}{Susanne Bødker}, \bibinfo{person}{Henrik
  Korsgaard}, \bibinfo{person}{Peter Lyle}, {and} \bibinfo{person}{Joanna
  Saad-Sulonen}.} \bibinfo{year}{2016}\natexlab{}.
\newblock \showarticletitle{Happenstance, Strategies and Tactics: Intrinsic
  Design in a Volunteer-Based Community}. In
  \bibinfo{booktitle}{\emph{Proceedings of the 9th Nordic Conference on
  Human-Computer Interaction}} (Gothenburg, Sweden)
  \emph{(\bibinfo{series}{NordiCHI '16})}. \bibinfo{publisher}{Association for
  Computing Machinery}, \bibinfo{address}{New York, NY, USA}, Article
  \bibinfo{articleno}{10}, \bibinfo{numpages}{10}~pages.
\newblock
\showISBNx{9781450347631}
\urldef\tempurl%
\url{https://doi.org/10.1145/2971485.2971564}
\showDOI{\tempurl}


\bibitem[\protect\citeauthoryear{Carrillo}{Carrillo}{2020}]%
        {Carrillo20}
\bibfield{author}{\bibinfo{person}{Karen~Juanita Carrillo}.}
  \bibinfo{year}{2020}\natexlab{}.
\newblock \bibinfo{booktitle}{\emph{How the Chicano Movement Championed
  Mexican-American Identity and Fought for Change}}.
\newblock
\urldef\tempurl%
\url{https://www.history.com/news/chicano-movement}
\showURL{%
Retrieved November 21, 2020 from \tempurl}


\bibitem[\protect\citeauthoryear{Castells}{Castells}{2016}]%
        {Castells16}
\bibfield{author}{\bibinfo{person}{Manuel Castells}.}
  \bibinfo{year}{2016}\natexlab{}.
\newblock \showarticletitle{75. Networks of Outrage and Hope: Social Movements
  in the Internet Age}.
\newblock \bibinfo{journal}{\emph{Democracy}} (\bibinfo{year}{2016}),
  \bibinfo{pages}{433–435}.
\newblock
\urldef\tempurl%
\url{https://doi.org/10.7312/blau17412-091}
\showDOI{\tempurl}


\bibitem[\protect\citeauthoryear{Castillo, Mendoza, and Poblete}{Castillo
  et~al\mbox{.}}{2011}]%
        {Castillo11}
\bibfield{author}{\bibinfo{person}{Carlos Castillo}, \bibinfo{person}{Marcelo
  Mendoza}, {and} \bibinfo{person}{Barbara Poblete}.}
  \bibinfo{year}{2011}\natexlab{}.
\newblock \showarticletitle{Information Credibility on Twitter}. In
  \bibinfo{booktitle}{\emph{Proceedings of the 20th International Conference on
  World Wide Web}} (Hyderabad, India) \emph{(\bibinfo{series}{WWW '11})}.
  \bibinfo{publisher}{Association for Computing Machinery},
  \bibinfo{address}{New York, NY, USA}, \bibinfo{pages}{675–684}.
\newblock
\showISBNx{9781450306324}
\urldef\tempurl%
\url{https://doi.org/10.1145/1963405.1963500}
\showDOI{\tempurl}


\bibitem[\protect\citeauthoryear{Chu and Yeo}{Chu and Yeo}{2020}]%
        {Chu20}
\bibfield{author}{\bibinfo{person}{Tsz~Hang Chu} {and} \bibinfo{person}{Tien
  Ee~Dominic Yeo}.} \bibinfo{year}{2020}\natexlab{}.
\newblock \showarticletitle{Rethinking mediated political engagement: social
  media ambivalence and disconnective practices of politically active youths in
  Hong Kong}.
\newblock \bibinfo{journal}{\emph{Chinese Journal of Communication}}
  \bibinfo{volume}{13}, \bibinfo{number}{2} (\bibinfo{year}{2020}),
  \bibinfo{pages}{148--164}.
\newblock
\urldef\tempurl%
\url{https://doi.org/10.1080/17544750.2019.1634606}
\showDOI{\tempurl}
\showeprint{https://doi.org/10.1080/17544750.2019.1634606}


\bibitem[\protect\citeauthoryear{Collier}{Collier}{2015}]%
        {Collier15}
\bibfield{author}{\bibinfo{person}{Andrea~King Collier}.}
  \bibinfo{year}{2015}\natexlab{}.
\newblock \bibinfo{booktitle}{\emph{The Black Panthers: Revolutionaries, Free
  Breakfast Pioneers}}.
\newblock
\urldef\tempurl%
\url{https://www.nationalgeographic.com/culture/food/the-plate/2015/11/04/the-black-panthers-revolutionaries-free-breakfast-pioneers/#close}
\showURL{%
Retrieved November 21, 2020 from \tempurl}


\bibitem[\protect\citeauthoryear{Cooper}{Cooper}{2015}]%
        {Cooper15}
\bibfield{author}{\bibinfo{person}{Lauren Cooper}.}
  \bibinfo{year}{2015}\natexlab{}.
\newblock \bibinfo{booktitle}{\emph{Native American Activism: 1960s to
  Present}}.
\newblock
\urldef\tempurl%
\url{https://www.zinnedproject.org/materials/native-american-activism-1960s-to-present/}
\showURL{%
Retrieved November 21, 2020 from \tempurl}


\bibitem[\protect\citeauthoryear{David K.~Yoo}{David K.~Yoo}{2016}]%
        {YooAzuma16}
\bibfield{author}{\bibinfo{person}{Eiichiro~Azuma David K.~Yoo}.}
  \bibinfo{year}{2016}\natexlab{}.
\newblock \bibinfo{booktitle}{\emph{The Oxford Handbook of Asian American
  History}}.
\newblock \bibinfo{publisher}{Oxford University Press}.
\newblock
\showISBNx{0199860475}


\bibitem[\protect\citeauthoryear{Dictionary.com}{Dictionary.com}{[n.d.]}]%
        {Infographics-Def}
\bibfield{author}{\bibinfo{person}{Dictionary.com}.}
  \bibinfo{year}{[n.d.]}\natexlab{}.
\newblock \bibinfo{booktitle}{\emph{infographic}}.
\newblock
\urldef\tempurl%
\url{https://www.dictionary.com/browse/infographic}
\showURL{%
Retrieved June 19, 2021 from \tempurl}


\bibitem[\protect\citeauthoryear{Donovan}{Donovan}{2018}]%
        {Donovan18}
\bibfield{author}{\bibinfo{person}{Joan Donovan}.}
  \bibinfo{year}{2018}\natexlab{}.
\newblock \showarticletitle{After the \#Keyword: Eliciting, Sustaining, and
  Coordinating Participation Across the Occupy Movement}.
\newblock   \bibinfo{volume}{4}, Article \bibinfo{articleno}{1}
  (\bibinfo{date}{Feb.} \bibinfo{year}{2018}).
\newblock
\urldef\tempurl%
\url{https://doi.org/10.1177/2056305117750720}
\showDOI{\tempurl}


\bibitem[\protect\citeauthoryear{Earl}{Earl}{2010}]%
        {Earl10}
\bibfield{author}{\bibinfo{person}{Jennifer Earl}.}
  \bibinfo{year}{2010}\natexlab{}.
\newblock \showarticletitle{The Dynamics Of Protest-Related Diffusion On The
  Web}.
\newblock \bibinfo{journal}{\emph{Information, Communication \& Society}}
  \bibinfo{volume}{13}, \bibinfo{number}{2} (\bibinfo{year}{2010}),
  \bibinfo{pages}{209–225}.
\newblock
\urldef\tempurl%
\url{https://doi.org/10.1080/13691180902934170}
\showDOI{\tempurl}


\bibitem[\protect\citeauthoryear{Earl and Kimport}{Earl and Kimport}{2013}]%
        {EarlKimport13}
\bibfield{author}{\bibinfo{person}{Jennifer Earl} {and}
  \bibinfo{person}{Katrina Kimport}.} \bibinfo{year}{2013}\natexlab{}.
\newblock \bibinfo{booktitle}{\emph{Digitally Enabled Social Change}}.
\newblock \bibinfo{publisher}{The MIT Press}, \bibinfo{address}{Cambridge, MA,
  USA}.
\newblock
\showISBNx{9780262015103}


\bibitem[\protect\citeauthoryear{Enjolras, Steen-Johnsen, and
  Wollebæk}{Enjolras et~al\mbox{.}}{2013}]%
        {Enjolras13}
\bibfield{author}{\bibinfo{person}{Bernard Enjolras}, \bibinfo{person}{Kari
  Steen-Johnsen}, {and} \bibinfo{person}{Dag Wollebæk}.}
  \bibinfo{year}{2013}\natexlab{}.
\newblock \showarticletitle{Social media and mobilization to offline
  demonstrations: Transcending participatory divides?}
\newblock \bibinfo{journal}{\emph{New Media \& Society}} \bibinfo{volume}{15},
  \bibinfo{number}{6} (\bibinfo{year}{2013}), \bibinfo{pages}{890--908}.
\newblock
\urldef\tempurl%
\url{https://doi.org/10.1177/1461444812462844}
\showDOI{\tempurl}
\showeprint{https://doi.org/10.1177/1461444812462844}


\bibitem[\protect\citeauthoryear{for Institutional~Diversity}{for
  Institutional~Diversity}{2019}]%
        {NCID19}
\bibfield{author}{\bibinfo{person}{National~Center for
  Institutional~Diversity}.} \bibinfo{year}{2019}\natexlab{}.
\newblock \bibinfo{title}{The Black Radical Tradition of Resistance}.
\newblock
\newblock
\urldef\tempurl%
\url{https://medium.com/national-center-for-institutional-diversity/the-black-radical-tradition-of-resistance-7277f09ef396}
\showURL{%
\tempurl}


\bibitem[\protect\citeauthoryear{Freelon, Mcilwain, and Clark}{Freelon
  et~al\mbox{.}}{2016}]%
        {Clark16}
\bibfield{author}{\bibinfo{person}{Deen Freelon}, \bibinfo{person}{Charlton~D.
  Mcilwain}, {and} \bibinfo{person}{Meredith~D Clark}.}
  \bibinfo{year}{2016}\natexlab{}.
\newblock \showarticletitle{Beyond the Hashtags: \#Ferguson,
  \#Blacklivesmatter, and the Online Struggle for Offline Justice}.
\newblock \bibinfo{journal}{\emph{SSRN Electronic Journal}}
  (\bibinfo{year}{2016}).
\newblock
\urldef\tempurl%
\url{https://doi.org/10.2139/ssrn.2747066}
\showDOI{\tempurl}


\bibitem[\protect\citeauthoryear{Garza}{Garza}{2014}]%
        {Garza14}
\bibfield{author}{\bibinfo{person}{Alicia Garza}.}
  \bibinfo{year}{2014}\natexlab{}.
\newblock \bibinfo{title}{A Herstory of the \#BlackLivesMatter Movement by
  Alicia Garza}.
\newblock
\newblock
\urldef\tempurl%
\url{https://thefeministwire.com/2014/10/blacklivesmatter-2/}
\showURL{%
\tempurl}


\bibitem[\protect\citeauthoryear{Gerbaudo}{Gerbaudo}{2012}]%
        {Gerbaudo12}
\bibfield{author}{\bibinfo{person}{Paolo Gerbaudo}.}
  \bibinfo{year}{2012}\natexlab{}.
\newblock \bibinfo{booktitle}{\emph{Tweets and the streets : social media and
  contemporary activism.}}
\newblock \bibinfo{publisher}{Pluto Press}.
\newblock
\showISBNx{9780745332499}


\bibitem[\protect\citeauthoryear{Ghoshal, Mendhekar, and Bruckman}{Ghoshal
  et~al\mbox{.}}{2020}]%
        {Ghoshal20}
\bibfield{author}{\bibinfo{person}{Sucheta Ghoshal}, \bibinfo{person}{Rishma
  Mendhekar}, {and} \bibinfo{person}{Amy Bruckman}.}
  \bibinfo{year}{2020}\natexlab{}.
\newblock \showarticletitle{Toward a Grassroots Culture of Technology
  Practice}.
\newblock \bibinfo{journal}{\emph{Proc. ACM Hum.-Comput. Interact.}}
  \bibinfo{volume}{4}, \bibinfo{number}{CSCW1}, Article
  \bibinfo{articleno}{054} (\bibinfo{date}{May} \bibinfo{year}{2020}),
  \bibinfo{numpages}{28}~pages.
\newblock
\urldef\tempurl%
\url{https://doi.org/10.1145/3392862}
\showDOI{\tempurl}


\bibitem[\protect\citeauthoryear{González-Bailón and Wang}{González-Bailón
  and Wang}{2016}]%
        {Gonzalez16}
\bibfield{author}{\bibinfo{person}{Sandra González-Bailón} {and}
  \bibinfo{person}{Ning Wang}.} \bibinfo{year}{2016}\natexlab{}.
\newblock \showarticletitle{Networked discontent: The anatomy of protest
  campaigns in social media}.
\newblock \bibinfo{journal}{\emph{Social Networks}}  \bibinfo{volume}{44}
  (\bibinfo{year}{2016}), \bibinfo{pages}{95–104}.
\newblock
\urldef\tempurl%
\url{https://doi.org/10.1016/j.socnet.2015.07.003}
\showDOI{\tempurl}


\bibitem[\protect\citeauthoryear{Gupta, Kumaraguru, Castillo, and Meier}{Gupta
  et~al\mbox{.}}{2014}]%
        {Gupta14}
\bibfield{author}{\bibinfo{person}{Aditi Gupta}, \bibinfo{person}{Ponnurangam
  Kumaraguru}, \bibinfo{person}{Carlos Castillo}, {and}
  \bibinfo{person}{Patrick Meier}.} \bibinfo{year}{2014}\natexlab{}.
\newblock \bibinfo{booktitle}{\emph{TweetCred: Real-Time Credibility Assessment
  of Content on Twitter}}.
\newblock \bibinfo{publisher}{Springer International Publishing},
  \bibinfo{address}{Cham}, \bibinfo{pages}{228--243}.
\newblock
\showISBNx{978-3-319-13734-6}
\urldef\tempurl%
\url{https://doi.org/10.1007/978-3-319-13734-6_16}
\showDOI{\tempurl}


\bibitem[\protect\citeauthoryear{Haroz, Kosara, and Franconeri}{Haroz
  et~al\mbox{.}}{2015}]%
        {Haroz15}
\bibfield{author}{\bibinfo{person}{Steve Haroz}, \bibinfo{person}{Robert
  Kosara}, {and} \bibinfo{person}{Steven~L. Franconeri}.}
  \bibinfo{year}{2015}\natexlab{}.
\newblock \showarticletitle{ISOTYPE Visualization: Working Memory, Performance,
  and Engagement with Pictographs}. In \bibinfo{booktitle}{\emph{Proceedings of
  the 33rd Annual ACM Conference on Human Factors in Computing Systems}}
  \emph{(\bibinfo{series}{CHI '15})}. \bibinfo{publisher}{ACM Press},
  \bibinfo{address}{New York, NY}, \bibinfo{pages}{1191--1200}.
\newblock
\urldef\tempurl%
\url{https://doi.org/10.1145/2702123.2702275}
\showDOI{\tempurl}


\bibitem[\protect\citeauthoryear{Hockin and Brunson}{Hockin and
  Brunson}{2018}]%
        {Hockin18}
\bibfield{author}{\bibinfo{person}{Sara~M. Hockin} {and}
  \bibinfo{person}{Rod~K. Brunson}.} \bibinfo{year}{2018}\natexlab{}.
\newblock \showarticletitle{The Revolution Might Not Be Televised (But It Will
  Be Lived Streamed): Future Directions for Research on Police–Minority
  Relations}.
\newblock \bibinfo{journal}{\emph{Race and Justice}} \bibinfo{volume}{8},
  \bibinfo{number}{3} (\bibinfo{year}{2018}), \bibinfo{pages}{199--215}.
\newblock
\urldef\tempurl%
\url{https://doi.org/10.1177/2153368716676320}
\showDOI{\tempurl}
\showeprint{https://doi.org/10.1177/2153368716676320}


\bibitem[\protect\citeauthoryear{Ince, Rojas, and Davis}{Ince
  et~al\mbox{.}}{2017}]%
        {Ince17}
\bibfield{author}{\bibinfo{person}{Jelani Ince}, \bibinfo{person}{Fabio Rojas},
  {and} \bibinfo{person}{Clayton~A. Davis}.} \bibinfo{year}{2017}\natexlab{}.
\newblock \showarticletitle{The social media response to Black Lives Matter:
  how Twitter users interact with Black Lives Matter through hashtag use}.
\newblock \bibinfo{journal}{\emph{Ethnic and Racial Studies}}
  \bibinfo{volume}{40}, \bibinfo{number}{11} (\bibinfo{year}{2017}),
  \bibinfo{pages}{1814–1830}.
\newblock
\urldef\tempurl%
\url{https://doi.org/10.1080/01419870.2017.1334931}
\showDOI{\tempurl}


\bibitem[\protect\citeauthoryear{Jones}{Jones}{2020}]%
        {Jones20}
\bibfield{author}{\bibinfo{person}{Josh Jones}.}
  \bibinfo{year}{2020}\natexlab{}.
\newblock \bibinfo{title}{The Radical Art of The Black Panther, the
  Revolution's Newspaper from 1967 to 1980}.
\newblock
\newblock
\urldef\tempurl%
\url{https://flashbak.com/the-radical-art-of-the-black-panther-the-revolutions-newspaper-from-1967-to-1980-429768/}
\showURL{%
\tempurl}


\bibitem[\protect\citeauthoryear{Laer and Aelst}{Laer and Aelst}{2010}]%
        {Laer10}
\bibfield{author}{\bibinfo{person}{Jeroen~Van Laer} {and}
  \bibinfo{person}{Peter~Van Aelst}.} \bibinfo{year}{2010}\natexlab{}.
\newblock \showarticletitle{INTERNET AND SOCIAL MOVEMENT ACTION REPERTOIRES}.
\newblock \bibinfo{journal}{\emph{Information, Communication \& Society}}
  \bibinfo{volume}{13}, \bibinfo{number}{8} (\bibinfo{year}{2010}),
  \bibinfo{pages}{1146--1171}.
\newblock
\urldef\tempurl%
\url{https://doi.org/10.1080/13691181003628307}
\showDOI{\tempurl}
\showeprint{https://doi.org/10.1080/13691181003628307}


\bibitem[\protect\citeauthoryear{Lee, Chen, and Chan}{Lee
  et~al\mbox{.}}{2017}]%
        {Lee17}
\bibfield{author}{\bibinfo{person}{Francis~L.F. Lee},
  \bibinfo{person}{Hsuan-Ting Chen}, {and} \bibinfo{person}{Michael Chan}.}
  \bibinfo{year}{2017}\natexlab{}.
\newblock \showarticletitle{Social media use and university students’
  participation in a large-scale protest campaign: The case of Hong Kong’s
  Umbrella Movement}.
\newblock \bibinfo{journal}{\emph{Telematics and Informatics}}
  \bibinfo{volume}{34}, \bibinfo{number}{2} (\bibinfo{year}{2017}),
  \bibinfo{pages}{457--469}.
\newblock
\showISSN{0736-5853}
\urldef\tempurl%
\url{https://doi.org/10.1016/j.tele.2016.08.005}
\showDOI{\tempurl}


\bibitem[\protect\citeauthoryear{Liao and Shi}{Liao and Shi}{2013}]%
        {Liao13}
\bibfield{author}{\bibinfo{person}{Qinying Liao} {and} \bibinfo{person}{Lei
  Shi}.} \bibinfo{year}{2013}\natexlab{}.
\newblock \showarticletitle{She Gets a Sports Car from Our Donation: Rumor
  Transmission in a Chinese Microblogging Community}. In
  \bibinfo{booktitle}{\emph{Proceedings of the 2013 Conference on Computer
  Supported Cooperative Work}} (San Antonio, Texas, USA)
  \emph{(\bibinfo{series}{CSCW '13})}. \bibinfo{publisher}{Association for
  Computing Machinery}, \bibinfo{address}{New York, NY, USA},
  \bibinfo{pages}{587–598}.
\newblock
\showISBNx{9781450313315}
\urldef\tempurl%
\url{https://doi.org/10.1145/2441776.2441842}
\showDOI{\tempurl}


\bibitem[\protect\citeauthoryear{Liu}{Liu}{2021}]%
        {Liu21}
\bibfield{author}{\bibinfo{person}{Jun Liu}.} \bibinfo{year}{2021}\natexlab{}.
\newblock \showarticletitle{Technology for Activism: Toward a Relational
  Framework}.
\newblock \bibinfo{journal}{\emph{Computer Supported Cooperative Work (CSCW)}}
  (\bibinfo{year}{2021}).
\newblock
\urldef\tempurl%
\url{https://doi.org/10.1007/s10606-021-09400-9}
\showDOI{\tempurl}


\bibitem[\protect\citeauthoryear{Lu, Wang, Lanir, Lanir, Zhao, Pfister,
  Cohen-Or, and Huang}{Lu et~al\mbox{.}}{2020}]%
        {Lu20}
\bibfield{author}{\bibinfo{person}{Min Lu}, \bibinfo{person}{Chufeng Wang},
  \bibinfo{person}{Joel Lanir}, \bibinfo{person}{Joel Lanir},
  \bibinfo{person}{Nanxuan Zhao}, \bibinfo{person}{Hanspeter Pfister},
  \bibinfo{person}{Daniel Cohen-Or}, {and} \bibinfo{person}{Hui Huang}.}
  \bibinfo{year}{2020}\natexlab{}.
\newblock \showarticletitle{Exploring Visual Information Flows in
  Infographics}. In \bibinfo{booktitle}{\emph{Proceedings of the 2020 CHI
  Conference on Human Factors in Computing Systems}}
  \emph{(\bibinfo{series}{CHI '20})}. \bibinfo{publisher}{ACM Press},
  \bibinfo{address}{New York, NY}, \bibinfo{pages}{1--12}.
\newblock
\urldef\tempurl%
\url{https://doi.org/10.1145/3313831.3376263}
\showDOI{\tempurl}


\bibitem[\protect\citeauthoryear{Maddock, Starbird, Al-Hassani, Sandoval,
  Orand, and Mason}{Maddock et~al\mbox{.}}{2015}]%
        {Maddock15}
\bibfield{author}{\bibinfo{person}{Jim Maddock}, \bibinfo{person}{Kate
  Starbird}, \bibinfo{person}{Haneen~J. Al-Hassani}, \bibinfo{person}{Daniel~E.
  Sandoval}, \bibinfo{person}{Mania Orand}, {and} \bibinfo{person}{Robert~M.
  Mason}.} \bibinfo{year}{2015}\natexlab{}.
\newblock \showarticletitle{Characterizing Online Rumoring Behavior Using
  Multi-Dimensional Signatures}. In \bibinfo{booktitle}{\emph{Proceedings of
  the 18th ACM Conference on Computer Supported Cooperative Work \& Social
  Computing}} (Vancouver, BC, Canada) \emph{(\bibinfo{series}{CSCW '15})}.
  \bibinfo{publisher}{Association for Computing Machinery},
  \bibinfo{address}{New York, NY, USA}, \bibinfo{pages}{228–241}.
\newblock
\showISBNx{9781450329224}
\urldef\tempurl%
\url{https://doi.org/10.1145/2675133.2675280}
\showDOI{\tempurl}


\bibitem[\protect\citeauthoryear{Massung, Coyle, Cater, Jay, and
  Preist}{Massung et~al\mbox{.}}{2013}]%
        {Massung13}
\bibfield{author}{\bibinfo{person}{Elaine Massung}, \bibinfo{person}{David
  Coyle}, \bibinfo{person}{Kirsten~F. Cater}, \bibinfo{person}{Marc Jay}, {and}
  \bibinfo{person}{Chris Preist}.} \bibinfo{year}{2013}\natexlab{}.
\newblock \showarticletitle{Using crowdsourcing to support pro-environmental
  community activism}. In \bibinfo{booktitle}{\emph{Proceedings of the SIGCHI
  Conference on Human Factors in Computing Systems}}
  \emph{(\bibinfo{series}{CHI '13})}. \bibinfo{publisher}{ACM Press},
  \bibinfo{pages}{371--380}.
\newblock
\urldef\tempurl%
\url{https://doi.org/10.1145/2470654.2470708}
\showURL{%
\tempurl}


\bibitem[\protect\citeauthoryear{Mckinley and Russonello}{Mckinley and
  Russonello}{2016}]%
        {Mckinley16}
\bibfield{author}{\bibinfo{person}{Angelica Mckinley} {and}
  \bibinfo{person}{Giovanni Russonello}.} \bibinfo{year}{2016}\natexlab{}.
\newblock \bibinfo{title}{Fifty Years Later, Black Panthers' Art Still
  Resonates}.
\newblock
\newblock
\urldef\tempurl%
\url{https://www.nytimes.com/2016/10/16/arts/fifty-years-later-black-panthers-art-still-resonates.html}
\showURL{%
\tempurl}


\bibitem[\protect\citeauthoryear{Mejorado}{Mejorado}{2020}]%
        {Mejorado20}
\bibfield{author}{\bibinfo{person}{Rebekah Mejorado}.}
  \bibinfo{year}{2020}\natexlab{}.
\newblock \bibinfo{booktitle}{\emph{“¡Printing the Revolution! The Rise and
  Impact of Chicano Graphics, 1965 to Now” Unites Historic Civil Rights-Era
  Prints With Works by Contemporary Graphic Artists}}.
\newblock
\urldef\tempurl%
\url{https://americanart.si.edu/press/printing-revolution-rise-and-impact-chicano-graphics}
\showURL{%
Retrieved December 29, 2020 from \tempurl}


\bibitem[\protect\citeauthoryear{Mendonça and Caetano}{Mendonça and
  Caetano}{2021}]%
        {Mendonca21}
\bibfield{author}{\bibinfo{person}{Ricardo~F. Mendonça} {and}
  \bibinfo{person}{Renato~Duarte Caetano}.} \bibinfo{year}{2021}\natexlab{}.
\newblock \showarticletitle{Populism as Parody: The Visual Self-Presentation of
  Jair Bolsonaro on Instagram}.
\newblock \bibinfo{journal}{\emph{The International Journal of Press/Politics}}
  \bibinfo{volume}{26}, \bibinfo{number}{1} (\bibinfo{year}{2021}),
  \bibinfo{pages}{210--235}.
\newblock
\urldef\tempurl%
\url{https://doi.org/10.1177/1940161220970118}
\showDOI{\tempurl}
\showeprint{https://doi.org/10.1177/1940161220970118}


\bibitem[\protect\citeauthoryear{Mitra, Wright, and Gilbert}{Mitra
  et~al\mbox{.}}{2017}]%
        {Mitra18}
\bibfield{author}{\bibinfo{person}{Tanushree Mitra}, \bibinfo{person}{Graham
  Wright}, {and} \bibinfo{person}{Eric Gilbert}.}
  \bibinfo{year}{2017}\natexlab{}.
\newblock \showarticletitle{Credibility and the Dynamics of Collective
  Attention}.
\newblock \bibinfo{journal}{\emph{Proc. ACM Hum.-Comput. Interact.}}
  \bibinfo{volume}{1}, \bibinfo{number}{CSCW}, Article \bibinfo{articleno}{80}
  (\bibinfo{date}{Dec.} \bibinfo{year}{2017}), \bibinfo{numpages}{17}~pages.
\newblock
\urldef\tempurl%
\url{https://doi.org/10.1145/3134715}
\showDOI{\tempurl}


\bibitem[\protect\citeauthoryear{Nittle}{Nittle}{2019}]%
        {Nittle19}
\bibfield{author}{\bibinfo{person}{Nadra~Kareem Nittle}.}
  \bibinfo{year}{2019}\natexlab{}.
\newblock \bibinfo{booktitle}{\emph{History of the Asian American Civil Rights
  Movement}}.
\newblock
\urldef\tempurl%
\url{https://www.thoughtco.com/asian-american-civil-rights-movement-history-2834596}
\showURL{%
Retrieved November 21, 2020 from \tempurl}


\bibitem[\protect\citeauthoryear{Noriega}{Noriega}{[n.d.]}]%
        {Noriega11}
\bibfield{author}{\bibinfo{person}{Chon~A. Noriega}.}
  \bibinfo{year}{[n.d.]}\natexlab{}.
\newblock \bibinfo{title}{The city of dreams...and shoes: Etc. Essay: Chicano
  art – Tate Etc}.
\newblock
\newblock
\urldef\tempurl%
\url{https://www.tate.org.uk/tate-etc/issue-23-autumn-2011/city-dreamsand-shoes}
\showURL{%
\tempurl}


\bibitem[\protect\citeauthoryear{Okamoto}{Okamoto}{2013}]%
        {Okamoto13}
\bibfield{author}{\bibinfo{person}{Dina Okamoto}.}
  \bibinfo{year}{2013}\natexlab{}.
\newblock \showarticletitle{Ethnic Movements}.
\newblock  (\bibinfo{date}{Jan.} \bibinfo{year}{2013}).
\newblock
\urldef\tempurl%
\url{https://doi.org/10.1002/9780470674871.wbespm278}
\showDOI{\tempurl}


\bibitem[\protect\citeauthoryear{Onuch}{Onuch}{2015}]%
        {Onuch15}
\bibfield{author}{\bibinfo{person}{Olga Onuch}.}
  \bibinfo{year}{2015}\natexlab{}.
\newblock \bibinfo{journal}{\emph{Studies in Ethnicity and Nationalism}}
  \bibinfo{volume}{15}, \bibinfo{number}{1} (\bibinfo{date}{April}
  \bibinfo{year}{2015}), \bibinfo{pages}{170--184}.
\newblock
\showISSN{1473-8481}
\urldef\tempurl%
\url{https://doi.org/10.1111/sena.12129}
\showDOI{\tempurl}


\bibitem[\protect\citeauthoryear{Parmelee and Roman}{Parmelee and
  Roman}{2019}]%
        {Parmelee19}
\bibfield{author}{\bibinfo{person}{John~H. Parmelee} {and}
  \bibinfo{person}{Nataliya Roman}.} \bibinfo{year}{2019}\natexlab{}.
\newblock \showarticletitle{Insta-Politicos: Motivations for Following
  Political Leaders on Instagram}.
\newblock \bibinfo{journal}{\emph{Social Media + Society}} \bibinfo{volume}{5},
  \bibinfo{number}{2} (\bibinfo{year}{2019}),
  \bibinfo{pages}{2056305119837662}.
\newblock
\urldef\tempurl%
\url{https://doi.org/10.1177/2056305119837662}
\showDOI{\tempurl}
\showeprint{https://doi.org/10.1177/2056305119837662}


\bibitem[\protect\citeauthoryear{Paulas}{Paulas}{2020}]%
        {Paulas20}
\bibfield{author}{\bibinfo{person}{Rick Paulas}.}
  \bibinfo{year}{2020}\natexlab{}.
\newblock \bibinfo{booktitle}{\emph{The Black Moms Who Occupied a Vacant House
  and Became Icons of the Homelessness Crisis}}.
\newblock
\urldef\tempurl%
\url{https://www.vice.com/en/article/bvgnmm/moms-4-housing-occupied-a-vacant-house-in-oakland-eviction}
\showURL{%
Retrieved January 9, 2021 from \tempurl}


\bibitem[\protect\citeauthoryear{Pavan}{Pavan}{2014}]%
        {Pavan14}
\bibfield{author}{\bibinfo{person}{Elena Pavan}.}
  \bibinfo{year}{2014}\natexlab{}.
\newblock \showarticletitle{Embedding Digital Communications Within Collective
  Action Networks: A Multidimensional Network Approach}.
\newblock \bibinfo{journal}{\emph{Mobilization: An International Quarterly}}
  \bibinfo{volume}{19}, \bibinfo{number}{4} (\bibinfo{year}{2014}),
  \bibinfo{pages}{441–455}.
\newblock
\urldef\tempurl%
\url{https://doi.org/10.17813/maiq.19.4.w24rl524u074126k}
\showDOI{\tempurl}


\bibitem[\protect\citeauthoryear{Pennock}{Pennock}{2017}]%
        {Pennock17}
\bibfield{author}{\bibinfo{person}{Pamela~E. Pennock}.}
  \bibinfo{year}{2017}\natexlab{}.
\newblock \bibinfo{booktitle}{\emph{The Rise of the Arab American Left}}.
\newblock \bibinfo{publisher}{University of North Carolina Press}.
\newblock
\showISBNx{9781469630991}


\bibitem[\protect\citeauthoryear{Pond and Lewis}{Pond and Lewis}{2019}]%
        {Pond19}
\bibfield{author}{\bibinfo{person}{Philip Pond} {and} \bibinfo{person}{Jeff
  Lewis}.} \bibinfo{year}{2019}\natexlab{}.
\newblock \showarticletitle{Riots and Twitter: connective politics, social
  media and framing discourses in the digital public sphere}.
\newblock \bibinfo{journal}{\emph{Information, Communication \& Society}}
  \bibinfo{volume}{22}, \bibinfo{number}{2} (\bibinfo{year}{2019}),
  \bibinfo{pages}{213--231}.
\newblock
\urldef\tempurl%
\url{https://doi.org/10.1080/1369118X.2017.1366539}
\showDOI{\tempurl}
\showeprint{https://doi.org/10.1080/1369118X.2017.1366539}


\bibitem[\protect\citeauthoryear{Reed}{Reed}{2005}]%
        {Reed05}
\bibfield{author}{\bibinfo{person}{T.V. Reed}.}
  \bibinfo{year}{2005}\natexlab{}.
\newblock \bibinfo{booktitle}{\emph{The Art of Protest: Culture and Activism
  from the Civil Rights Movement to the Streets of Seattle}}.
\newblock \bibinfo{publisher}{Thunder’s Mouth Press}.
\newblock
\showISBNx{0816637709}


\bibitem[\protect\citeauthoryear{Rotman, Vieweg, Yardi, Chi, Preece,
  Shneiderman, Pirolli, and Glaisyer}{Rotman et~al\mbox{.}}{2011}]%
        {Rotman11}
\bibfield{author}{\bibinfo{person}{Dana Rotman}, \bibinfo{person}{Sarah
  Vieweg}, \bibinfo{person}{Sarita Yardi}, \bibinfo{person}{Ed Chi},
  \bibinfo{person}{Jenny Preece}, \bibinfo{person}{Ben Shneiderman},
  \bibinfo{person}{Peter Pirolli}, {and} \bibinfo{person}{Tom Glaisyer}.}
  \bibinfo{year}{2011}\natexlab{}.
\newblock \showarticletitle{From slacktivism to activism: participatory culture
  in the age of social media}. In \bibinfo{booktitle}{\emph{Extended Abstracts
  on Human Factors in Computing Systems}} \emph{(\bibinfo{series}{CHI EA
  '11})}. \bibinfo{publisher}{ACM Press}, \bibinfo{pages}{819--822}.
\newblock
\urldef\tempurl%
\url{https://doi.org/10.1145/1979742.1979543}
\showURL{%
\tempurl}


\bibitem[\protect\citeauthoryear{Rudolfo~Anaya}{Rudolfo~Anaya}{2017}]%
        {Aztlan}
\bibfield{author}{\bibinfo{person}{Enrique R.~Lamadrid Rudolfo~Anaya, Francisco
  A.~Lomelí}.} \bibinfo{year}{2017}\natexlab{}.
\newblock \bibinfo{booktitle}{\emph{Aztlán: Essays on the Chicano Homeland.
  Revised and Expanded Edition.}}
\newblock \bibinfo{publisher}{University of New Mexico Press}.
\newblock
\showISBNx{0826356761}


\bibitem[\protect\citeauthoryear{Segerberg and Bennett}{Segerberg and
  Bennett}{2011}]%
        {Segerberg11}
\bibfield{author}{\bibinfo{person}{Alexandra Segerberg} {and}
  \bibinfo{person}{W.~Lance Bennett}.} \bibinfo{year}{2011}\natexlab{}.
\newblock \showarticletitle{Social Media and the Organization of Collective
  Action: Using Twitter to Explore the Ecologies of Two Climate Change
  Protests}.
\newblock \bibinfo{journal}{\emph{The Communication Review}}
  \bibinfo{volume}{14}, \bibinfo{number}{3} (\bibinfo{year}{2011}),
  \bibinfo{pages}{197--215}.
\newblock
\urldef\tempurl%
\url{https://doi.org/10.1080/10714421.2011.597250}
\showDOI{\tempurl}
\showeprint{https://doi.org/10.1080/10714421.2011.597250}


\bibitem[\protect\citeauthoryear{Selander and Jarvenpaa}{Selander and
  Jarvenpaa}{2016}]%
        {Selander16}
\bibfield{author}{\bibinfo{person}{Lisen Selander} {and}
  \bibinfo{person}{Sirkka~L. Jarvenpaa}.} \bibinfo{year}{2016}\natexlab{}.
\newblock \showarticletitle{Digital Action Repertoires and Transforming a
  Social Movement Organization}.
\newblock \bibinfo{journal}{\emph{MIS Q.}} \bibinfo{volume}{40},
  \bibinfo{number}{2} (\bibinfo{date}{June} \bibinfo{year}{2016}),
  \bibinfo{pages}{331–352}.
\newblock
\showISSN{0276-7783}
\urldef\tempurl%
\url{https://doi.org/10.25300/MISQ/2016/40.2.03}
\showDOI{\tempurl}


\bibitem[\protect\citeauthoryear{Shahin and Ng}{Shahin and Ng}{2021}]%
        {Shahin21}
\bibfield{author}{\bibinfo{person}{Saif Shahin} {and} \bibinfo{person}{Yee
  Man~Margaret Ng}.} \bibinfo{year}{2021}\natexlab{}.
\newblock \showarticletitle{Connective action or collective inertia? Emotion,
  cognition, and the limits of digitally networked resistance}.
\newblock \bibinfo{journal}{\emph{Social Movement Studies}}
  \bibinfo{volume}{0}, \bibinfo{number}{0} (\bibinfo{year}{2021}),
  \bibinfo{pages}{1--19}.
\newblock
\urldef\tempurl%
\url{https://doi.org/10.1080/14742837.2021.1928485}
\showDOI{\tempurl}
\showeprint{https://doi.org/10.1080/14742837.2021.1928485}


\bibitem[\protect\citeauthoryear{Siddarth and Pal}{Siddarth and Pal}{2020}]%
        {Siddarth20}
\bibfield{author}{\bibinfo{person}{Divya Siddarth} {and}
  \bibinfo{person}{Joyojeet Pal}.} \bibinfo{year}{2020}\natexlab{}.
\newblock \showarticletitle{Engaging the Crowd: Social Movement Building via
  Online Bystander Mobilization}. In \bibinfo{booktitle}{\emph{Proceedings of
  the 2020 International Conference on Information and Communication
  Technologies and Development}} (Guayaquil, Ecuador)
  \emph{(\bibinfo{series}{ICTD2020})}. \bibinfo{publisher}{Association for
  Computing Machinery}, \bibinfo{address}{New York, NY, USA}, Article
  \bibinfo{articleno}{5}, \bibinfo{numpages}{13}~pages.
\newblock
\showISBNx{9781450387620}
\urldef\tempurl%
\url{https://doi.org/10.1145/3392561.3394633}
\showDOI{\tempurl}


\bibitem[\protect\citeauthoryear{Stanley}{Stanley}{2020}]%
        {Stanley20}
\bibfield{author}{\bibinfo{person}{Phiona Stanley}.}
  \bibinfo{year}{2020}\natexlab{}.
\newblock \showarticletitle{Unlikely hikers? Activism, Instagram, and the queer
  mobilities of fat hikers, women hiking alone, and hikers of colour}.
\newblock \bibinfo{journal}{\emph{Mobilities}} \bibinfo{volume}{15},
  \bibinfo{number}{2} (\bibinfo{year}{2020}), \bibinfo{pages}{241--256}.
\newblock
\urldef\tempurl%
\url{https://doi.org/10.1080/17450101.2019.1696038}
\showDOI{\tempurl}
\showeprint{https://doi.org/10.1080/17450101.2019.1696038}


\bibitem[\protect\citeauthoryear{Starbird, Arif, and Wilson}{Starbird
  et~al\mbox{.}}{2019}]%
        {Starbird19}
\bibfield{author}{\bibinfo{person}{Kate Starbird}, \bibinfo{person}{Ahmer
  Arif}, {and} \bibinfo{person}{Tom Wilson}.} \bibinfo{year}{2019}\natexlab{}.
\newblock \showarticletitle{Disinformation as Collaborative Work: Surfacing the
  Participatory Nature of Strategic Information Operations}.
\newblock \bibinfo{journal}{\emph{Proc. ACM Hum.-Comput. Interact.}}
  \bibinfo{volume}{3}, \bibinfo{number}{CSCW}, Article \bibinfo{articleno}{127}
  (\bibinfo{date}{Nov.} \bibinfo{year}{2019}), \bibinfo{numpages}{26}~pages.
\newblock
\urldef\tempurl%
\url{https://doi.org/10.1145/3359229}
\showDOI{\tempurl}


\bibitem[\protect\citeauthoryear{Starbird and Palen}{Starbird and
  Palen}{2012}]%
        {Starbird12}
\bibfield{author}{\bibinfo{person}{Kate Starbird} {and} \bibinfo{person}{Leysia
  Palen}.} \bibinfo{year}{2012}\natexlab{}.
\newblock \showarticletitle{(How) will the revolution be retweeted?:
  Information diffusion and the 2011 Egyptian uprising}. In
  \bibinfo{booktitle}{\emph{Proceedings of the ACM 2012 conference on Computer
  Supported Cooperative Work}} \emph{(\bibinfo{series}{CSCW '12})}.
  \bibinfo{publisher}{ACM Press}, \bibinfo{address}{New York, NY},
  \bibinfo{pages}{7--16}.
\newblock
\urldef\tempurl%
\url{https://doi.org/10.1145/2145204.2145212}
\showDOI{\tempurl}


\bibitem[\protect\citeauthoryear{Sudbanthad}{Sudbanthad}{2008}]%
        {Sudbanthad08}
\bibfield{author}{\bibinfo{person}{Pitchaya Sudbanthad}.}
  \bibinfo{year}{2008}\natexlab{}.
\newblock \bibinfo{booktitle}{\emph{Emory Douglas}}.
\newblock
\urldef\tempurl%
\url{https://www.aiga.org/design-journeys-emory-douglas}
\showURL{%
Retrieved December 29, 2020 from \tempurl}


\bibitem[\protect\citeauthoryear{Suwana}{Suwana}{2020}]%
        {Suwana20}
\bibfield{author}{\bibinfo{person}{Fiona Suwana}.}
  \bibinfo{year}{2020}\natexlab{}.
\newblock \showarticletitle{Digital Activism in Bali: The ForBALI Movement}. In
  \bibinfo{booktitle}{\emph{Security, Democracy, and Society in Bali}}
  \emph{(\bibinfo{series}{Vandenberg A., Zuryani N.})}.
  \bibinfo{publisher}{Palgrave Macmillan, Singapore},
  \bibinfo{pages}{253--284}.
\newblock
\urldef\tempurl%
\url{https://doi.org/10.1007/978-981-15-5848-1_11}
\showURL{%
\tempurl}


\bibitem[\protect\citeauthoryear{Theocharis, Lowe, van Deth, and
  García-Albacete}{Theocharis et~al\mbox{.}}{2015}]%
        {Theocharis15}
\bibfield{author}{\bibinfo{person}{Yannis Theocharis}, \bibinfo{person}{Will
  Lowe}, \bibinfo{person}{Jan~W. van Deth}, {and} \bibinfo{person}{Gema
  García-Albacete}.} \bibinfo{year}{2015}\natexlab{}.
\newblock \showarticletitle{Using Twitter to mobilize protest action: online
  mobilization patterns and action repertoires in the Occupy Wall Street,
  Indignados, and Aganaktismenoi movements}.
\newblock \bibinfo{journal}{\emph{Information, Communication \& Society}}
  \bibinfo{volume}{18}, \bibinfo{number}{2} (\bibinfo{year}{2015}),
  \bibinfo{pages}{202--220}.
\newblock
\urldef\tempurl%
\url{https://doi.org/10.1080/1369118X.2014.948035}
\showDOI{\tempurl}
\showeprint{https://doi.org/10.1080/1369118X.2014.948035}


\bibitem[\protect\citeauthoryear{Tillery}{Tillery}{2019}]%
        {Tillery19}
\bibfield{author}{\bibinfo{person}{Alvin~B. Tillery}.}
  \bibinfo{year}{2019}\natexlab{}.
\newblock \showarticletitle{What Kind of Movement is Black Lives Matter? The
  View from Twitter}.
\newblock \bibinfo{journal}{\emph{The Journal of Race, Ethnicity, and
  Politics}} \bibinfo{volume}{4}, \bibinfo{number}{2} (\bibinfo{year}{2019}),
  \bibinfo{pages}{297–323}.
\newblock
\urldef\tempurl%
\url{https://doi.org/10.1017/rep.2019.17}
\showDOI{\tempurl}


\bibitem[\protect\citeauthoryear{Tufekci}{Tufekci}{2017}]%
        {Tufekci17}
\bibfield{author}{\bibinfo{person}{Zeynep Tufekci}.}
  \bibinfo{year}{2017}\natexlab{}.
\newblock \bibinfo{booktitle}{\emph{Twitter and Tear Gas}}.
\newblock \bibinfo{publisher}{Yale University Press}.
\newblock
\urldef\tempurl%
\url{https://doi.org/doi:10.12987/9780300228175}
\showDOI{\tempurl}


\bibitem[\protect\citeauthoryear{Twyman, Keegan, and Shaw}{Twyman
  et~al\mbox{.}}{2017}]%
        {Twyman16}
\bibfield{author}{\bibinfo{person}{Marlon Twyman}, \bibinfo{person}{Brian~C.
  Keegan}, {and} \bibinfo{person}{Aaron Shaw}.}
  \bibinfo{year}{2017}\natexlab{}.
\newblock \showarticletitle{Black Lives Matter in Wikipedia: Collaboration and
  Collective Memory around Online Social Movements}. In
  \bibinfo{booktitle}{\emph{Proceedings of the 2017 ACM Conference on Computer
  Supported Cooperative Work and Social Computing}}
  \emph{(\bibinfo{series}{CSCW '17})}. \bibinfo{publisher}{ACM Press},
  \bibinfo{address}{New York, NY}, \bibinfo{pages}{1400--1412}.
\newblock
\urldef\tempurl%
\url{https://doi.org/10.1145/2998181.2998232}
\showDOI{\tempurl}


\bibitem[\protect\citeauthoryear{Vaidya, Votipka, Mazurek, and Sherr}{Vaidya
  et~al\mbox{.}}{2019}]%
        {Tavish19}
\bibfield{author}{\bibinfo{person}{Tavish Vaidya}, \bibinfo{person}{Daniel
  Votipka}, \bibinfo{person}{Michelle~L. Mazurek}, {and} \bibinfo{person}{Micah
  Sherr}.} \bibinfo{year}{2019}\natexlab{}.
\newblock \bibinfo{booktitle}{\emph{Does Being Verified Make You More Credible?
  Account Verification's Effect on Tweet Credibility}}.
\newblock \bibinfo{publisher}{Association for Computing Machinery},
  \bibinfo{address}{New York, NY, USA}, \bibinfo{pages}{1–13}.
\newblock
\showISBNx{9781450359702}
\urldef\tempurl%
\url{https://doi.org/10.1145/3290605.3300755}
\showURL{%
\tempurl}


\bibitem[\protect\citeauthoryear{Wang, Sundin, Murray-Rust, and Bach}{Wang
  et~al\mbox{.}}{2020}]%
        {Wang20}
\bibfield{author}{\bibinfo{person}{Zezhong Wang}, \bibinfo{person}{Lovisa
  Sundin}, \bibinfo{person}{Dave Murray-Rust}, {and} \bibinfo{person}{Benjamin
  Bach}.} \bibinfo{year}{2020}\natexlab{}.
\newblock \showarticletitle{Cheat Sheets for Data Visualization Techniques}. In
  \bibinfo{booktitle}{\emph{Proceedings of the 2020 CHI Conference on Human
  Factors in Computing Systems}} \emph{(\bibinfo{series}{CHI '20})}.
  \bibinfo{publisher}{ACM Press}, \bibinfo{address}{New York, NY},
  \bibinfo{pages}{1--13}.
\newblock
\urldef\tempurl%
\url{https://doi.org/10.1145/3313831.3376271}
\showDOI{\tempurl}


\bibitem[\protect\citeauthoryear{Wei}{Wei}{1993}]%
        {Wei93}
\bibfield{author}{\bibinfo{person}{William Wei}.}
  \bibinfo{year}{1993}\natexlab{}.
\newblock \bibinfo{booktitle}{\emph{The Asian American Movement}}.
\newblock \bibinfo{publisher}{Temple University Press}.
\newblock
\showISBNx{1439903743}


\bibitem[\protect\citeauthoryear{Wipf and Plante}{Wipf and Plante}{2020}]%
        {WipfLaPlante20}
\bibfield{author}{\bibinfo{person}{Carly Wipf} {and}
  \bibinfo{person}{Mauricio~La Plante}.} \bibinfo{year}{2020}\natexlab{}.
\newblock \bibinfo{booktitle}{\emph{San Jose Mayor Sam Liccardo drops his
  ‘strong mayor’ plan}}.
\newblock
\urldef\tempurl%
\url{https://sanjosespotlight.com/san-jose-mayor-sam-liccardo-drops-his-strong-mayor-plan/}
\showURL{%
Retrieved January 11, 2021 from \tempurl}


\end{thebibliography}

\appendix

\section{Figure Content Description} In this section, we include content descriptions for figure images. 

\subsection{Figure 1} 

\small Left to right: (1) \textit{Text:} COVID-19 Financial Relief. While Congress did nothing for 6 months, here's what the rest of the world is doing: Money distributed by countries during the pandemic: Japan: Up to 100\% of your salary. Norway: Up to 90\% of your salary. Germany: Up to 87\% of your salary. France: Up to 84\% of your salary. U.K.: Up to 80\% of your salary. Italy: Up to 80\% of your salary. Canada: \$2000 a month since April. U.S.A.: \$700 one-time payment, first relief in 6 months. \textit{Visuals:} Teal and yellow table with a row for each country. (2) \textit{Text:} Build Black. Buy Black. Bank Black. \textit{Visuals:} Black background and white text, once in bold, filled text and once in white-outlined text. Accompanied by a yellow shopping card icon. (3) \textit{Text:} The fight for trans Black Lives is not over. Let's continue to rise up! \textit{Visuals: } Text is white and outlined in black. The background is comprised of pink, blue, and white wavy rows. (4) \textit{Text: } The Black Panther Party. Founded October 1966. What did they give us? \textit{Visuals: } The background includes a picture of a black panther in a white circle over a yellow-tinted image of a Black Panther protest. \cite{} (5) \textit{Text: } Black Women \& Femmes Are Divine: A Reclamation in the Name of Breonna Taylor. Saturday, 10/3 12PM PST. \textit{Visuals: } A pink-tinted image of four women/femmes in white T-shirts. (6) \textit{Text: } BREATHE Day. IG Take Over. From the Streets to Congress: A BREATHE Day conversation with Jamaal Bowman. Monifa Bandele, Jamaal Bowman (jamaalbowmanny). September 29th 11:00AM ET. The BREATHE Act. M4BL. \textit{Visuals: } Orange, cyan, and purple patterns in the background. Images of Bowman and Bandele in the center. (7) \textit{Text: } Seven Years of Growth: BLM's Founder and incoming Executive Director reflects on the movement. \textit{Visuals: } The background is a purple, red, and yellow tinted collage of posters and protesting individuals, including a Black Lives Matter Sign and a BLM sign with a yellow fist. (8) \textit{Text: } Here is how we \#DefundThePolice. First. Demand that lawmakers support reparations for all families of those killed and survivors of police violence. Second. Demand that State, city and mincipality spend less on law enforcement and incarceration. Period. Finally. Demand investment into Black communities. It is not enough to defund the police, we need to put in place systems to uplift and protect. That means divesting from: police in schools, criminalizing mental health, military weapons against citizens and investing in teachers \& counselors, mental health \& restorative services, community-led harm reduction. \textit{Visuals: } Yellow and black background with black, white, yellow text throughout. Headers have brushstroke-like highlighting. (9) \textit{Text: } Call to Action. Continue sharing her story, posting her face, and saying her name. PLEASE USE \#BreonnaTaylor \& \#JusticeForBre. Tag all responsible parties. [Twitter:] LMPD, MayorGregFisher, GovAndyBeshear. [Instagram:] LMPD.KY, MayorGregFisher, GovAndyBeshear. Make calls \& send emails for Breonna to the investigative agencies, institutions and individuals in charge and make the demands known!. FightForBreonna.org. \textit{Visuals: } Yellow box outline with white and yellow text. The background is a black-and-white image of a protest. (10) \textit{Text: } Crisis Response: Sign the Petitions. We need more. We demand more. \textit{Visuals: } The text is in bold white and "Sign the Petitions" has a blue highlight. The background is black with red coronavirus speckled throughout. (11) \textit{Text: } We celebrate the Life and Legacy of Martin Luther King Jr. We will continue to fight. \textit{Visuals: } Gray grainy background with a yellow circle. Two black and white pictures that appear to be taped on. The left one is of MLK at a gathering and the right one is of a protest. (12) \textit{Text: } Black Futures Month. Global Pop Up: Torontoa, Atlanta, Los Angeles. February 17-March 3. Featuring Karen Miranda Augustine, Rodney Diverlus, Jasmyn Fyffe, Kim Ninkuru, Kike Otuije, Pete Owusu-Ansah, Gloria Swain, Diseiye Thompson, Kyisha Williams, Ravyn Wngz. February 17, 2019 7-10 PM Opening Reception ASL Provided. Access Info: There are three steps to the entrance. ASL provided for opening event. 76 Geary Ave. infoblacklivesmatter.ca. \textit{Visuals: } White background with black patterned and green-shaded borders. Green highlighting, circles, and text accents. (13) \textit{Text: } What Issues Matter to You? \#WhatMattersin2020. \textit{Visuals: } Dark gray image of protesters in background with poster that says "Demilitarize the Police." "Matter" is in yellow cursive and text is in white and yellow. (14) \textit{Text: } Court Support Needed!! BLMLA Organizer, Melina Abdullah is being prosecuted for her activism. When they come for one of us, they come for all of us. Join us!! Friday 9/28 8:30 AM 210 W. Temple St. (Criminal Courts Building) Dept. 48. **Wear your \#BlackLivesMatter gear, if you have it.** \textit{Visuals: } Background is a black and white image of Abdullah with a megaphone along with yellow, black, and white patterns and shapes. (15) \textit{Text: } End Prison Slavery! Nationwide prison strike Aug 21-Sept 9, 2018. \textit{Visuals: } The head and shoulders of two men is pictured in front of a pink, orange, and yellow gradient. In front, a smaller individual holds a flag reading "Strike" atop a layer of orange and pink roses. (16) \textit{Text: } \#SayHerName. A week of action to end violence against Black women and girls. June 11th - 17th. bit.ly/sayhername2018. \textit{Visuals: } Dark gray background with yellow outlined circle enclosing the text. Word accents in green and red. A protester with a megaphone is pictured at the bottom. (17) \textit{Text: } APTP/BLM Sacramento and Justice Network Call For A National Day of Action. 50th Anniversary of MLK's Assasination. DA's Office, 901 G ST, Wed Apr 4, 3-5:30. We have nothing to lose but our chains!. We love you Stephon. The time is always right to do what is right. \textit{Visuals:} Purple background, text in yellow and white. Graphical yellow and white images of Assata Shakur, Stephon, and MLK with the quotes listed respectively. (18) \textit{Text: } Flint. Flint will have clean and free water. There will be a day when watter warriors will happily retire. \textit{Visuals: } An individual with white face paint and a turquoise shirt stands in front of a wall of the "Flint will have clean and free water" quote repeated in white text, with "Flint" listed large and in white at the top. (19) \textit{Text: } Beyond the Moment. Day of Political Education. National Prep Call. Tues. March 28th 8PM EST / 5PM PST. Join a broad mutli-racial, cross-sector coalition of organizers from around the country as we share plans for a mass day of poltiical education that will fall ont he 50th anniversary of Dr. King's "Beyond Vietnam" speec on april 4th. Find out how to plug in and organize your own. \textit{Visuals: } The "Beyond the Moment" header is slanted and a black and white image of protesters with the American flag is at the top. Background is gray and turquoise. (20) \textit{Text: } \$40K in 4 Days. For \#4YearsofBLM. Please donate at celebrateBLM.com. "When Black people get free, we all get free." \textit{Visuals: } Large black and gray text atop a black and white image of a protest.

\subsection{Figure 2} Left to right: (1) \textit{Text: } Allyship is a constant process of actions, not beliefs. \textit{Visuals: } Black background with pink bold text. (2) \textit{Text: } Use Your Platform. However large or small, you have an opportunity to reach people. Use it. Share news stories, and updates---especially if your timeline is not very diverse. Work to diversify your timeline. Raise awareness amongst those who may otherwise not see. Understand that your voices can travel to places that ours can't. But, remember---posting is not taking action. Emails, phone calls, letters, donations, conversations are actions. \textit{Visuals: } Pink background with white and black bold text. (3) \textit{Text: } What to Say When People Deny the Reality of What's Happening Right Now Part 2. Another quick guide of possible responses to common remarks made in regard to recent6 protests sparked by George Floyd's murder, most of which are made by, you guessed it, white people. \textit{Visuals: } Beige background with black text. (4) \textit{Text: } 7 Tips to Maintain Momentum \& Stay Involved. Get Involved and Wield Your Dollar. Protest \& Aid Protestors. Help Vulnerable Population and the Homeless. Stay Connected to the Movement, Organizations, and Activists. Support Black Businesses, Friends, Family, and Colleagues. Use What You're Good At. Hold Racists and Racist Institutions Accountable. \textit{Visuals: } Overlayed dark gray, green, and maroon squares. White text.

\subsection{Figure 3} Left to right: (1) \textit{Text: } You can spell pretty without exotic. \textit{Visuals: } Peach background with an eye in the center of the text. Text in maroon, brown, red, teal, and yellow. (2) So... Sanrio's your new aesthetic? \textit{Visuals: } Pink background. Image of Hello Kitty picking cherries from a tree. Text is in red with yellow highlight. (3) \textit{Text: } Stop fetishizing Asians. \textit{Visuals: } Peach and beige waves in the background. A brown silhouette of a person accompanies the text in brown. (4) \textit{Text: } Voguish. Problematic. \textit{Visuals: } Peach background. Two eyes framed by two cupped hands. Yellow flowers at the top. "Problematic" is enclosed in brackets and red and "Voguish" is in mild yellow at the top. (5) \textit{Text: } They can't burn us all. \textit{Visuals: } Three individuals with fists raised frame the text, which is black with yellow highlight. Red background. (6) \textit{Text: } Trump is banning WeChat. \textit{Visuals: } Mild yellow background with a person holding a phone. The phone has an icon of Trump and an icon of the WeChat logo. Text is in dark red. 

\subsection{Figure 5} Four images are on the left and four are on the right. For the 2014 images on the left, from left to right: (1) A crowd is seated in a church. (2) Two individuals hold a poster with a person's face and the text "My Life Matters." (3) An individual is being arrested on a street, surrounded by police. (4) An individual is painting art on concrete. For the 2020 images on the right, from left to right: (1) Black Voters Showed Up! This time in Georgia! Thank you for your work: Fair Fight and New Georgia Project! (2) \textit{Text: } Black Women \& Femmes Are Divine: A Reclamation in the Name of Breonna Taylor. Saturday, 10/3 12PM PST. \textit{Visuals: } A pink-tinted image of four women/femmes in white T-shirts. (3) \textit{Text: } BREATHE Day. IG Take Over. From the Streets to Congress: A BREATHE Day conversation with Jamaal Bowman. Monifa Bandele, Jamaal Bowman (jamaalbowmanny). September 29th 11:00AM ET. The BREATHE Act. M4BL. \textit{Visuals: } Orange, cyan, and purple patterns in the background. Images of Bowman and Bandele in the center. (4) \textit{Text: } Seven Years of Growth: BLM's Founder and incoming Executive Director reflects on the movement. \textit{Visuals: } The background is a purple, red, and yellow tinted collage of posters and protesting individuals, including a Black Lives Matter Sign and a BLM sign with a yellow fist.

\subsection{Figure 6} Left to right: (1) \textit{Text: } Free Angela! Free All of Us! Angela Davis Finally Acquitted of False Charges. Vote for Survival: Bobby Seale for Mayor! Elaine Brown for City Council! \textit{Visuals: } An image of Davis's face on the right and Davis standing on the left. Bright pink background with bold black text \cite{PrattLibrary}. (2) \textit{Text: } Murdered. Sam Napier. One word is a thousand words to a thousand ears. Sam Napier spread the people's word. Even now we hear it resound ten thousand times. "Circulate to Educate." Sam Napier, Black Panther Party, Intercommunal News Service, Circulation Manager, Murdered by Fascists. \textit{Visuals: } "Murdered" is in bold text at the top. Napiers face is pictured large in the center with rays extending from it. Light blue background \cite{Libcom}. (3) \textit{Text: } Caustion: Surviving is Criminal: Events in Two Black Men's Lives Dramatize Why. Prison Camps U.S.A. A-000000000. The Unknown Slaves. \textit{Visuals: } Two heads appear with shackles about the neck. Title is in blue text and Prison Camp text appears as a label at the bottom \cite{SummerLove17}. (4) \textit{Text: } Our Fight is Not in Vietnam. Free the GI's. A crying individual appears with a helmet and weaponry. In their helmet is an image of war and destruction. The text is on a protest sign \cite{Mckinley16}. (5) \textit{Text: } Afro-American solidarity with the oppressed People of the world. \textit{Visuals: } An individual with a spear and gun in a pink and teal jacket stares forward. The background is comprised of pink and gray alternating rays \cite{Jones20}. (6) \textit{Text:} Free the NY 21. And All Political Prisoners. \textit{Visuals: } Orange background. A picture of protesters with fists raised lines the bottom of the image \cite{Jones20}. (7) \textit{Text: } Political Prisoners of USA Fascism. \textit{Visuals: } Two individuals in leather jackets, and one with a gun stand \cite{Jones20}. The text is in bold blue. (8) \textit{Text: } Free Angela. \textit{Visuals: } A headshot of Davis speaking. Text is in green \cite{Jones20}.

\subsection{Figure 8} Left to right: (1) Farmworkers stand distraught behind barbed wire in front of farmhouses  \cite{Noriega11}. (2) Large text reads "We are NOT a minority!!" An individual to the right of the text points to the viewer. Background is a blue sky over a horizon with greenery at the bottom  \cite{Acosta}.

\end{document}